\documentclass[runningheads]{llncs}

\usepackage{url}
\usepackage[pdftex]{graphicx}
\usepackage{amsmath,amssymb,xcolor,wrapfig,paralist}
\usepackage{algorithmicx,algorithm}
\usepackage[noend]{algpseudocode}
\usepackage{multirow}
\usepackage{bbm}
\usepackage{comment}
\usepackage{csquotes}
\usepackage{subcaption}
\usepackage[section]{placeins}

\usepackage{hyperref}
\hypersetup{
    colorlinks=true,
    linkcolor=blue,
    urlcolor=blue,
    }

\usepackage{mathtools}  
\usepackage{xparse}  
\usepackage{xifthen}  

\usepackage{booktabs}
\usepackage{proof}
\usepackage{tikz}
\usepackage{colortbl}
\usepackage{pdflscape}
\usetikzlibrary{shapes,positioning,arrows,calc,automata,matrix,fit}
\tikzstyle{state}+=[minimum size = 8mm, inner sep=0,outer sep=1]
\tikzset{->,>=stealth'}

\definecolor{wwhite}{gray}{1}
\usepackage[T1]{fontenc}

\usepackage{tabularx}
\newcolumntype{L}{>{\raggedright\arraybackslash}p{1.6cm}}
\newcolumntype{C}{>{\centering\arraybackslash}p{1.6cm}}
\newcolumntype{R}[1]{>{\raggedleft\arraybackslash}p{#1}}
\allowdisplaybreaks

\newcommand{\thmhelperpre}[2]{\newcommand{\theoremlike}[1]{\par\medskip\penalty-250\refstepcounter{theorem}{\bfseries\noindent##1 \ref{#1}.}\itshape}\theoremlike{#2}}
\newcommand{\thmhelperpost}{\par\medskip%
 \renewcommand{\theoremlike}[1]{\par\medskip\penalty-250\refstepcounter{theorem}{\bfseries\noindent##1 \thesection .\thetheorem.}\itshape}%
}

\newcommand{\track}[1]{{\textcolor{red}{#1}}}

\newcommand{\todo}[1]{
\begin{tikzpicture}[remember picture, baseline=-0.75ex]
\node [coordinate] (inText) {};
\end{tikzpicture}
\marginpar{
\begin{tikzpicture}[remember picture, font=\scriptsize]
\draw node[draw=red, text width = 3.5cm, inner sep=0.3mm] (inNote){#1};
\end{tikzpicture}
}
\begin{tikzpicture}[remember picture, overlay]
\draw[draw]
([yshift=-0.2cm] inText)
-| (inNote.south);
\end{tikzpicture}}

\renewcommand{\todo}[1]{}

\newcommand{\rewformat}[1]{\textbf{#1}}

\renewcommand{\algorithmicrequire}{\textbf{Input: }}
\renewcommand{\algorithmicensure}{\textbf{Output: }}
\newcommand{\qee}{\hfill$\triangle$} 
\newcommand{\highlight}[1]{\colorbox{black!15}{$\displaystyle#1$}}
\newcommand{\zug}[1]{( #1  )}
\newcommand{\stam}[1]{}
\newcommand{\realpos}{\mathbb{R}_{\geq 0}}
\newcommand{\unif}{\Mdp_{C}}

\newcommand{\pr}{\mathbb P}
\newcommand{\expected}{\mathbb E}

\newcommand{\pmin}{p_{\mathsf{min}}}

\newcommand{\upperbound}{u}
\newcommand{\lowerbound}{l}
\newcommand{\lu}{\lowerbound,\upperbound}

\newcommand{\markovChain}{\ensuremath{\mathcal{M}} }

\newcommand{\UPDATE}{\mathsf{UPDATE}}
\newcommand{\SIMULATE}{\mathsf{SIMULATE}}
\newcommand{\STUCK}{\mathsf{LOOPING}}
\newcommand{\SUREEC}{\mathsf{\deltasure~EC}}
\newcommand{\INITIALIZE}{\mathsf{INITIALIZE\_VI\_BOUNDS}}

\newcommand{\simulateMecStandard}{\mathsf{SIMULATE\_MEC\_STANDARD}}
\newcommand{\simulateMecDirAccess}{\mathsf{SIMULATE\_MEC\_DIRECT\_ACCESS}}
\newcommand{\simulateMecHeuristic}{\mathsf{SIMULATE\_MEC}}

\newcommand{\tsucc}{\mathsf{t}}

\renewcommand{\path}{\pi}

\newcommand{\deltasure}{\delta_{TP}\textit{-sure}}

\newcommand{\Naturals}{\mathbb{N}}
\DeclarePairedDelimiter{\delimfun}{(}{)}
\NewDocumentCommand{\fun}{smm}{\IfBooleanTF{#1}{{#2}\delimfun{#3}}{{#2}\delimfun*{#3}}}
\NewDocumentCommand{\funMacro}{smm}{\IfNoValueTF{#3}{#1}{\fun{#2}{#3}}}
\NewDocumentCommand{\abs}{sm}{\IfBooleanTF{#1}{\delimabs{#2}}{\delimabs*{#2}}}
\DeclarePairedDelimiter{\delimabs}{\lvert}{\rvert}
\DeclarePairedDelimiter{\delimset}{\lbrace}{\rbrace}

\newcommand{\Reals}{\mathbb{R}}
\newcommand{\Distributions}{\mathcal{D}}
\NewDocumentCommand{\distributions}{d()}{\funMacro{\mathcal{D}}{#1}}
\DeclareMathOperator{\supp}{supp}
\newcommand{\Mdp}{\mathcal{M}}

\newcommand{\unionSym}{\cup}
\newcommand{\unionBin}{\mathbin{\unionSym}}
\newcommand{\union}{\unionBin}
\newcommand{\intersectionSym}{\cap}
\newcommand{\intersectionBin}{\mathbin{\intersectionSym}}
\newcommand{\intersection}{\intersectionBin}

\newcommand{\UnionSym}{\bigcup}
\newcommand{\Union}{\UnionSym}
\NewDocumentCommand{\set}{sm}{\IfBooleanTF{#1}{\delimset{#2}}{\delimset*{#2}}}

\DeclareDocumentCommand{\states}{D<>{} O{}  t'}{\mathsf{S}_{#1}^{\IfBooleanTF{#3}{\prime~#2}{#2}}}
\DeclareDocumentCommand{\state}{D<>{} O{}  t'}{\mathsf{s}_{#1}^{\IfBooleanTF{#3}{\prime #2}{#2}}}
\DeclareDocumentCommand{\action}{D<>{} O{} t'}{\mathsf{a}_{#1}^{\IfBooleanTF{#3}{\prime#2}{#2}}}
\DeclareDocumentCommand{\trans}{D<>{} O{} t' D(){} D(){}}{\mathbb{T}{#1}^{\IfBooleanTF{#3}{\prime}{}#2}\ifthenelse{\isempty{#4}}{}{(#4)}\ifthenelse{\isempty{#5}}{}{(#5)}}
\DeclareDocumentCommand{\Av}{D<>{} O{} t' D(){}}{\mathsf{Av}_{#1}^{\IfBooleanTF{#3}{\prime}{}#2}\ifthenelse{\isempty{#4}}{}{(#4)}}
\newcommand{\transR}{\mathsf{R}}
\DeclareDocumentCommand{\post}{D<>{} O{} D(){}}{\mathsf{Post}_{#1}^{#2}\ifthenelse{\isempty{#3}}{}{(#3)}}
\newcommand{\initstate}{s_\textsf{init}}
\newcommand{\av}{\mathsf{Av}}
\newcommand{\rew}{r}
\NewDocumentCommand{\actions}{d()}{{\IfNoValueTF{#1}{\mathsf{Act}}{\fun{\mathsf{Act}}{#1}}}}
\newcommand{\mec}{\mathsf{MEC}}

\DeclareDocumentCommand{\ub}{D<>{} O{}  D(){} t'}{\mathsf{U}_{#1}^{\IfBooleanTF{#4}{\prime #2}{#2}}\ifthenelse{\isempty{#3}}{}{(#3)}}
\DeclareDocumentCommand{\lb}{D<>{} O{}  D(){} t'}{\mathsf{L}_{#1}^{\IfBooleanTF{#4}{\prime #2}{#2}}\ifthenelse{\isempty{#3}}{}{(#3)}}
\newcommand{\transdelta}{\delta_{\trans}}
\newcommand{\stepsUntilSure}{\textit{requiredSamples}}
\DeclareMathOperator{\leaves}{\mathbin{\mathop{exits}}}
\newcommand{\rmax}{r_{\max}}

\newcommand*{\infpath}{\rho}
\newcommand{\fpath}{w}
\newcommand{\straa}{\pi}
\newcommand{\straas}{\Pi}
\newcommand{\gain}{v} 

\colorlet{safecellcolor}{yellow!5}                                                                     \colorlet{goodcellcolor}{green!10}                                                                     \colorlet{badcellcolor}{blue!10}

\newcommand{\tpImprecision}{\text{TP-imprecision }}
\newcommand{\mpImprecision}{\text{MP-imprecision }}
\newcommand{\tpInconfidence}{\text{TP-inconfidence }}
\newcommand{\mpInconfidence}{\text{MP-inconfidence }}

\newcommand{\symbolTpImprecision}{\varepsilon_{TP}}
\newcommand{\symbolMpImprecision}{\varepsilon_{MP}}
\newcommand{\symbolTpInconfidence}{\delta_{TP}}
\newcommand{\symbolMpInconfidence}{\delta_{MP}}

\newcommand{\alphaR}{\alpha_{R}}
\newcommand{\deltaR}{\delta_{R}}
\newcommand{\deltaMPOne}{\delta_{MP1}}
\newcommand{\deltaMPTwo}{\delta_{MP2}}

\begin{document}

\title{
PAC Statistical Model Checking of Mean Payoff in Discrete- and Continuous-Time MDP\thanks{This work has been partially supported by the DST-SERB project SRG/2021/000466 \emph{Zero-sum and Nonzero-sum Games for Controller Synthesis of Reactive Systems} and by the German Research Foundation (DFG) projects 427755713 (KR 4890/3-1) \emph{Group-By Objectives in Probabilistic Verification (GOPro)} and 383882557 (KR 4890/2-1) \emph{Statistical Unbounded Verification (SUV)}.}
}

\author{Chaitanya Agarwal\inst{1} \and
Shibashis Guha\inst{2}
\and
Jan Křetínský\inst{3} \and
M. Pazhamalai\inst{4}}

\authorrunning{C. Agarwal et al.}

\institute{New York University, New York, USA \and
Tata Institute of Fundamental Research, Mumbai, India \and
Technical University of Munich, Munich, Germany \and
Chennai Mathematical Institute, Chennai, India}

\maketitle


\begin{abstract}
Markov decision processes (MDP) and continuous-time MDP (CTMDP) are the fundamental models for non-deterministic systems with probabilistic uncertainty. Mean payoff (a.k.a. long-run average reward) is one of the most classic objectives considered in their context. We provide the first algorithm to compute mean payoff probably approximately correctly in unknown MDP; further, we extend it to unknown CTMDP. We do not require any knowledge of the state space, only a lower bound on the minimum transition probability, which has been advocated in literature. In addition to providing probably approximately correct (PAC) bounds for our algorithm, we also demonstrate its practical nature by running experiments on standard benchmarks.
\end{abstract}

\section{Introduction} \label{sec:intro}
\emph{Markov decision process} (MDP) \cite{Puterman,Bertsekas95,Sennott99} is a basic model for systems featuring both probabilistic and non-deterministic behaviour.
They come in two flavours: \emph{discrete-time} MDP (often simply MDP) and \emph{continuous-time} MDP (CTMDP).
While the evolution of MDP happens in discrete steps, their natural real-time extension CTMDP additionally feature random time delays governed by exponential probability distributions. 
Their application domain ranges across a wide spectrum, e.g.
operations research~\cite{DBLP:journals/jacm/BrunoDF81,Feinberg04}, power management and scheduling~\cite{DBLP:journals/tcad/QiuQP01}, networked and distributed systems~\cite{DBLP:conf/srds/HaverkortHK00,DBLP:conf/sosp/GhemawatGL03}, or communication protocols \cite{DBLP:conf/qest/KwiatkowskaNP12}, to name a few.
One of the key aspects of such systems is their performance, often formalized as \emph{mean payoff} (also called long-run average reward), one of the classic and most studied objectives on (CT)MDP \cite{Puterman} with numerous applications \cite{feinberg2012handbook}.
In this context, probabilistic model checking and performance evaluation intersect \cite{DBLP:journals/cacm/BaierHHK10}.
While the former takes the verification perspective of the worst-case analysis and the latter the perspective of optimization for the best case, they are mathematically dual and thus algorithmically the same.

The range of analysis techniques provided by literature is very rich, encompassing linear programming, policy iteration, or value iteration. 
However, these are applicable only in the setting where the (CT)MDP is known (\emph{whitebox} setting).
In order to handle the \emph{blackbox} setting, where the model is unknown or only partially known, \emph{statistical model checking} (SMC) \cite{DBLP:conf/cav/YounesS02} relaxes the requirement of the hard guarantees on the correctness (claimed precision) of the result.
Instead it replaces it with \emph{probably approximately correct} (PAC) analysis, which provides essentially a confidence interval on the result: with probability (confidence) at least $1-\delta$, the result of the analysis is $\varepsilon$-close to the true value.
This kind of analysis may be applicable to those systems for which we do not have exclusive access to their internal functionalities, but we can still observe their behaviour.

In this paper, we provide the \emph{first algorithm with PAC bounds on the mean payoff in blackbox MDP}.
We treat both the discrete-time and continuous-time MDP, and the SMC algorithm not only features PAC bounds (returning the result with prescribed precision and confidence), but is also in the form of an anytime algorithm (gradually improving the result and, if terminated prematurely, can return the current approximation with its precision and the current confidence).

The difficulty with blackbox models is that we do not know the exact transition probabilities, not even the number of successors for an action from a state.
The algorithm thus must simulate the MDP to obtain any information.
The visited states can be augmented to a model of the MDP and statistics used to estimate the transition probabilities.
The estimates can be used to compute mean payoff precisely on the model.
The results of \cite{Chatterjee12} and \cite{Solan03} then provide a method for estimating the number of times each state-action pair needs to be visited in an MDP to obtain a PAC bound on the expected mean-payoff value of the original MDP.
However, notice  that this requires that the topology be learnt perfectly, for which we either need some knowledge of the state space or recent development in the spirit of \cite{AKW19}.
On the one hand, this simple algorithm thus follows in a straightforward way from the recent results in the literature (although to the best of our knowledge it has not been presented as such yet).
On the other hand, the required number of samples using these bounds is \emph{prohibitively large}, and therefore, giving guarantees with such analysis is not feasible at all in practice.
In fact, the numbers are astronomic already for Markov chains with a handful of states \cite{DacaHenzingerKretinskyPetrov2016}.
We discuss further drawbacks of such a na\"ive solution in Section~\ref{sec:overview}.
Our main contribution in this paper is a \emph{practical algorithm}.
It takes the most promising actions from every state and uses the \emph{on-demand value iteration} \cite{cav17}, not even requiring an exhaustive exploration of the entire MDP.
Using techniques of \cite{DacaHenzingerKretinskyPetrov2016,AKW19}, we can show that the partial model captures enough information.
Most importantly, instead of using \cite{Chatterjee12,Solan03}, the PAC bounds are derived directly from the concrete confidence intervals, reflecting the width of each interval and the topology of the model, in the spirit of the practical SMC for reachability \cite{AKW19}.

\emph{Our contribution} can be summarized as follows:
\begin{itemize}
    \item We provide the first algorithm with PAC bounds on the mean payoff in blackbox MDP (Sec.~\ref{sec:algo}) and its extension to blackbox CTMDP (Sec.~\ref{sec:ctmdp2}).
    \item We discuss the drawbacks of a possible more straightforward solution and how to overcome them (in Sec.~\ref{sec:overview} on the conceptual level, before we dive into the technical algorithms in the subsequent sections).
    \item We evaluate the algorithm on the standard benchmarks of MDP and CTMDP and discuss the effect of heuristics, partial knowledge of the model, and variants of the algorithms (Sec.~\ref{sec:results}).
\end{itemize}

 \subsubsection*{Related work}
SMC of unbounded-horizon properties of MDPs was first considered in
\cite{DBLP:conf/sac/LassaigneP12,DBLP:conf/qest/HenriquesMZPC12} for reachability.
\cite{DBLP:conf/tacas/HahnPSSTW19} gives a model-free algorithm for $\omega$-regular properties, which is convergent but provides no bounds on the current error.
Several approaches provide SMC for MDPs and unbounded-horizon properties with \emph{PAC guarantees}.
Firstly, the algorithm of \cite{DBLP:conf/rss/FuT14} requires (1) the mixing time $T$ of the MDP, (2) the ability to restart simulations also in non-initial states, (3) visiting \emph{all} states sufficiently many times, and thus (4) the knowledge of the size of the state space $|S|$.
Secondly, \cite{atva}, based on delayed Q-learning \cite{DBLP:conf/icml/StrehlLWLL06}, lifts the assumptions (2) and (3) and instead of (1) requires only (a bound on) the minimum transition probability $\pmin$.
Thirdly, \cite{AKW19} additionally lifts the assumption (4), keeping only $\pmin$, as in this paper.
In \cite{DacaHenzingerKretinskyPetrov2016}, it is argued that while unbounded-horizon properties cannot be analysed without any information on the system, knowledge of (a~lower bound on) the minimum transition probability $\pmin$ is a relatively light and realistic assumption in many scenarios, in particular compared to the knowledge of the whole topology.
In this paper, we thus adopt this assumption.

In contrast to SMC that uses possibly more (re-started) runs of the system, there are \emph{online learning} approaches, where the desired behaviour is learnt for the single run.
Model-based learning algorithms for mean payoff have been designed both for minimizing regret~\cite{joa10,ao06} as well as for PAC online learning \cite{concur18,uai20}.

\section{Preliminaries} \label{sec:prelims}
 A \emph{probability distribution} on a finite set $X$ is a mapping $\rho: X\mapsto [0,1]$, such that $\sum_{x\in X} \rho(x) = 1$. We denote by $\Distributions(X)$ the set of all probability distributions on $X$.
%


\begin{definition}[MDP]
	A \emph{Markov decision process} is a tuple of the form $\Mdp = (\states, \initstate, \actions, \av, \trans, \rew)$, where $\states$ is a finite set of states, $\initstate \in \states$ is the initial state, $\actions$ is a finite set of actions, $\av: \states \to 2^{\actions}$ assigns to every state a set of available actions, $\trans: \states \times \actions \to \distributions(\states)$ is a transition function that given a state $s$ and an action $a\in \av(s)$ yields a probability distribution over successor states, and $\rew : \states \to \Reals^{\geq 0}$ is a reward function, assigning rewards to states.
\end{definition}
%
For ease of notation, we write $\trans(s, a, t)$ instead of $\trans(s, a)(t)$.
We denote by $\post(s,a)$, the set of states that can be reached from $s$ through action $a$.
Formally, $\post(s,a) = \{t \:|\: \trans(s, a, t) > 0\}$.

The choices of actions are resolved by strategies, generally taking history into account and possibly randomizing.
However, for mean payoff it is sufficient to consider \emph{positional} strategies of the form $\straa: \states \to \actions$.
The semantics of an MDP with an initial state $\initstate$ is given in terms of each strategy $\sigma$ inducing a Markov chain $\Mdp^\sigma_{\initstate}$ with the respective probability space and unique probability measure $\pr^{\Mdp^\sigma_{\initstate}}$, and the expected value $\expected^{\Mdp^\sigma_{\initstate}}[F]$ of a random variable $F$ (see e.g. \cite{BaierBook}).
We drop $\Mdp^\sigma_{\initstate}$ when it is clear from the context.

\paragraph{End components}
An \emph{end-component} (EC) $M = (T,A)$, with $\emptyset \neq T \subseteq \states$ and $A:T \rightarrow 2^{\actions}$ of an MDP $\Mdp$ is a \emph{sub-MDP} of $\Mdp$ such that: for all $s \in T$, we have that $A(s)$ is a subset of the actions available from $s$; for all $a \in A(s)$, we have $\post(s,a) \subseteq T$; and, it's underlying graph is strongly connected. A \emph{maximal end-component} (MEC) is an EC that is not included in any other EC.
%
Given an MDP $\Mdp$, the set of its MECs is denoted by $\mec(\Mdp)$.
For $\mec(\Mdp) = \{(T_1, A_1), \dots, (T_n, A_n)\}$, we define $\mec_{\states} = \Union_{i=1}^n T_i$ as the set of all states contained in some MEC.

\begin{definition}[continuous-time MDP (CTMDP)]
	A \emph{continuous-time Markov decision process} is a tuple of the form $\Mdp = (\states, \initstate, \actions, \av, \transR, \rew)$, where $\states$ is a finite set of states, $\initstate \in \states$ is the initial state, $\actions$ is a finite set of actions, $\av: \states \to 2^{\actions}$ assigns to every state a set of available actions, $\transR: \states \times \actions \times \states \to \Reals_{\geq 0}$ is a transition rate matrix that given a state $s$ and an action $a\in \av(s)$ yields a probability distribution over successor states, and $\rew : \states \to \Reals_{\geq 0}$ is a reward rate function, assigning a reward function to a state denoting the reward obtained for spending unit time in $s$.
\end{definition}
A strategy in a CTMDP decides immediately after entering a state which action needs to be chosen from the current state.
For a given state $s \in \states$, and an action $a \in \av(s)$, we denote by $\lambda(s, a) = \sum_{t} \transR(s,a,t) > 0$ the \emph{exit rate} of $a$ in $s$.
The \emph{residence time} for action $a$ in $s$ is exponentially distributed with mean $\frac{1}{\lambda(s, a)}$.
An equivalent way of looking at CTMDP is that in state $s$, we wait for a time which is exponentially distributed with mean $\lambda(s, a)$, and then with probability $\Delta(s,a,t)=\transR(s,a,t) / \lambda(s,a)$, we make a transition to state $t$.
The reward 
accumulated for spending time ${\sf t}$ in $s$ is $r(s) \cdot {\sf t}$.

\paragraph{Uniformization}
A \emph{uniform} CTMDP has a constant exit rate $C$ for all state-action pairs i.e, $\lambda(s,a) = C$ for all states $s \in \states$ and actions $a \in \av(\state)$. The procedure of converting a non-uniform CTMDP into a uniform one is called \emph{uniformization}. Consider a non-uniform CTMDP $\Mdp$. Let $C \in \realpos$ such that $C \geqslant \lambda(s,a)$ for all $s \in \states$ and $a \in \actions$. We can obtain a uniform CTMDP $\unif$ by assigning the new rates.
\begin{align}
    \transR'(s,a,t) = \begin{cases} 
    \transR(s,a,t) & \text{if } s \neq t \\
    \transR(s,a,t)+C - \lambda(s,a) & \text{if } s=t 
    \end{cases} \label{eq:uniform_eqn}
\end{align}
For every action $a \in \av(s)$ from each state $s$ in the new CTMDP we have a self loop  if $\lambda(s,a) < C$. 
Due to a constant transition rate, the mean interval time between two any two actions is constant.
An example of uniformization is given in Figure~\ref{fig:P1}.

\paragraph{Mean Payoff}
In this work, we consider the (maximum) \emph{mean payoff} (or \emph{long-run average reward}) of an MDP $\Mdp$, which intuitively describes the (maximum) average reward per step we expect to see when simulating the MDP for time going to infinity.
Formally, let $S_i,A_i,R_i$ be random variables giving the state visited, action played, and reward obtained in step $i$, and for CTMDP, $T_i$ the time spent in the state appearing in step $i$.
For MDP, $R_i:=r(S_i)$, whereas for CTMDP, $R_i:=r(S_i)\cdot T_i$; consequently,  for a CTMDP and a strategy $\pi$, we have $\expected^\straa_s(R_i) = \frac{r(S_i)}{\lambda(S_i, A_i)}$.
%

Thus given a strategy $\straa$, the $n$-step average reward is
\begin{equation*}
\gain^\straa_n(s) := \expected^\straa_s \left (\frac1n\sum_{i=0}^{n-1} R_i \right) = \frac1n\sum_{i=0}^{n-1} \frac{r(S_i)}{\lambda(S_i, A_i)}.
\end{equation*}
with the latter equality holding for CTMDP.
For both MDP and CTMDP, the \emph{mean payoff}  is then 
\begin{equation*}
	\gain(s) := \max_{\straa}\liminf_{n\to\infty} \gain^\straa_n,
\end{equation*}
where the maximum over all strategies can also be without loss of generality restricted to the set of positional strategies $\straas^{\mathsf{MD}}$.
A well-known alternative characterization we use in this paper is
\begin{equation}
	\gain(s) = \max_{\straa \in \straas^{\mathsf{MD}}} \sum_{M \in \mec(\Mdp)} \pr^\straa_s[\Diamond\Box M] \cdot \gain_M, \label{eq:decomp}
\end{equation}
where $\Diamond$ and $\Box$ respectively denote the standard LTL operators \emph{eventually} and \emph{always} respectively. Further, $\Diamond\Box M$ denotes the set of paths that eventually remain forever within $M$ and $\gain_M$ is the unique value achievable in the (CT)MDP restricted to the MEC $M$.
Note that $\gain_M$ does not depend on the initial state chosen for the restriction. 

We consider algorithms that have a limited information about the MDP.

\begin{definition}[Blackbox and greybox]\label{def:limit}
An algorithm inputs an MDP or a CTMDP as \emph{blackbox} if
\begin{itemize}
	\item it knows $\initstate$,
	\item for a given state,\footnote{In contrast to practical setups in monitoring, our knowledge of the current state is complete, i.e., the previously visited states can be uniquely identified.} an oracle returns its available actions,
	\item given a state $s$ and action $a$, it can sample a successor $t$ according to $\trans(s,a)$,

    \item it knows $\pmin \leqslant \min_{\substack{s \in \states,a \in \Av(s)\\ t \in \post(s,a)}} \trans(s,a,t)$, an under-approximation of the minimum transition probability. 
\end{itemize}
When input as \emph{greybox}, it additionally knows the number $\abs{\post(s,a)}$ of successors for each state $s$ and action $a$.
Note that the exact probabilities on the transitions in an MDP or the rates in a CTMDP are unknown for both blackbox and greybox learning settings.
\end{definition}

\section{Overview of Our Approach} \label{sec:overview}
	
	Since no solutions are available in the literature and our solution consists of multiple ingredients, we present it in multiple steps to ease the understanding.
	First, we describe a more \emph{na\"ive} solution and pinpoint its drawbacks.
	Second, we give an \emph{overview} of a more sophisticated solution, eliminating the drawbacks.
	Third, we fill in its \emph{details} in the subsequent sections.
	Besides, each of the three points is first discussed on discrete-time MDPs and then on continuous-time MDPs.
	The reason for this is twofold: the separation of concerns simplifies the presentation; and the algorithm for discrete-time MDP is equally important and deserves a standalone  description.
	
	\subsection{Na\"ive solution}
	
	We start by suggesting a conceptually simple solution.
	We can learn mean payoff $MP$ in an MDP $\Mdp$ as follows:
	\begin{enumerate}
		\item[(i)] Via simulating the MDP $\Mdp$, we learn a model ${\Mdp}^\prime$ of $\Mdp$, i.e., we obtain confidence intervals\todo{confidence interval imprecision inconsistency} on the \emph{t}ransition \emph{p}robabilities of $\Mdp$ (of some given width $\symbolTpImprecision$, called \emph{TP-imprecision}, and confidence $1-\symbolTpInconfidence$, where $\symbolTpInconfidence$ is called \emph{TP-inconfidence}).
		\item[(ii)] We compute the mean payoff $\widehat{MP}$ on the (imprecise) model ${\Mdp}^\prime$.
		\item[(iii)] We compute the \emph{MP-imprecision} $\varepsilon_{MP}=|\widehat{MP}-MP|$ of the \emph{m}ean \emph{p}ayoff from the TP-imprecision by the ``robustness'' theorem \cite{krish} which quantifies how mean payoff can change when the system is perturbed with a given maximum perturbation.
		Further, we compute the overall \emph{MP-inconfidence} $\delta_{MP}$ from the TP-inconfidence $\delta_{TP}$; in particular, we can simply accumulate all the uncertainty and set $\delta_{MP}=|\trans|\cdot\delta_{TP}$, where $|\trans|$ is the number of transitions.
		The result is then probably approximately correct, being $\varepsilon_{MP}$-precise with confidence $1-\delta_{MP}$.
		(Inversely, from a desired $\varepsilon_{MP}$ we can also compute a sufficient $\varepsilon_{TP}$ to be used in the first step.)
	\end{enumerate}

	Learning the model, i.e.\ the transition probabilities, can be easily done by observing the simulation runs and collecting, for each state-action pair $(s,a)$, a statistics of which states occur right after playing $a$ in $s$.
	The frequency of each successor $t$ among all successors then estimates the transition probability $\trans(s,a,t)$.
	This is the standard task of estimating the generalized Bernoulli variable (a fixed distribution over finitely many options) with confidence intervals.
	We stop simulating when \emph{each} transition probability has a precise enough confidence interval (with $\varepsilon_{TP}$ and $\delta_{TP}$ yielded by the robustness theorem from the desired overall precision).\footnote{Several non-trivial questions are dealt with later on: how to resolve the action choices during simulations; when to stop each simulation run and start a new one; additionally, in the black-box setting, when do we know that all successors of each transition have been observed. In particular, the last one is fundamental for the applicability of the robustness theorem. While the literature typically assumes the greybox setting or even richer information, to allow for such an algorithm with PAC bounds, our approach only needs $\pmin$.}
	The drawbacks are \emph{(D1: uniform importance)} that even transitions with little to no impact on the mean payoff have to be estimated precisely (with $\varepsilon_{TP}$ and $\delta_{TP}$); and \emph{(D2: uniform precision required)} that, even restricting our attention to ``important'' transitions, it may take a long time before the last one is estimated precisely (while others are already estimated overly precisely).
	
	Subsequently, using standard algorithms the mean payoff $\widehat{MP}$ can be computed precisely by linear programming \cite{Puterman} or precisely enough by value iteration \cite{cav17}.
	The respective $MP$ can then be estimated by the robustness theorem \cite{krish}, which yields for a given maximum perturbation of transition probabilities (in our case, $\varepsilon_{TP}$/2) an upper bound on the respective perturbation of the mean payoff $\varepsilon_{MP}/2$.
	The drawbacks are \emph{(D3: uniform precision utilized)} that more precise confidence intervals for transitions (obtained due to D2) are not utilized, only the maximum imprecision is taken into account; and  \emph{(D4: a-priori bounds)} that the theorem is extremely conservative.
	Indeed, it reflects neither the topology of the MDP nor how impactful each transition is and thus provides an a-priori bound, extremely loose compared to the possible values of mean payoff that can be actually obtained for concrete values within the confidence intervals.
	This is practically unusable beyond a handful of states even for Markov chains \cite{DacaHenzingerKretinskyPetrov2016}.
	
	For CTMDP $\Mdp$, we additionally need to estimate the rates (we show below how to do this).
	Subsequently, we can uniformize the learnt CTMDP ${\Mdp}^\prime$.
	Mean payoff of the uniformized CTMDP is then equal to the mean payoff of its embedded MDP\footnote{An embedded MDP of a CTMDP is obtained by considering for every state $s$, actions $a \in \av(s)$, and transitions $t\in \post(s,a)$, such that $\trans(s,a,t)=\Delta(s,a,t)$, and by disregarding the transition rate matrix.}.
	Consequently, we can proceed as before but we also have to compute (i) confidence intervals for the rates from finitely many observations, and (ii) the required precision and confidence of these intervals so that  the respective induced error on the mean payoff is not too large.
	Hence all the drawbacks are inherited and, additionally, also applied to the estimates of the rates.
	Besides, \emph{(D5: rates)} while imprecisions of rates do not increase MP-imprecision too much, the bound obtained via uniformization and the robustness theorem is very loose.
	Indeed, imprecise rates are reflected as imprecise self-loops in the uniformization, which themselves do not have much impact on the mean payoff, but can increase the TP-imprecision and thus hugely the MP-imprecision from the robustness theorem.
	
	Finally, note that for both types of MDP, \emph{(D6: not anytime)} this na\"ive algorithm is not an anytime algorithm\footnote{An anytime algorithm can, at every step, return the current estimate with its imprecision, and this bound converges to 0 in the limit.} since it works with pre-computed $\varepsilon_{TP}$ and $\delta_{TP}$.
	Instead it returns the result with the input precision if given enough time; if not given enough time, it does not return anything (also, if given more time, it does not improve the precision).
	
	\subsection{Improved solution}
	
	Now we modify the solution so that the drawbacks are eliminated.
	The main ideas are 
	(i) to allow for differences in TP-imprecisions ($\varepsilon_{TP}$ can vary over transitions) and even deliberately ignore less important transitions and instead improve precision for transitions where more information is helpful the most; 
	(ii) rather than using the a-priori robustness theorem, to utilize the precision of each transition to its maximum; and
	(iii) to give an anytime algorithm that reflects the current confidence intervals and, upon improving them, can efficiently improve the mean-payoff estimate without recomputing it from scratch.
	There are several ingredients used in our approach.
	
	Firstly, \cite{cav17} provides an anytime algorithm for approximating  mean payoff in a fully known MDP.
	The algorithm is a version of value iteration, called \emph{on-demand}, performing improvements (so called Bellman updates) of the mean-payoff estimate in each state.
	Moreover, the algorithm is simulation-based, performing the updates in the visited states, biasing towards states where a more precise estimate is helpful the most (``on demand'').
	This matches well our learning setting.
	However, the approach assumes precise knowledge of the transition probabilities and, even more importantly, heavily relies on the knowledge of MECs.
	Indeed, it decomposes the mean-payoff computation according to Eq.~\ref{eq:decomp} into computing mean payoff within MECs and optimizing (weighted) reachability of the MECs (with weights being their mean payoffs).
	When the MECs are unknown, none of these two steps can be executed.
	
	Secondly, \cite{AKW19} provides an efficient way of learning reachability probabilities (in the grey-box and black-box settings).
	Unfortunately, since it considers TP-inconfidence to be the same for all transitions, causing different TP-imprecisions, the use of robustness theorem in~\cite{AKW19} makes the learning algorithm used there practically unusable in many cases.
	On a positive note, the work identifies the notion of $\delta_{TP}$-sure EC, which reflects how confident we are, based on the simulations so far, that a set of states is an EC.
	This notion will be crucial also in our algorithm.

	Both approaches are based on ``bounded value iteration'', which computes at any moment of time both a lower and an upper bound on the value that we are approximating (mean payoff or reachability, respectively).
	This yields anytime algorithms with known imprecision, the latter---being a learning algorithm on an incompletely known MDP---only with some confidence.
	Note that the upper bound converges only because ECs are identified and either collapsed (in the former) or deflated~\cite{KKKW18} (in the latter), meaning their upper bounds are decreased in a particular way to ensure correctness.
	\smallskip
	
	Our algorithm on (discrete-time) MDP $\Mdp$ performs, essentially, the following.
	It simulates $\Mdp$ in a similar way as \cite{AKW19}.
	With each visit of each state, not only it updates the model (includes this transition and improves the estimate of the outgoing transition probabilities), but also updates the estimate of the mean payoff by a Bellman update.
	Besides, at every moment of time, the current model yields a hypothesis what the actual MECs of $\Mdp$ are and the respective confidence.
	While we perform the Bellman updates on all visited states deemed transient, the states deemed to be in MECs are updated separately, like in \cite{cav17}.
	However, in contrast to \cite{cav17}, where every MEC is fully known and can thus be collapsed, and in contrast to the ``bounded'' quotient of \cite{AKW19} (see Appendix~\ref{app:def}), we instead introduce a special action \textsf{stay} in each of its states, which simulates remaining in the (not fully known) MEC and obtaining its mean-payoff estimate via reachability:
	
\begin{definition}[{\sf stay}-augmented MDP] \label{def:stay_mec}
	Let $\Mdp = (\states, \initstate, \actions, \av, \trans, \rew)$ be an MDP and $\lu: \mec(\Mdp) \to [0,1]$ be real functions on MECs.
	We augment the {\sf stay} action to $\Mdp$ to obtain ${\Mdp}^\prime = ({\states}^\prime, \initstate, {\actions}^\prime, {\av}^\prime, {\trans}^\prime, {\rew}^\prime)$, where
	\begin{itemize}
	    \item ${\states}^\prime = \states \uplus \{s_+, s_-, s_?\}$,
	    \item ${\actions}^\prime = \actions \uplus \{\sf stay\}$,
	    \item ${\av}^\prime(s) = \begin{cases}
	    \av(s)& \text{for }s\in S\setminus\bigcup\mec(\Mdp)\\
	    \av(s)\cup\{\sf stay\}& \text{for }s\in \bigcup\mec(\Mdp)\\
	    \{\sf stay\}& \text{for } s \in \{s_+, s_-, s_?\}
	    \end{cases}$
	    \item ${\trans}^\prime$ extends $\trans$ by ${\trans}^\prime(s, {\sf stay}) = \set*{s_+ \mapsto \lowerbound(M), s_- \mapsto 1 - \upperbound(M), s_? \mapsto \upperbound(M) - \lowerbound(M)}$ on $s \in M\in \mec(\Mdp)$ and by ${\trans}^\prime(s,{\sf stay},s)=1$ for $s \in \{s_+, s_-, s_?\}$.
	    \item ${\rew}^\prime$ extends ${\rew}$ by ${\rew}^\prime(s_+) = {\rew}^\prime(s_?) = {\rew}^\prime(s_-) = 0$.\footnote{Intuitively, a higher transition probability to $s_+$ indicates that the MEC has high value, a higher transition probability to $s_?$ indicates high uncertainty in the value of the MEC, while a higher transition probability to $s_-$ indicates that the MEC has low value.}
	\end{itemize}
\end{definition}
	\begin{corollary}
	If $\lu$ are valid lower and upper bounds on the mean-payoff within MECs of $\Mdp$ then 
	$\max_\sigma\pr^{M^\sigma}[\Diamond \{s_+\}] \leqslant v(\initstate)\leqslant max_\sigma\pr^{M^\sigma}[\Diamond\{s_+,s_?\}]$
	\footnote{For simplicity of the presentation, we assume the rewards are between 0 and 1, for all states. If they are not, we can always rescale them to [0,1] by dividing them by the maximum reward observed so far and correspondingly adjust $\trans(\cdot,\mathsf{stay},\cdot)$.} where, $\max_\sigma\pr^{M^\sigma}[\Diamond S]$ gives the maximum probability of reaching some state in $S$ over all strategies.
	\end{corollary}
	
	This turns the problem into reachability, and thus allows for deflating (defined for reachability in \cite{AKW19}) and an algorithm combining \cite{AKW19} and \cite{cav17}.
	Concrete details are explained in the subsequent section.
	To summarize, (D1) and (D2) are eliminated by not requiring uniform TP-imprecisions; (D3) and (D4) are eliminated via updating lower and upper bounds (using deflating) instead of using the robustness theorem.
	
	Concerning CTMDP, in Section \ref{sec:ctmdp2} we develop a confidence interval computation for the rates.
	Further, we design an algorithm deriving the MP-imprecision resulting from the rate imprecisions, that acts directly on the level of the CTMDP and not on the embedded MDP of the uniformization.
	This effectively removes (D5).


\section{Algorithm for Discrete-Time MDP} \label{sec:algo}
Now that we explained the difficulties of a na\"ive approach, and the concepts from literature together with novel ideas to overcome them, we describe the actual algorithm for the discrete-time setting. Following a general outline of the algorithm, we give detailed explanations behind the components and provide the statistical guarantees the algorithm gives.
\paragraph*{Overall Algorithm and Details}
Our version of an on-demand value iteration for mean payoff in black-box MDP is outlined in Algorithm 1. 
Initially, the input MDP $\mathcal{M}$ is augmented with terminal states $(\{s_+, s_-, s_?\})$ to obtain the {\sf stay}-augmented MDP ${\Mdp}^\prime$.
We learn a {\sf stay}-augmented MDP ${\Mdp}^\prime = ({\states}^\prime, \initstate, {\actions}^\prime, {\av}^\prime,$ ${\trans}^\prime, {r}^\prime)$ by collecting samples through several simulation runs (Lines~\ref{line:Ni}-\ref{line:conf}).
Over the course of the algorithm, we identify MECs with $\delta_{TP}$ confidence (Line~\ref{line:find_mec}) and gradually increase precision on their respective values (Lines~\ref{line:check_sink}-\ref{line:update_mec}). 
As stated earlier, these simulations are biased towards actions that lead to MECs potentially having higher rewards. 
Values for MECs are encoded using the $\mathsf{stay}$ action (Line~\ref{line:update_stay}) and propagated throughout the model using bounded value iteration (Lines~\ref{line:initialize}-\ref{line:vi_termination}).
In Line~\ref{line:initialize}, we reinitialize the values of the states in the partial model since new MECs may be identified and also existing MECs may change.
Finally, we claim that the probability estimates ${\trans}^\prime$ are correct with confidence $\delta_{MP}$ and if the bounds on the value are precise enough, we terminate the algorithm. Otherwise, we repeat this overall process with improved bounds (Line~\ref{line:termination-test}).


%

\paragraph{Simulation} The $\SIMULATE$\footnote{For technical details on the $\SIMULATE$ procedure, see Algorithm~\ref{alg:simulateNewMDP} in Appendix \ref{app:algo}.} function simulates a run over the input blackbox MDP $\Mdp$ and returns the visited states in order.
The simulation of $\Mdp'$ is executed by simulating $\Mdp$ together with a random choice if action {\sf stay} is taken.
Consequently, a simulation starts from $\initstate$ and ends at one of the terminal states $(\{s_+, s_-, s_?\})$.
During simulation, we enhance our estimate of ${\Mdp}^\prime$ by visiting new states, exploring new actions and improving our estimate of ${\trans}^\prime$ with more samples.
When states are visited for the first time, actions are chosen at random, and subsequently, actions with a higher potential reward are chosen. 
If a simulation is stuck in a loop, we check for the presence of an MEC with $\delta_{TP}$ confidence.
If a $\delta_{TP}$-sure MEC is found, we add a {\sf stay} action with $l,u = 0,1$, otherwise we keep simulating until the required confidence is achieved. After that, we take the action with the highest upper bound that is leaving the MEC to continue the simulation.
We do several such simulations to build a large enough model before doing value iteration in the next steps.

\begin{algorithm}[t]
    \caption{Mean-payoff learning for black-box MDP}\label{alg:general}
    \algorithmicrequire MDP $\Mdp$, imprecision $\varepsilon_{MP} > 0$, $\mpInconfidence$ $\symbolMpInconfidence > 0$, 
    lower bound $\pmin$ on transition probabilities in $\Mdp$  \\
    \textbf{Parameters:} revisit threshold $k \geq 2$, episode length $n\geq 1$\\
    \algorithmicensure upon termination $\symbolMpImprecision$-precise estimate of the maximum mean payoff for $\Mdp$ with confidence $1-\symbolMpInconfidence$, i.e.\ $(\symbolMpImprecision,1-\symbolMpInconfidence)$-PAC estimate
    \begin{algorithmic}[1]
        \Procedure{$\mathsf{ON\_DEMAND\_BVI}$}{}
        
            \emph{//Initialization} 
            \State Set $\lb(s_+)=\ub(s_+)=\ub(s_?)=1$, $\lb(s_-)=\ub(s_-)=\lb(s_?)=0$ \Comment{Augmentation}
            \State ${\states}^\prime = \emptyset$ \Comment{States of learnt model}
            \Repeat
            
            \emph{//Get $n$ simulation runs and update MP of MECs where they end up}
            \For{$n$ times} \label{line:Ni}
                \State $\fpath \gets \SIMULATE(k)$ \Comment{Path taken by the simulation}
                \State ${\states}^\prime \leftarrow {\states}^\prime \cup w$ \Comment{Add states to the model}\label{line:update_model}
                \State $\symbolTpInconfidence \gets \frac{\symbolMpInconfidence\cdot \pmin}{\lvert\{{a} \vert\, s \in {\states}^\prime \wedge {a} \in {\av}^\prime(s) \}\rvert}$ \Comment{Split inconfidence among all transitions} \label{line:conf}
                \If{last state of $\fpath$ is $s_+$ or $s_?$} \Comment{Probably entered a good MEC $M$}\label{line:check_sink}
                    \State $M \gets$ MEC from which we entered the last state of $\fpath$ 
                    \State $\mathsf{UPDATE\_MEC\_VALUE}(M)$ \label{line:update_mec} \Comment{Increase precision using more VI}
                    \State Update ${\trans}^\prime(s, \mathsf{stay})$ according to Definition \ref{def:stay_mec} for all $s\in M$ \label{line:update_stay}
                \EndIf
            \EndFor
            
            \emph{//Identify $\delta_{TP}$-sure MECs and propagate their MP by VI for reachability}
            \State $\mathit{ProbableMECs}\gets\mathsf{FIND\_MECS}$  \Comment{$\delta_{TP}$-sure MECs}  \label{line:find_mec}
            \State $\INITIALIZE$ \label{line:initialize} \Comment{Reinitialize $\lb,\ub$ for all states}
            \Repeat     
                \State $\UPDATE({\states}^\prime)$  
                \Comment{One Bellman update per state}
                \For{ $T \in \mathit{ProbableMECs}$}
                    \State $\mathsf{DEFLATE}(T)$ \Comment{Ensure safe but converging $\ub$}
                \EndFor
            \Until{$\lb$ and $\ub$ close to their respective fixpoints} \label{line:vi_termination}
            \Until{$\ub(\initstate)$ - $\lb(\initstate) < \frac{2\symbolMpImprecision}{\rmax}
            $}\Comment{$\symbolMpImprecision$ is the absolute error; we use ``$<\frac{2\symbolMpImprecision}{\rmax}$'' for relative difference between upper and lower values, where $\rmax = \max\limits_{s\in{\states}^\prime} r(s)$.
            }\label{line:termination-test}
        \EndProcedure
    \end{algorithmic}
\end{algorithm}

\paragraph*{Estimating transition probabilities}
\cite{AKW19} gives an analysis to estimate bounds on transition probabilities for reachability objective in MDPs.
For completeness, we briefly restate it here.
Given an \mpInconfidence$\symbolMpInconfidence$, we distribute the inconfidence over all individual transitions as $$\symbolTpInconfidence := \dfrac{\symbolMpInconfidence\cdot \pmin}{\lvert\{{a} \vert\, \state \in {\states}^\prime \wedge {a} \in {\av}^\prime(s) \}\rvert},$$
\todo{The following can be removed if there is a space issue.}where $\frac{1}{\pmin}$ gives an upper bound on the maximum number of possible successors for an available action from a state\footnote{If we additionally know the maximum number of possible successors, $\max\limits_{s \in \states,a \in \Av(s)} |\post(s,a)|$, we can use that instead of $\frac{1}{\pmin}$ to obtain a slightly smaller TP-imprecision. A more detailed analysis is presented in Appendix~\ref{app:pminandmaxsuccessorcomparison}.}.
The Hoeffding's inequality gives us a bound on the number of times an action $a$ needs to be sampled from state $s$, denoted $\#(s,a)$, to achieve a \tpImprecision $\symbolTpImprecision \leqslant \sqrt{\dfrac{\ln \delta_{TP}}{-2\#(s, a)}}$ on $\trans(s,a,t)$, such that

\begin{equation*}
\widehat{\mathbb{T}}(s,a,t) := \max (0, \dfrac{\#(s, a, t)}{\#(s, a)} - \symbolTpImprecision)
\end{equation*}
where, $\#(s, a, t)$ is the number of times $t$ is sampled when action $a$ is chosen from $s$.
\paragraph{Updating mean-payoff values}
Using $\widehat{\mathbb{T}}(s,a,t)$, we compute estimates of the upper and lower bounds of the values corresponding to every action from a state visited in the partial model that is constructed so far.
We use the following \emph{modified} Bellman equations~\cite{AKW19}:

\begin{equation*}
\widehat{\lb}(s,a) := \sum\limits_{t:\#(s,a,t)>0} \widehat{\mathbb{T}}(s,a,t) \cdot \lb(t)
\end{equation*}
\begin{equation*}
\widehat{\ub}(s,a) := \sum\limits_{t:\#(s,a,t)>0} \widehat{\mathbb{T}}(s,a,t) \cdot \ub(t) + (1-\sum\limits_{t:\#(s,a,t)>0} \widehat{\mathbb{T}}(s,a,t)),
\end{equation*}
where $\lb(t) = \max\limits_{{a} \in \Av(t)} \widehat{\lb}(t,a)$ and $\ub(t) = \max\limits_{{a} \in \Av(t)}\widehat{\ub}(t,a)$ are bounds on the value of from a state, $v(s)$.
When a state is discovered for the first time during the simulation, and is added to the partial model, we initialize $\lb(s)$, and $\ub(s)$ to 0, and 1, respectively.
Note that $\sum\limits_{t:\#(s,a,t) > 0} \widehat{\mathbb{T}}(s,a,t) < 1$. We attribute the remaining probability to unseen successors and assume their value to be 0 (1) to safely under-(over-)approximate the lower (upper) bounds.
We call these \emph{blackbox Bellman update} equations, since it assumes that all the successors of a state-action pair may not have been visited.
\paragraph*{Estimating values of end-components}
End-components are identified with an inconfidence of $\symbolTpInconfidence$.
As observed in \cite{DacaHenzingerKretinskyPetrov2016}, assuming an action has been sampled $n$ times, the probability of missing a transition for that action is at most $(1-\pmin)^n$.
Thus, for identifying ($T,A$) as a $\deltasure$ MEC, every action in $A$ that is available from a state $s \in T$ needs to be sampled at least $\frac{\ln \symbolTpInconfidence}{\ln (1-\pmin)}$ times.

Once a $\deltasure$ MEC $M$ is identified, we estimate its upper ($\gain^u_M$) and lower ($\gain^l_M$) bounds using value iteration.\footnote{Note that one requires the ECs to be aperiodic for the VI to converge. ~\cite{Puterman} suggests a way that deals with this.}
While running value iteration, we assume, with a small inconfidence, that there are no unseen outgoing transitions.
So we use the following modified Bellman update equations inside the MEC where we under-(over-)approximate the lower(upper) bound to a much lesser degree.
\begin{equation*}
\widehat{\lb}(s,a) := \sum\limits_{t:\#(s, a, t)>0} \widehat{\mathbb{T}}(s, a, t) \cdot \lb(t) + \min\limits_{t:\#(s, a, t)>0}\lb(t) \cdot (1-\sum\limits_{t:\#(s, a, t)>0} \widehat{\mathbb{T}}(s, a, t))
\end{equation*}
\begin{equation*}
\widehat{\ub}(s, a) := \sum\limits_{t:\#(s, a, t)>0} \widehat{\mathbb{T}}(s, a, t) \cdot \ub(t) + \max\limits_{t:\#(s, a, t)>0}\ub(t) \cdot (1-\sum\limits_{t:\#(s, a, t)>0} \widehat{\mathbb{T}}(s, a, t))
\end{equation*}
Following the assumption, we call these \emph{greybox} (See Definition~\ref{def:limit}) Bellman update equations.
The value iteration algorithm further gives us bounds on $\gain^u_M$ and $\gain^l_M$. We say that the upper estimate of $\gain^u_M$ ($\widehat{\gain}^{u}_M$) and the lower estimate of $\gain^l_M$ ($\widehat{\gain}^{l}_M$) are the overall upper and lower bounds of the mean-payoff value of $M$, respectively. To converge the overall bounds, we need value iteration to return more precise estimates of $\gain^l_M$ and $\gain^u_M$, and we need to sample the actions inside $M$ many times to reduce the difference between $\gain^l_M$ and $\gain^u_M$. 
We call this procedure, $\mathsf{UPDATE\_MEC\_VALUE}$\footnote{This is outlined in technical detail in Appendix~\ref{app:algo}}.

Now, some MECs might have very low values or may not be reachable from $\initstate$ with high probability. In such cases, no optimal strategy might visit these MECs, and it might not be efficient to obtain very precise mean-payoff values for every MEC that may be identified in an MDP. We follow the on-demand heuristic~\cite{cav17} where we progressively increase the precision on mean-payoff values as an MEC seems more likely to be a part of an optimal strategy. The {\sf stay} action on MECs helps in guiding simulation towards those MECs that have a higher lower bound of the mean-payoff value. In particular, whenever the simulation ends up in $s_+$ or $s_?$, we run $\mathsf{UPDATE\_MEC\_VALUE}$ with higher precision on the MEC that led to these states. 
If the simulation ends up in these states through a particular MEC more often, it indicates that the MEC is likely to be a part of an optimal strategy, and it would be worth increasing the precision on its mean-payoff value.

\paragraph{Deflate operation}
Unlike in the case of computation of mean payoff for whitebox models~\cite{AKW19} where a MEC is collapsed following the computation of its value, for blackbox learning, once a set of states is identified as a $\deltasure$ MEC, we cannot collapse them.
This is because collapsing would prevent a proper future analysis of those states, which is undesirable in a blackbox setting. However, this leads to other problems.
To illustrate this, we consider an MDP that only has a single MEC $M$ and one outgoing action from every individual state.
\todo{I don't think we need to establish a relation with mean-payoff here. Describing in the context of reachability only can be enough. It saves 3-4 lines of space.}
Recall from Equation~\ref{eq:decomp} that we compute the mean-payoff by reducing it to a reachability problem.
Once the mean-payoff for the MEC, and the probabilities corresponding to {\sf stay} action in Line~\ref{line:update_stay} are computed, to compute the reachability probability, the upper and lower bounds of all states in the MECs are initialized to $1$ and $0$ respectively.
Now suppose that the sum of probabilities to $s_+$ and $s_?$ be $p$ denoting the upper bound on the value of the mean-payoff to be $p \cdot r_{\max}$.
Clearly, the upper bound on the reachability value of this MDP is $p$.
Now, when we do BVI to calculate this value, from every state in $M$, there would be \emph{at least} two action choices, one that stays inside the MEC, and one that corresponds to the {\sf stay} action.
Initially, all states, except the terminal states, would have upper and lower values set to 0 and 1, respectively.
Thus, among the two action choices, one would have upper value $p$, while the other would have upper value $1$, and hence, the Bellman update assigns the upper value of the state to $1$.
As one can see, this would go on, and convergence wouldn't happen, and hence the true mean-payoff value will not be propagated to the initial state of the MDP.
To avoid this, we need the deflate operation which lowers the upper reachability value to the best outgoing action, i.e. in this case, the {\sf stay} action with value $p$.
More technical details are provided in Appendix~\ref{app:algo}.

\paragraph{Statistical guarantees}
The following theorem shows that the mean-payoff value learnt by Algorithm~\ref{alg:general} is PAC on an input blackbox MDP.
\begin{theorem}\label{thm:algo}
Algorithm~\ref{alg:general} has the property that when it stops, it returns an interval for the mean-payoff value of the MDP that is PAC for the given MP-inconfidence
$\symbolMpInconfidence$ and the \mpImprecision $\symbolMpImprecision$.
\end{theorem}
\todo{Such lines can be written in, say, the footnotes to save some space.}The proof of this theorem appears in Appendix~\ref{app:mdp_proof}.

\paragraph{Anytime algorithm} As a direct consequence, we obtain an anytime algorithm from Algorithm~\ref{alg:general} by (1) dropping the termination test on Line~\ref{line:termination-test}, i.e. replacing it with \textbf{until false}, and (2) upon query (or termination) by the user, we output $(\ub(\initstate)+\lb(\initstate))/2$ as the estimate and, additionally, we output $(\ub(\initstate)$ - $\lb(\initstate))/2$ as the current imprecision.

\paragraph{Using greybox update equations during blackbox learning}
We also consider the variant where we use greybox update equations to estimate the mean-payoff values.
However, assuming we keep the \tpImprecision unchanged, the overall \tpInconfidence now has to include the probability of missing some successor of a state $s$ for an action $a$\footnote{Assuming $\#(s,a)$ to be as small as $200$, and $\pmin=0.05$, the probability of missing a transition is $3.5 \cdot 10^{-5}$.}.
%
%
Given a number of samples $\#(s,a)$, the probability that we miss a particular successor is at most $(1-{\pmin})^{\#(s, a)}$, and hence the overall \tpInconfidence corresponding to using greybox equations for blackbox learning increases to $\symbolTpInconfidence + (1-{\pmin})^{\#(s, a)}$.
We also note that the use of greybox update equations on estimating the transition probabilities also gives us a PAC guarantee but with an increased MP-Inconfidence resulting from an increased TP-inconfidence. See Appendix~\ref{app:error_tolerance} for more discussion on this.

\section{Algorithm for Continuous-Time MDP} \label{sec:ctmdp2}
\todo{Check Appendix}In this section, we describe an algorithm to learn blackbox CTMDP models for mean-payoff objective while respecting the PAC guarantees.
As in the case of MDPs, we reduce the mean-payoff problem to a reachability problem.
We follow the same overall framework as in MDPs, where we compute the probability to reach the end-components under an optimal strategy, and we compute their respective mean-payoff values.
Computing reachability probabilities in a CTMDP is the same as computing reachability probabilities in the underlying embedded MDP. 
Similar to estimating $\trans(s,a,t)$ in Section~\ref{sec:algo} for MDPs, we estimate $\Delta(s,a,t)$\footnote{Recall that an estimate of $\Delta(s,a,t)$ is the ratio between $\#(s,a,t)$ and $\#(s,a)$, and is the probability with which we go to state $t$ from $s$ when action $a$ is chosen from $s$.} for CTMDPs, and follow the simulation-based procedure in Algorithm~\ref{alg:general} to compute reachability probabilities.
However, unlike MECs in MDPs, where the mean-payoff value depends solely on the transition probabilities, the mean-payoff value in a CTMDP also depends on the rates $\lambda(s,a)$ for $s \in T$ and $a \in A(s)$ for an MEC $M=(T,A)$.
Thus to compute the value of an MEC, we also estimate the rates of the state-action pairs. 
Once we get the estimates of the rates, we uniformize the CTMDP to obtain a uniform CTMDP that can be treated as an MDP by disregarding the rates while preserving the mean-payoff value~\cite{Puterman}.

\paragraph{Estimating rates}
Recall that for an action $a$, the time spent in $s$ is exponentially distributed with a parameter $\lambda(s,a)$, and $\frac{1}{\lambda(s,a)}$ is the mean of this distribution.
During the simulation of a CTMDP, for every state $s$ reached and action $a$ chosen from $s$, we construct a sequence $\tau_{s,a}$ of the time difference between the entry and the corresponding exit from $s$ when action $a$ is chosen.
Then, the average over the sequence $\tau_{s,a}$ gives us an estimate $\frac{1}{\widehat{\lambda}(s,a)}$ of $\frac{1}{\lambda(s,a)}$ (Abbreviated to $\frac{1}{\lambda}$ from now on when $(s,a)$ is clear from the context.).

Assuming a multiplicative error $\alpha_R$ on our estimates of $\frac{1}{\lambda}$, the lemma below uses Chernoff bounds\footnote{Since $\lambda$ is not bounded, we cannot use Hoeffding's inequality as in the case of estimating the transition probabilities.} to give the number of samples that need to be collected from an exponential distribution so that the estimated mean $\frac{1}{\widehat{\lambda}}$ is at most $\alpha_R$-fraction away from the actual mean $\frac{1}{\lambda}$ with probability at least $1-\delta_R$, where $\alpha_R, \delta_R \in (0, 1)$. Further by Cramer's theorem~\cite{DZ10}, it follows that this is the tightest possible bound for the number of samples collected.

\begin{lemma} \label{lem:ctmdp-pac}
Let $X_1, \dots, X_n$ be exponentially distributed i.i.d. random variables with mean $\frac{1}{\lambda}$. Then we have that
\[
\pr \Big [\abs{\frac{1}{\widehat{\lambda}} - \frac{1}{\lambda} \geqslant \frac{1}{\lambda} \cdot \alpha_R)} \Big ] \leqslant \displaystyle{\inf_{-\lambda < t < 0}}
\Big (\frac{\lambda}{\lambda +t}\Big)^n \cdot e^{\frac{tn}{\lambda}(1+\alpha_R)} + \displaystyle{\inf_{ t > 0}} 
\Big (\frac{\lambda}{\lambda + t}\Big)^n \cdot e^{\frac{tn}{\lambda}(1-\alpha_R)},
\]
where $\frac{1}{n}\sum_{i=1}^n X_i = \frac{1}{\widehat{\lambda}}$.
\end{lemma}
Assuming the right-side of the inequality is at most $\delta_R$, we have that $\lambda \in [\hat{\lambda}(1-\alpha_R), \hat{\lambda}(1+\alpha_R)]$, or $\widehat{\lambda} \in [\frac{\lambda}{1+\alpha_R}, \frac{\lambda}{1-\alpha_R}]$ with probability at least $1-\delta_R$.
Table~\ref{tab:nSamplesTable} shows the number of samples required for various values of $\alpha_R$ and $\delta_R$\footnote{In Appendix~\ref{app:ctmdp_proofs2}, we show an example of how we compute the number of samples for one of the entries.}.
The proof of Lemma~\ref{lem:ctmdp-pac} appears in Appendix~\ref{app:ctmdp_proofs2}.
\begin{table}[t]
\begin{center}
\begin{tabular}{ | m{1.1cm} | m{1cm}| m{1cm} | m{1cm} | m{1.6cm} | } 
\hline
$\alpha_R$ \textbackslash $\delta_R$ & 10\% & 5\% & 0.01\% & 0.00001\%\\ 
\hline
3\% & 7000 & 9000 & 23000 & 60000\\
\hline
5\% & 2500 & 3100 & 8000 & 13400 \\ 
\hline
\end{tabular}
\caption{Lookup table for number of samples based on $\alpha_R$ and $\deltaR$}
\label{tab:nSamplesTable}
\end{center}
\vspace*{-2em}
\end{table}

Given a maximum multiplicative error $\alpha_R$ on the mean of the exponential distributions of the state-action pairs in a CTMDP, we say that the rate $\lambda$ is known \emph{$\alpha_R$-precisely} if $\widehat{\lambda} \in [\frac{\lambda}{1+\alpha_R}, \frac{\lambda}{1-\alpha_R}]$. We now quantify the bounds on the estimated mean-payoff value.
Let $\Mdp$ be a CTMDP, $v_{\Mdp}$ be its actual mean-payoff value, and let $\widehat{v}_{\Mdp}$ denote its mean-payoff when the rates of the state-action pairs are known $\alpha_R$-precisely.
Then we have the following.
\begin{lemma} \label{lem:CTMDP_bounds}
Given a CTMDP $\Mdp$ with rates known $\alpha_R$-precisely, with transition probabilities known precisely, and with maximum reward per unit time over all states $r_{max}$, we have $v_{\Mdp}(\frac{1-\alpha_R}{1+\alpha_R}) \leq \widehat{v}_{\Mdp} \leq v_{\Mdp}(\frac{1+\alpha_R}{1-\alpha_R})$ and $|\widehat{v}_{\Mdp} - v_{\Mdp}| \leq r_{max} \frac{2 \alpha_R}{1 - \alpha_R}$. 
\end{lemma}

The proof of this lemma can be found in Appendix~\ref{app:ctmdp_proofs2}.

\paragraph{Estimating mean-payoff values of MECs}
Using our bounds on the rates of the transitions, we now compute bounds on the mean-payoff values of MECs in CTMDPs. We first show that the mean payoff is maximized or minimized at the boundaries of the estimates of the rates.
Intuitively, to maximise the mean-payoff value, for a state $s_i$ with a high reward, we would like to maximise the time spent in $s_i$ or equivalently, minimise the rate $\lambda(s_i,a)$ for every outgoing action $a$ from $s_i$.
We do the opposite when we want to find a lower bound on the mean-payoff value in the MEC.
Consider an MEC $M$ having states $T=\{s_1, ..., s_m\}$. Assume that $\lambda_i$ is the rate of an action $a$ from state $s_i$, such that a positional mean-payoff maximizing strategy $\sigma$ chooses $a$ from $s_i$. Then, the expected mean-payoff value of $M$ is given by, 
\begin{equation}
    v_{M} = \frac{\sum\limits_{s_{i} \in T} \frac{r\left(s_{i}\right) \pi_{i}}{\lambda_{i}}}{\sum\limits_{s_{i} \in T} \frac{\pi_{i}}{\lambda_{i}}}, \label{eq:mean_payoff_ctmdp}
\end{equation}
where $\pi_{i}$ denotes the expected fraction of total time spent in $s_{i}$ under $\sigma$.

Now, we have estimates $\frac{1}{\widehat{\lambda}_i}$ of $\frac{1}{\lambda}$, such that, $\lambda_{i} \in \left[\widehat{\lambda}_{i} \left(1-\alpha_R\right), \widehat{\lambda}_{i} \left(1+\alpha_R\right)\right]$ with high probability. Let $\lambda_{i}^{l}=\widehat{\lambda}_{i} \left(1-\alpha_R\right)$ and $\lambda_{i}^{u}=\widehat{\lambda}_{i} \left(1+\alpha_R\right)$. 
\begin{proposition}
In Equation~\ref{eq:mean_payoff_ctmdp}, the maximum and the minimum values of $v_{M}$ occur at the boundaries of the estimates of $\lambda_i$ for each $1 \leqslant i \leqslant m$.
\end{proposition}
In particular, $v_{M}$ is maximized when,
\begin{equation}
    \lambda_{i} = 
    \begin{cases}
    \lambda_{i}^{l}, & \text{if}\ r(s_{i}) \geq v_{M} \\
    \lambda_{i}^{u}, & \text{otherwise}
    \end{cases}
    \label{eq:lambda_value_max}
\end{equation}

Once we fix the rates for each of the states in $M$, we uniformize $M$ to obtain a uniform CTMDP $M_C$ which is an MEC and can be treated as an MDP for computing its mean-payoff value~\cite{Puterman}.
Let for a state-action pair, the rate be $\lambda(s,a)$, and the uniformization constant be $C$.
For a successor $t$ from $s$ under action $a$ such that $t\neq s$, we have $\Delta(s,a,t)=\frac{\#(s,a,t)}{\#(s,a)} \cdot \frac{\lambda(s,a)}{C}$, and $\Delta(s,a,s)=1-\sum\limits_{t \neq s}\Delta(s,a,t)$.
Finally, value iteration on $M_C$ with appropriate confidence width gives us the lower and the upper estimates of the mean-payoff value of the MEC $M$.

We now describe an iterative procedure to identify those states of the MEC for which the upper bound on the estimates of the rates are assigned, and those states for which the lower bound on the estimates of the rates are assigned in order to maximize or minimize the mean-payoff value of the MEC.
Assume w.l.o.g. that the states $s_1, \dots, s_m$ are sorted in decreasing order of their rewards $r(s_i)$.
In iteration $j$, we set $\lambda_i=\lambda_i^l$ for $1 \leqslant i \leqslant j$, and we set $\lambda_i=\lambda_i^u$ for the remaining states and recompute $v_M$. The maximum value of $v_M$ across all iterations gives the upper bound on $v_M$. Similarly we can find the lower bound on $v_{M}$. A pseudocode for the algorithm appears in Algorithm~\ref{alg:mecMeanPayoffBoundsTrueAlgorithm}.
Overall, value iteration is done $2|T|$ times\footnote{In our experiments, we use a heuristic to estimate $v_M$ that provides good approximate bounds and is more efficient. 
We first compute an initial estimate of $\widehat{v}_{M}$
using our current estimates, $\widehat{\lambda}$. We then compute the upper bound by assigning the rates as in Equation~\ref{eq:lambda_value_max} where $v_M$ is replaced with $\widehat{v}_{M}$. Similarly, the lower bound can also be found. A detailed pseudocode of this algorithm is described in Algorithm~\ref{alg:maximumMecMeanPayoffPracticalAlgorithm}.}.

\paragraph{Overall Algorithm}
As stated in the beginning of this section, an algorithm for computing the mean payoff in blackbox CTMDP models largely follows the same overall framework as stated in Section~\ref{sec:algo}.
By sampling the actions, we obtain estimates of the rates and the transition probabilities.
The reachability probabilities to the MECs of the CTMDP are estimated using the estimates of the transition probabilities while the mean-payoff values of MECs are estimated using uniformization as decribed above.
The confidence widths on the transition probabilities in a uniformized MEC are assigned based on the number of samples $\#(s,a)$ for a state-action pair $(s,a)$.
A detailed pseudocode of this procedure along with the overall algorithm is provided in Algorithm~\ref{alg:ctmdp2} in Appendix~\ref{app:ctmdp_algo2}.

\paragraph{Statistical guarantees}
Let $\symbolTpInconfidence$ and $\delta_R$ be the $\tpInconfidence$ and the inconfidence on individual transition rates, respectively. Further, let $\delta_{MP1}$ and $\delta_{MP2}$ be the overall inconfidence on the transition probabilities and transition rates, respectively.
Then, $\symbolTpInconfidence := \dfrac{{\symbolMpInconfidence}_{1}\cdot \pmin}{\lvert\{{a} \vert\, \state \in \widehat{S} \wedge {a} \in \av(s) \}\rvert}$, and $\delta_R := \dfrac{{\symbolMpInconfidence}_{2}}{\lvert\{{a} \vert\, \state \in \widehat{S} \wedge {a} \in \av(s) \}\rvert}$.
Thus, we have that the overall inconfidence on the mean-payoff value, $\delta_{MP} = \deltaMPOne+\deltaMPTwo$.
Thus, to achieve a given inconfidence on the mean-payoff value, we fix $\symbolTpInconfidence$ and $\delta_R$, and adjust the imprecisions $\varepsilon_{TP}$ and $\alpha_R$ accordingly.\footnote{See Appendix~\ref{app:numberSamplesCTMDP2} for a more detailed calculation of the number of samples required to make transition probabilities and the rates precise.}

As in the case of MDPs, our learning algorithm for blackbox CTMDP models is an anytime algorithm that is PAC for the given \mpInconfidence $\symbolMpInconfidence$.

\section{Experimental Results} \label{sec:results}
\todo{Add a distribution of number of times MECs are explored.}
We implemented our algorithms as an extension of {\sc Prism}~\cite{PRISM} and tested it on $15$ MDP benchmarks and $10$ CTMDP benchmarks. Several of these benchmarks were selected from the Quantitative Verification Benchmark Set~\cite{HKPQR19}\footnote{The CTMDP benchmarks are available as Markov automata models that were converted to CTMDP models using a tool developed in the thesis~\cite{Butkova20}. More details about our benchmarks and the respective parameters we use can be found in Appendix~\ref{app:moreExp}.}. The results for MDP and CTMDP blackbox learning are shown in Table~\ref{tab:results} and Table~\ref{tab:ctmdpResults} respectively. 
Here, we scale the upper and lower bounds to 1 and 0, and show the average values taken over 10 experiments. 
The experiments were run on a desktop machine with an Intel $i5$ $3.2$ GHz quad core processor and $16$ GB RAM.
The $\mpImprecision$ $\symbolMpImprecision$ is set to $10^{-2}$, {\sf revisitThreshold} $k$ is set to $6$, $\mpInconfidence$ $\symbolMpInconfidence$ is set to $0.1$ and $n$ is set to $10000$. We further use a timeout of $30$ minutes.
In the case of a timeout, the reported upper and lower bounds on the mean payoff still correspond to the input MP-inconfidence $\symbolMpInconfidence$, although the $\mpImprecision$may not be the desired one.

\begin{figure}[t]
    \centering
    \begin{subfigure}{.4\textwidth}
        \includegraphics[width=1\linewidth]{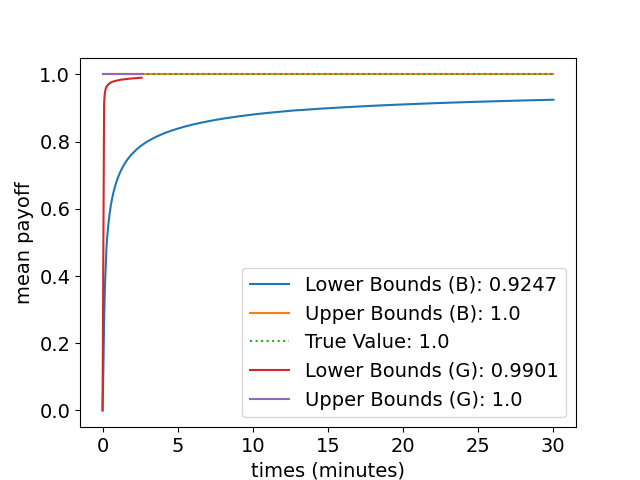}
        \caption{zeroconf}
        \label{fig:zeroconf}
    \end{subfigure}%
    \begin{subfigure}{.4\textwidth}
        \includegraphics[width=1\linewidth]{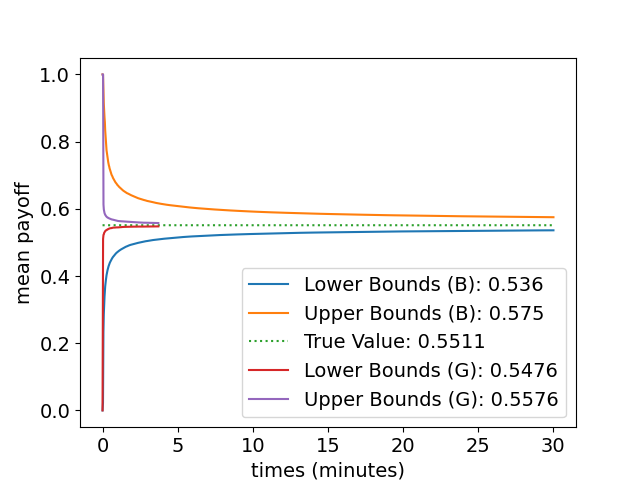}
        \caption{pacman}
        \label{fig:pacman}
    \end{subfigure}
    \begin{subfigure}{.4\textwidth}
        \includegraphics[width=1\linewidth]{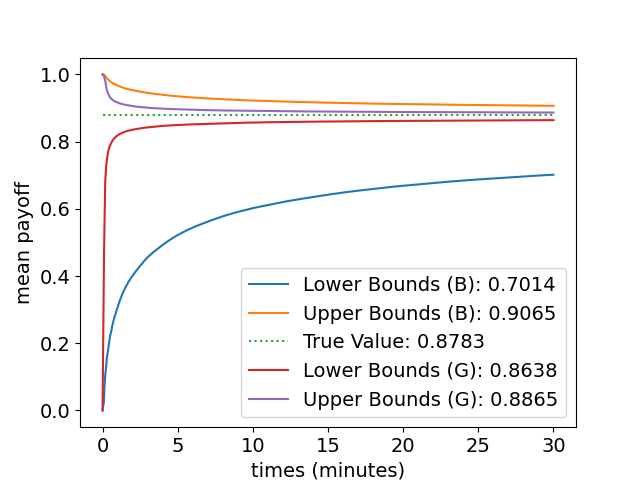}
        \caption{QueuingSystem}
        \label{fig:QueuingSystem}
    \end{subfigure}
    \begin{subfigure}{.4\textwidth}
        \includegraphics[width=1\linewidth]{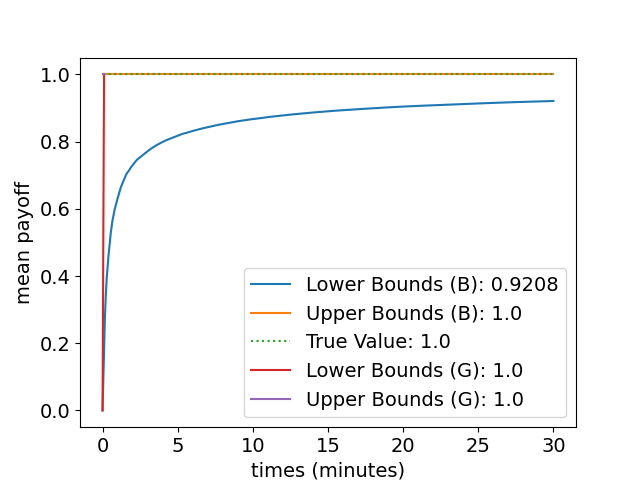}
        \caption{SJS3}
        \label{fig:SJS-procn_6_jobn_2_sctmdp}
    \end{subfigure}
\captionsetup{justification=centering}
\caption{Convergence of lower and upper bounds for blackbox update and greybox update equations. More plots can be found in Appendix~\ref{app:more_plots_mdp_benchmarks} and \ref{app:more_plots_ctmdp_benchmarks}}
\end{figure}

\paragraph{Blackbox learning for MDPs} We see that in Table~\ref{tab:results} for blackbox learning, $9$ out of $15$ benchmarks converge well, such that the precision is within $0.1$.
In fact, for many of these $9$ benchmarks, a precision of $0.1$ is achieved much before the timeout (TO).
In particular, Figure~\ref{fig:zeroconf} and Figure~\ref{fig:pacman}, show this for {\sf zeroconf} and {\sf pacman}. More such plots for MDP benchmarks can be found in Appendix~\ref{app:more_plots_mdp_benchmarks}.
{\sf zeroconf} has a large transient part and a lot of easily reachable single state MECs. Since it has a true value of $1$, the upper and the lower values converge after exploring only a few MECs. We note that our algorithm only needed to explore a very small percentage of the states to attain the input precision.
{\sf cs\_nfail} has many significant MECs, and the learning algorithm needs to explore each of these MECs, while in {\sf sensor} there is a relatively large MEC of around $30$ states, and the simulation inside this MEC takes considerable amount of time.

\begin{table}[t]
\scalebox{0.9}{
\centering
\begin{tabular}{|l|l|l|llll|llll|}
\hline
\multicolumn{1}{|c|}{\multirow{2}{*}{Benchmarks}} &
  \multicolumn{1}{c|}{\multirow{2}{*}{\begin{tabular}[c]{@{}c@{}}Number\\ of states\footnotemark\end{tabular}}} &
  \multicolumn{1}{c|}{\multirow{2}{*}{Value}} &
  \multicolumn{4}{c|}{Blackbox} &
  \multicolumn{4}{c|}{\begin{tabular}[c]{@{}c@{}}Blackbox with \\ greybox update equations\end{tabular}} \\ \cline{4-11} 
\multicolumn{1}{|c|}{} &
  \multicolumn{1}{c|}{} &
  \multicolumn{1}{c|}{} &
  \multicolumn{1}{l|}{\begin{tabular}[c]{@{}l@{}}States \\ explored\end{tabular}} &
  \multicolumn{1}{l|}{\begin{tabular}[c]{@{}l@{}}Lower \\ bound\end{tabular}} &
  \multicolumn{1}{l|}{\begin{tabular}[c]{@{}l@{}}Upper \\ bound\end{tabular}} &
  \begin{tabular}[c]{@{}l@{}}Time \\ (s)\end{tabular} &
  \multicolumn{1}{l|}{\begin{tabular}[c]{@{}l@{}}States\\ explored\end{tabular}} &
  \multicolumn{1}{l|}{\begin{tabular}[c]{@{}l@{}}Lower\\ bound\end{tabular}} &
  \multicolumn{1}{l|}{\begin{tabular}[c]{@{}l@{}}Upper \\ bound\end{tabular}} &
  \begin{tabular}[c]{@{}l@{}}Time\\ (s)\end{tabular} \\ \hline
virus &
  809 &
  0 &
  \multicolumn{1}{l|}{809} &
  \multicolumn{1}{l|}{0.0} &
  \multicolumn{1}{l|}{0.5319} &
  TO &
  \multicolumn{1}{l|}{809} &
  \multicolumn{1}{l|}{0.0} &
  \multicolumn{1}{l|}{0.008} &
  273.01 \\ \hline
cs\_nfail &
  184 &
  0.333 &
  \multicolumn{1}{l|}{184} &
  \multicolumn{1}{l|}{0.3275} &
  \multicolumn{1}{l|}{0.3618} &
  TO &
  \multicolumn{1}{l|}{184} &
  \multicolumn{1}{l|}{0.332} &
  \multicolumn{1}{l|}{0.337} &
  126.77 \\ \hline
investor &
  6688 &
  0.95 &
  \multicolumn{1}{l|}{6284} &
  \multicolumn{1}{l|}{0.8458} &
  \multicolumn{1}{l|}{0.9559} &
  TO &
  \multicolumn{1}{l|}{5835} &
  \multicolumn{1}{l|}{0.945} &
  \multicolumn{1}{l|}{0.954} &
  620.23 \\ \hline
zeroconf &
  3001911 &
  TO &
  \multicolumn{1}{l|}{487} &
  \multicolumn{1}{l|}{0.923} &
  \multicolumn{1}{l|}{1.0} &
  TO &
  \multicolumn{1}{l|}{360} &
  \multicolumn{1}{l|}{0.990} &
  \multicolumn{1}{l|}{1.0} &
  116.04 \\ \hline
sensors &
  189 &
  0.333 &
  \multicolumn{1}{l|}{189} &
  \multicolumn{1}{l|}{0.3299} &
  \multicolumn{1}{l|}{0.3513} &
  TO &
  \multicolumn{1}{l|}{189} &
  \multicolumn{1}{l|}{0.332} &
  \multicolumn{1}{l|}{0.336} &
  64.64 \\ \hline
consensus &
  272 &
  0.1083 &
  \multicolumn{1}{l|}{272} &
  \multicolumn{1}{l|}{0.093} &
  \multicolumn{1}{l|}{0.1605} &
  TO &
  \multicolumn{1}{l|}{272} &
  \multicolumn{1}{l|}{0.103} &
  \multicolumn{1}{l|}{0.113} &
  190.32 \\ \hline
ij10 &
  1023 &
  1 &
  \multicolumn{1}{l|}{1023} &
  \multicolumn{1}{l|}{0.3626} &
  \multicolumn{1}{l|}{1.0} &
  TO &
  \multicolumn{1}{l|}{1023} &
  \multicolumn{1}{l|}{0.999} &
  \multicolumn{1}{l|}{1.0} &
  26.822 \\ \hline
ij3 &
  7 &
  1 &
  \multicolumn{1}{l|}{7} &
  \multicolumn{1}{l|}{0.990} &
  \multicolumn{1}{l|}{1.0} &
  15.92 &
  \multicolumn{1}{l|}{7} &
  \multicolumn{1}{l|}{0.999} &
  \multicolumn{1}{l|}{1.0} &
  0.7127 \\ \hline
pacman &
  498 &
  0.5511 &
  \multicolumn{1}{l|}{496} &
  \multicolumn{1}{l|}{0.5356} &
  \multicolumn{1}{l|}{0.5754} &
  TO &
  \multicolumn{1}{l|}{496} &
  \multicolumn{1}{l|}{0.5477} &
  \multicolumn{1}{l|}{0.5577} &
  215.36 \\ \hline
wlan &
  2954 &
  1 &
  \multicolumn{1}{l|}{2954} &
  \multicolumn{1}{l|}{0.6577} &
  \multicolumn{1}{l|}{1.0} &
  TO &
  \multicolumn{1}{l|}{2935} &
  \multicolumn{1}{l|}{1.0} &
  \multicolumn{1}{l|}{1.0} &
  16.924 \\ \hline
blackjack &
  3829 &
  0 &
  \multicolumn{1}{l|}{3829} &
  \multicolumn{1}{l|}{0.0} &
  \multicolumn{1}{l|}{0.3014} &
  TO &
  \multicolumn{1}{l|}{3829} &
  \multicolumn{1}{l|}{0.0} &
  \multicolumn{1}{l|}{0.006} &
  91.503 \\ \hline
counter &
  8 &
  0.5 &
  \multicolumn{1}{l|}{8} &
  \multicolumn{1}{l|}{0.4998} &
  \multicolumn{1}{l|}{0.5} &
  30.37 &
  \multicolumn{1}{l|}{8} &
  \multicolumn{1}{l|}{0.4999} &
  \multicolumn{1}{l|}{0.5} &
  15.215 \\ \hline
recycling &
  5 &
  0.727 &
  \multicolumn{1}{l|}{5} &
  \multicolumn{1}{l|}{0.726} &
  \multicolumn{1}{l|}{0.727} &
  1.309 &
  \multicolumn{1}{l|}{5} &
  \multicolumn{1}{l|}{0.726} &
  \multicolumn{1}{l|}{0.727} &
  0.927 \\ \hline
busyRing &
  1912 &
  1 &
  \multicolumn{1}{l|}{1733} &
  \multicolumn{1}{l|}{0.706} &
  \multicolumn{1}{l|}{1.0} &
  TO &
  \multicolumn{1}{l|}{1542} &
  \multicolumn{1}{l|}{0.999} &
  \multicolumn{1}{l|}{1.0} &
  34.86 \\ \hline
busyRingMC &
  2592 &
  1 &
  \multicolumn{1}{l|}{2574} &
  \multicolumn{1}{l|}{0.969} &
  \multicolumn{1}{l|}{1.0} &
  TO &
  \multicolumn{1}{l|}{2507} &
  \multicolumn{1}{l|}{0.999} &
  \multicolumn{1}{l|}{1.0} &
  114.50 \\ \hline
\end{tabular}
}
\caption{\label{tab:results} Results on MDP benchmarks.}
\end{table}

\todo{To save space, we can omit this paragraph.}
{\sf virus} consists of a single large MEC of more than $800$ states, and its true value is $0$.
As we simulate the MEC more and more, the \tpImprecision on the transition probabilities decreases and the upper bound on the mean-payoff reduces over time.
{\sf ij10} contains one MEC with $10$ states in it. The value converges faster and reaches a value of $1$, during blackbox learning. This model has relatively high number of actions, more than 5, for many of its states outside the MEC. This leads to a higher TP-imprecision. Further, due to the conservative nature of the blackbox update equations, the upper and the lower values converge very slowly. 

\footnotetext{The number of states and the values are computed using the probabilistic model-checker {\sc Storm}~\cite{DJKV17}}
{\sf consensus, ij10, ij3, pacman, wlan} were used in \cite{AKW19} for learning policies for reachability objectives.
The target states in these benchmarks are sink states with self loops, and we add a reward of $1$ on these target states so that the rechability probability becomes the same as the mean payoff.
Running these modified benchmarks with mean-payoff objective gives similar upper and lower bounds as the bounds reported for reachability probability in~\cite{AKW19}, and our experiments also take similar time as reported in~\cite{AKW19}.


The {\sf blackjack} model~\cite{SB98} is similar to {\sf zeroconf} model. It has $3829$ states and $2116$ MECs. It has a large transient part and a lot of single state MECs. \todo{It's not clear from the text, why this results in slower convergence.} But unlike, {\sf zeroconf} all of the MECs have a value of $0$. Thus, simulation takes more time as the \tpImprecision reduces slowly. 


\paragraph{Blackbox learning with greybox update equations for MDPs} We show the results of these experiments in the right side of Table~\ref{tab:results}.
As observed, convergence is much faster here for all the benchmarks.
In fact, all our benchmarks converged correctly within a few seconds to a few minutes.
Hence for a small degradation in \mpInconfidence use of greybox update equations works well in practice.
We show the effect on \mpInconfidence in more detail in Table~\ref{tab:errorProb2} in Appendix~\ref{app:greyboxerrorprobability}.

\begin{table}[]
\centering
\scalebox{0.9}{
\begin{tabular}{|l|l|l|llll|llll|}
\hline
\multicolumn{1}{|c|}{\multirow{2}{*}{Benchmarks}} &
  \multicolumn{1}{c|}{\multirow{2}{*}{\begin{tabular}[c]{@{}c@{}}Number\\ of states\end{tabular}}} &
  \multicolumn{1}{c|}{\multirow{2}{*}{Value}} &
  \multicolumn{4}{c|}{Blackbox} &
  \multicolumn{4}{c|}{\begin{tabular}[c]{@{}c@{}}Blackbox with \\ greybox update equations\end{tabular}} \\ \cline{4-11} 
\multicolumn{1}{|c|}{} &
  \multicolumn{1}{c|}{} &
  \multicolumn{1}{c|}{} &
  \multicolumn{1}{c|}{\begin{tabular}[c]{@{}c@{}}States\\ explored\end{tabular}} &
  \multicolumn{1}{c|}{\begin{tabular}[c]{@{}c@{}}lower\\ bound\end{tabular}} &
  \multicolumn{1}{c|}{\begin{tabular}[c]{@{}c@{}}upper\\ bound\end{tabular}} &
  \multicolumn{1}{c|}{\begin{tabular}[c]{@{}c@{}}Time\\ (s)\end{tabular}} &
  \multicolumn{1}{c|}{\begin{tabular}[c]{@{}c@{}}States\\ explored\end{tabular}} &
  \multicolumn{1}{c|}{\begin{tabular}[c]{@{}c@{}}lower\\ bound\end{tabular}} &
  \multicolumn{1}{c|}{\begin{tabular}[c]{@{}c@{}}upper\\ bound\end{tabular}} &
  \multicolumn{1}{c|}{\begin{tabular}[c]{@{}c@{}}Time\\ (s)\end{tabular}} \\ \hline
DynamicPM &
  816 &
  1.0 &
  \multicolumn{1}{l|}{816} &
  \multicolumn{1}{l|}{0.436} &
  \multicolumn{1}{l|}{1.0} &
  TO &
  \multicolumn{1}{l|}{816} &
  \multicolumn{1}{l|}{0.998} &
  \multicolumn{1}{l|}{1.0} &
  37.68 \\ \hline
ErlangStages &
  508 &
  1.0 &
  \multicolumn{1}{l|}{508} &
  \multicolumn{1}{l|}{0.962} &
  \multicolumn{1}{l|}{1.0} &
  TO &
  \multicolumn{1}{l|}{508} &
  \multicolumn{1}{l|}{0.999} &
  \multicolumn{1}{l|}{1.0} &
  8.118 \\ \hline
PollingSystem1 &
  16 &
  0.922 &
  \multicolumn{1}{l|}{16} &
  \multicolumn{1}{l|}{0.811} &
  \multicolumn{1}{l|}{0.937} &
  TO &
  \multicolumn{1}{l|}{16} &
  \multicolumn{1}{l|}{0.816} &
  \multicolumn{1}{l|}{0.937} &
  TO \\ \hline
PollingSystem2 &
  348 &
  0.999 &
  \multicolumn{1}{l|}{348} &
  \multicolumn{1}{l|}{0.637} &
  \multicolumn{1}{l|}{0.999} &
  TO &
  \multicolumn{1}{l|}{348} &
  \multicolumn{1}{l|}{0.998} &
  \multicolumn{1}{l|}{0.999} &
  21.893 \\ \hline
PollingSystem3 &
  1002 &
  0.999 &
  \multicolumn{1}{l|}{1002} &
  \multicolumn{1}{l|}{0.232} &
  \multicolumn{1}{l|}{1.0} &
  TO &
  \multicolumn{1}{l|}{1002} &
  \multicolumn{1}{l|}{0.99} &
  \multicolumn{1}{l|}{1.0} &
  864.05 \\ \hline
QueuingSystem &
  266 &
  0.8783 &
  \multicolumn{1}{l|}{266} &
  \multicolumn{1}{l|}{0.703} &
  \multicolumn{1}{l|}{0.906} &
  TO &
  \multicolumn{1}{l|}{266} &
  \multicolumn{1}{l|}{0.865} &
  \multicolumn{1}{l|}{0.886} &
  TO \\ \hline
SJS1 &
  17 &
  1.0 &
  \multicolumn{1}{l|}{17} &
  \multicolumn{1}{l|}{0.999} &
  \multicolumn{1}{l|}{1.0} &
  133.96 &
  \multicolumn{1}{l|}{17} &
  \multicolumn{1}{l|}{0.997} &
  \multicolumn{1}{l|}{1.0} &
  1.05 \\ \hline
SJS2 &
  7393 &
  0.999 &
  \multicolumn{1}{l|}{7341} &
  \multicolumn{1}{l|}{0.02} &
  \multicolumn{1}{l|}{1.0} &
  TO &
  \multicolumn{1}{l|}{7268} &
  \multicolumn{1}{l|}{0.936} &
  \multicolumn{1}{l|}{1.0} &
  TO \\ \hline
SJS3 &
  433 &
  1.0 &
  \multicolumn{1}{l|}{433} &
  \multicolumn{1}{l|}{0.919} &
  \multicolumn{1}{l|}{1.0} &
  TO &
  \multicolumn{1}{l|}{432} &
  \multicolumn{1}{l|}{0.999} &
  \multicolumn{1}{l|}{0.999} &
  5.3814 \\ \hline
toy &
  12 &
  1.0 &
  \multicolumn{1}{l|}{12} &
  \multicolumn{1}{l|}{0.99} &
  \multicolumn{1}{l|}{1.0} &
  5.6 &
  \multicolumn{1}{l|}{12} &
  \multicolumn{1}{l|}{0.999} &
  \multicolumn{1}{l|}{1.0} &
  1.112 \\ \hline
\end{tabular}
}
\caption{Results on CTMDP benchmarks}
\label{tab:ctmdpResults}
\end{table}
\footnotetext{The number of states and the true mean-payoff values are computed by first uniformizing the CTMDP, and then using {\sc Storm} on the underlying MDP.}

\paragraph{Blackbox learning for CTMDPs} In Table~\ref{tab:ctmdpResults} we show the results for CTMDP benchmarks.
%
The number of states in these benchmarks vary from as low as $12$ to more than $7000$. All the models that we use here have a lot of small end-components in them. 
We observe that the upper and the lower values take more time to converge as the size of the model grows.
Figure~\ref{fig:QueuingSystem} and Figure~\ref{fig:SJS-procn_6_jobn_2_sctmdp} show the convergence of lower and upper bounds for {\sf QueuingSystem} and {\sf SJS3}. Similar plots for other CTMDP benchmarks can be found in Appendix~\ref{app:more_plots_ctmdp_benchmarks}.
As in the case of MDPs, using \emph{greybox update equations} speeds up the learning process significantly.

\paragraph{Additional experimental results} \todo{We already write this in footnote 12. We should remove the repetition.}We performed experiments on our benchmarks to compare the results if $\max\limits_{s \in \states,a \in \Av(s)} |\post(s,a)|$, is known.
The results of these experiments show a small improvement in convergence when this information is available (see Appendix~\ref{app:pminandmaxsuccessorcomparison} for details).

We also ran experiments for \emph{greybox learning}.
Recall from Definition~\ref{def:limit} that in greybox learning, for every state-action pair, we know the number of successors of the state for the given action. 
As expected, their convergence is much faster than that for blackbox learning, but the convergence is comparable to the case where we do blackbox learning with greybox update equations.
The details of the greybox learning experiments can be found in Appendix~\ref{app:greybox}. 

\section{Conclusion} \label{sec:conc}
We presented the first PAC SMC algorithm for computing mean payoff in unknown MDPs and CTMDPs, where the only information needed is a lower bound on minimum transition probability, as advocated in \cite{DacaHenzingerKretinskyPetrov2016}.
In contrast to a naive algorithm, which follows in a quite straightforward way from the literature, our algorithm is practically applicable, overcoming the astronomic number of simulation steps required.
To this end, in particular, the inconfidence had to be distributed in non-uniformly over the transitions and then imprecision propagated by value iteration with precision guarantees.
In future, we would like to thoroughly analyse how well weakening the PAC bounds can be traded for a yet faster convergence.
On the practical side, applying importance sampling and importance splitting could further improve the efficiency.

\subsubsection*{Acknowledgements}
The second author would like to thank Subhajit Goswami for insightful discussions on learning transition rate matrix in a CTMDP and for pointing to useful references.

\bibliographystyle{plain}
\bibliography{ref}


\newpage
\appendix
\begin{center}
    {\Large Appendix}
\end{center}
\section{Additional Definitions, Examples} \label{app:def}

Here we provide additional detailed definitions used in the full proofs in the appendix, and also provide examples to illustrate some of the definitions used in the paper.

\subsection{About MEC Quotients}
Here we provide the definition of various kinds of MEC quotients which is a standard tool to compute mean payoff in MDPs when they are completely known.

To obtain a MEC quotient~\cite{DeAlfaro1997}, each MEC is merged into a single representative state, while transitions between MECs are preserved.
Intuitively, this abstracts the MDP to its essential infinite time behaviour.
The quotient also includes for every MEC $M$ the self loops labelled by those actions for which there are transitions within the MEC $M$. The probabilities over such transitions in the self-loops are preserved.

During blackbox learning, we will only have $\deltasure$ EC, and as we do more exploration, previously identified end components may change, in which case it will be hard to update these representations. So instead we use, Stay-augumented MDP as described in Section~\ref{sec:overview}.

For an MDP $\Mdp$, its MEC quotient is denoted by $\widehat{\Mdp}$, and the merged representatives of the MECs $M_1, \dots, M_n$ are denoted by $\widehat{s}_1, \dots, \widehat{s}_n$ respectively.

\begin{definition}[MEC quotient~\cite{DeAlfaro1997}] \label{def:mq}
	Let $\Mdp = (\states, \initstate, \actions, \av, \trans, \rew)$ be an MDP with MECs $\mec(\Mdp) = \{(T_1, A_1), \dots, (T_n, A_n)\}$.
	Further, define $\mec_{\states} = \Union_{i=1}^n T_i$ as the set of all states contained in some MEC.
	The \emph{MEC quotient of} $\Mdp$ is defined as the MDP $\widehat\Mdp = (\widehat\states, \widehat{s}_\textrm{init}, \widehat\actions, \widehat\av, \widehat\trans, \widehat\rew)$, where:
	\begin{itemize}
		\item $\widehat\states = \states \setminus \mec_{\states} \union \set{\widehat{s}_1, \dots, \widehat{s}_n}$,
		\item if for some $T_i$ we have $\initstate\in T_i$, then $\widehat{s}_\textrm{init} = \widehat{s}_i$,  otherwise $\widehat{s}_\textrm{init} = \initstate$,
		\item $\widehat\actions =  \set{(s,a) \mid s\in \states, a\in \av(s)}$,
		\item the available actions $\widehat\av$ are defined as
		\begin{align*}
			\forall s \in \states \setminus \mec_{\states}.~& \widehat\av(s) = \set{(s,a) \mid a \in \av(s)} \\
			\forall 1 \leqslant i \leqslant n.~& \widehat\av(\widehat{s}_i) = \set{(s,a) \mid s \in T_i \land a \in \av(s) \setminus A_i},
		\end{align*}
		\item the transition function $\widehat\trans$ is defined as follows.
		Let $\widehat{s} \in \widehat{S}$ be some state in the quotient and $(s, a) \in \av(\widehat{s})$ an action available in $\widehat{s}$.
		Then
		\begin{equation*}
			\widehat{\trans}(\widehat{s}, (s, a), \widehat{s}') = \begin{dcases*}
				{\sum}_{s' \in T_j} \trans(s, a, s') & if $\widehat{s}' = \widehat{s}_j$, \\
				\trans(s, a, \widehat{s}') & otherwise, i.e.\ $\widehat{s}' \in \states \setminus \mec_{\states}$.
			\end{dcases*}
		\end{equation*}
		For the sake of readability, we omit the added self-loop transitions of the form $\trans(\widehat{s}_i, (s, a), \widehat{s}_i)$ with $s \in T_i$ and $a \in A_i$ from all figures.
		\item Finally, for $\widehat{s} \in \widehat{\states}$, $(s, a) \in \widehat{\av}(\widehat{s})$, we define $\widehat\rew(s, (s, a)) = \rew(s, a)$.
	\end{itemize}
	Furthermore, we refer to $\widehat{s}_1, \dots, \widehat{s}_n$ as \emph{collapsed states} and identify them with the corresponding MECs.
\end{definition}
For the sake of readability, we omit the added self-loop transitions of the form $\trans(\widehat{s}_i, (s, a), \widehat{s}_i)$ with $s \in T_i$ and $a \in A_i$ from all figures.

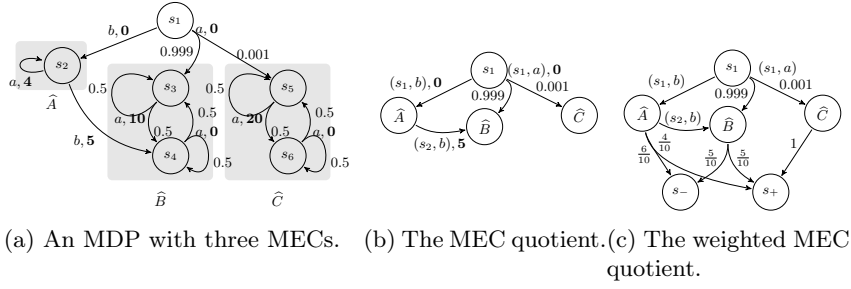
\begin{figure}[t]
	\centering
	\subfloat[An MDP with three MECs.]{\label{fig:example}
		\scalebox{0.6}{
			\begin{tikzpicture}[auto]
				\node[state] (s1) at (-3,0.5) {$s_1$};
				\node[state] (s6) at (-0.5,-2.5) {$s_6$};
				\node[state] (s5) at (-0.5,-1) {$s_5$};
				\node[state] (s3) at (-3.1,-1) {$s_3$};
				\node[state] (s4) at (-3.1,-2.5) {$s_4$};
				\node[state] (s2) at (-5.5,-0.5) {$s_2$};
				\node[] at (-2.3,0.3) {$a, \rewformat{0}$};
				\node[] at ($(s4) + (0.8,0.5)$) {$a,\rewformat{0}$};
				\node[] at ($(s6) + (0.8,0.5)$) {$a, \rewformat{0}$};
				\node[] at ($(s3) + (-0.9,-0.7)$) {$a, \rewformat{10}$};
				\node[] at ($(s5) + (-0.9,-0.7)$) {$a, \rewformat{20}$};
			
				\path[->]
					(s1) edge node[right]{$0.001$} (s5)
					(s1) edge[out=-30,in=45] node[left]{$0.999$} (s3)
					(s1) edge node[above] {$b, \rewformat{0}$} (s2)
					(s5) edge[out=225,in=135] node[below right]{0.5} (s6)
					(s5) edge[out=225,in=155,looseness=8] coordinate (e5loop) node[above left]{$0.5$} (s5)
					(s6) edge[out=45,in=-35,looseness=4] coordinate (e6loop) node[below right] {$0.5$} (s6)
					(s6) edge[out=45,in=-45] node[above right] {$0.5$} (s5)
					(s3) edge[out=225,in=135] node[below right]{0.5} (s4)
					(s3) edge[out=225,in=155,looseness=8] coordinate (e3loop) node[above left]{$0.5$} (s3)
					(s4) edge[out=45,in=-35,looseness=4] coordinate (e4loop) node[below right] {$0.5$} (s4)
					(s4) edge[out=45,in=-45] node[above right] {$0.5$} (s3)
					(s2) edge[loop left] coordinate (e2loop) node[below, inner sep=7pt]{$a, \rewformat{4}$} (s2)
					(s2) edge[out=-80,in=180,bend right] node[below left] {$b, \rewformat{5}$} (s4);
				
				\node[rectangle,rounded corners=3pt,draw=none,fill=black,fill opacity=0.1,fit=(s2) (e2loop)] (rectA) {};
				\node[rectangle,rounded corners=3pt,draw=none,fill=black,fill opacity=0.1,fit=(s3) (s4) (e3loop) (e4loop)] (rectB) {};
				\node[rectangle,rounded corners=3pt,draw=none,fill=black,fill opacity=0.1,fit=(s5) (s6) (e5loop) (e6loop)] (rectC) {};
				\node[] at ($(rectA) + (0,-0.8)$) {$\widehat{A}$};
				\node[] at ($(rectB) + (0,-1.7)$) {$\widehat{B}$};
				\node[] at ($(rectC) + (0,-1.7)$) {$\widehat{C}$};
			\end{tikzpicture}
		}
	}
	\subfloat[The MEC quotient.]{\label{fig:quo}
		\scalebox{0.6}{
			\begin{tikzpicture}[auto, align=center]
				\node[state] (s0) at (-3,0.5) {$s_1$};
				\node[state] (s2) at (-1,-0.5) {$\widehat C$};
				\node[state] (s3) at (-3.1,-0.75) {$\widehat B$};
				\node[state] (s5) at (-5,-0.5) {$\widehat A$};
				
				\node[] at (-2,0.5) {$(s_1,a), \rewformat{0}$};
				
				\path[->]
				(s0) edge node[anchor=south west,inner sep=1pt] {$0.001$} (s2)
				(s0) edge[out=-30,in=45] node[left]{$0.999$} (s3)
				(s0) edge node[above left, inner sep=1pt] {$(s_1, b), \rewformat{0}$} (s5)
				(s5) edge[out=-30,in=190] node[below] {$(s_2, b), \rewformat{5}$} (s3);
				
				\node[] at (-2,-2.5) {}; 
			\end{tikzpicture}
		}
	}
	\subfloat[The weighted MEC quotient.]{\label{fig:wquo}
		\scalebox{0.6}{
			\begin{tikzpicture}[auto, align=center]
			\node[state] (s0) at (-3,0.5) {$s_1$};
			\node[state] (s2) at (-1,-0.5) {$\widehat C$};
			\node[state] (s3) at (-3.1,-0.75) {$\widehat B$};
			\node[state] (s5) at (-5,-0.5) {$\widehat A$};
			\node[state] (sp) at (-2.15,-2.25) {$s_+$};
			\node[state] (sm) at (-4.15,-2.25) {$s_-$};
			
			\node[] at (-2,0.5) {$(s_1,a)$};

			\path[->]
			(s0) edge node[anchor=south west,inner sep=1pt] {$0.001$} (s2)
			(s0) edge[out=-30,in=45] node[left]{$0.999$} (s3)
			(s0) edge node[above left, inner sep=1pt] {$(s_1, b)$} (s5)
			(s5) edge[out=-30,in=190] node[above] {$(s_2, b)$} (s3)
			(s5) edge[out=-80,in=170,looseness=0.75] node[anchor=south, inner sep=5pt, pos=0.3]{$\frac{4}{10}$} (sp)
			(s5) edge[out=-80,in=120,looseness=0.5] node[left]{$\frac{6}{10}$} (sm)
			(s2) edge node[left,pos=0.3] {$1$} (sp)
			(s3) edge[out=-90,in=150] node[anchor=south west,inner sep=0pt]{$\frac{5}{10}$} (sp)
			(s3) edge[out=-90,in=30]  node[anchor=south east,inner sep=0pt]{$\frac{5}{10}$} (sm);
			\end{tikzpicture}
		}
	}
	\caption{An example of how the MEC quotient is constructed.
		By $a, \rewformat{r}$ we denote that the action $a$ yields a reward of $\rewformat{r}$.}
\end{figure}

\begin{example}
	Figure~\ref{fig:example} shows an MDP with three MECs, $\widehat{A} = (\set{s_2}, \set{a}), \widehat{B} = (\set{s_3,s_4}, \set{a}), \widehat{C} = (\set{s_5,s_6}, \set{a}))$.
	Its MEC quotient is shown in Figure~\ref{fig:quo}.
	\qed
\end{example}

To define weighted MEC quotient, we define a function $f$ as the normalized approximated value, i.e.\ for some MEC $M_i$ we set $f(\hat{s}_i) = \frac{1}{\rmax} w(M_i)$, so that it takes values in $[0, 1]$.
Then, the probability of reaching $s_+$ upon taking the $\mathsf{stay}$ action in $\hat{s}_i$ is defined as $f(\hat{s}_i)$ and dually the transition to $s_-$ is assigned $1 - f(\hat{s}_i)$ probability.
If for example some MEC $M$ had a value $v(M) = \frac{2}{3} \rmax$, we would have that $\trans(\hat{s}, \mathsf{stay}, s_+) = \frac{2}{3}$.
This way, we can interpret reaching $s_+$ as obtaining the maximal possible reward, and reaching $s_-$ to obtaining no reward.

\begin{definition}[Weighted MEC quotient] \label{def:wmq}
	Let $\widehat\Mdp = (\widehat\states, \hat{s}_\textrm{init}, \widehat\actions, \widehat\av, \widehat\trans, \widehat\rew)$ be the MEC quotient of an MDP $\Mdp$ and let $\mec_{\widehat{S}} = \set{\hat{s}_1, \dots, \hat{s}_n}$ be the set of collapsed states.
	Further, let $f :\mec_{\widehat{S}} \to [0,1]$ be a function assigning a value to every collapsed state.
	We define the \emph{weighted MEC quotient of $\Mdp$ and $f$} as the MDP ${\Mdp}^f = ({\states}^f, {\initstate}^f, \widehat\actions \union \set{\mathsf{stay}}, {\av}^f, {\trans}^f, {\rew}^f)$, where
	\begin{itemize}
		\item ${\states}^f = \widehat\states \union \set{s_+, s_-}$,
		\item $\initstate^f = \hat{s}_\textrm{init}$,
		\item $\av^f$ is defined as
		\begin{align*}
			\forall \hat{s} \in \widehat\states.~& \av^f(\hat{s}) = \begin{dcases*}
				\widehat\av(\hat{s}) \union \set{\mathsf{stay}} &if $\hat{s} \in \mec_{\widehat{S}},$ \\
				\widehat\av(\hat{s}) &otherwise,
			\end{dcases*} \\
			& \av^f(s_+) = \av^f(s_-) = \emptyset,
		\end{align*}
		\item ${\trans}^f$ is defined as
		\begin{align*}
			\forall \hat{s} \in \widehat\states, \hat{a} \in \widehat\actions \setminus \set{\mathsf{stay}}.~& {\trans}^f(\hat{s}, \hat{a}) = \widehat\trans(\hat{s}, \hat{a}) \\
			\forall \hat{s}_i \in \mec_{\widehat{S}}.~& {\trans}^f(\hat{s}_i, \mathsf{stay}) = \set{s_+ \mapsto f(\hat{s}_i), s_- \mapsto 1 - f(\hat{s}_i)},
		\end{align*}
		\item and the reward function $\rew^f(\hat{s}, \hat{a})$ is chosen arbitrarily (e.g.\ $0$ everywhere), since we only consider a reachability problem on $\Mdp^f$.
	\end{itemize}
\end{definition}
\begin{example}
	Consider the MDP in Figure~\ref{fig:example}.
	The average rewards of the MECs are $\gain = \set*{\widehat{A} \mapsto 4, \widehat{B} \mapsto 5, \widehat{C} \mapsto 10}$.
	With $f$ defined as in Theorem~\ref{th:quo}, Figure~\ref{fig:wquo} shows the weighted MEC quotient $\Mdp^f$.
	\qee
\end{example}

\begin{theorem} \label{th:quo}
	Given an MDP $\Mdp$ with MECs $\mec(\Mdp) = \set*{M_1, \dots, M_n}$, define $f(\hat{s}_i) = \tfrac{1}{\rmax} \gain(M_i)$ the function mapping each MEC $M_i$ to its value.
	Moreover, let $\Mdp^f$ be the weighted MEC quotient of $\Mdp$ and $f$.
	Then
	\begin{equation*}
		\gain(\initstate) = \rmax \cdot \sup_{\straa \in \straas} \pr^\straa_{\Mdp^f, \initstate^f}(\Diamond s_+).
	\end{equation*}
\end{theorem}

We formally define bounded MEC quotient from~\cite{cav17} below.
\begin{definition}[Bounded MEC quotient] \label{def:bquo}
	Let $\widehat\Mdp = (\widehat\states, \hat{s}_\textrm{init}, \widehat\actions, \widehat\av, \widehat\trans, \widehat\rew)$ be the MEC quotient of an MDP $\Mdp$ with collapsed states $\mec_{\widehat{\states}} = \set{\hat{s}_1, \dots, \hat{s}_n}$ and let $\lu: \set{\hat{s}_1, \dots, \hat{s}_n} \to [0,1]$ be functions that assign a lower and upper bound, respectively, to every collapsed state in $\widehat\Mdp$.
	The \emph{bounded MEC quotient $\Mdp^{\lu}$ of $\Mdp$ and $\lu$} is 
	similar to the weighted MEC quotient with the following changes.
	\begin{itemize}
		\item ${\states}^{\lu} = \widehat\states \union \set{s_?}$,
		\item $\av^{\lu}(s_?) = \emptyset$,
		\item $\forall \hat{s} \in \mec_{\widehat{S}}.~{\trans}^{\lu}(\hat{s}, \mathsf{stay}) = \set*{s_+ \mapsto \lowerbound(\hat{s}), s_- \mapsto 1 - \upperbound(\hat{s}), s_? \mapsto \upperbound(\hat{s}) - \lowerbound(\hat{s})}$.
	\end{itemize}
\end{definition}

\subsection{Markov Chain}
\begin{definition} [Markov Chain]
A \emph{Markov chain} (MC, for short) is a tuple $\mathcal{M} = \zug{S,E,\pr}$, where $S$ is a set of states, $E \subseteq S \times S$ is a set of edges (we assume in this paper that the set $E(s)$ of outgoing edges from $s$ is nonempty and finite for all $s \in S$), and $\pr: S \to \Distributions(E)$ assigns a probability distribution  -- on the set $E(s)$ of outgoing edges from $s$ -- to all states $s \in S$. In the following, $\pr(s,(s,s'))$ is denoted $\pr(s,s')$, for all $s \in S$. The Markov chain $\markovChain$ is \emph{finite} if $S$ is finite.
\end{definition}

\subsection{Strategies}
Earlier, in the main body of the paper, we defined positional strategies.
We now define strategies in general for completeness and for some of the proofs that appear later in the appendix.

A \emph{finite path} $\fpath = s_0 a_0 s_1 a_1 \dots s_n \in (\states \times \actions)^* \times \states$ is a finite prefix of an
infinite path.

A \emph{strategy} on an MDP is a function $\straa: (\states \times \actions)^*\times \states \to \distributions(\actions)$, which given a finite path $\fpath = s_0 a_0 s_1 a_1 \dots s_n$ yields a probability distribution $\straa(\fpath) \in \distributions(\av(s_n))$ on the actions to be taken next.
We call a strategy \emph{memoryless randomized} (or \emph{stationary}) if it is of the form $\straa: \states \to \distributions(\actions)$, and \emph{memoryless deterministic} (or \emph{positional}) if it is of the form $\straa: \states \to \actions$.
We denote the set of all strategies of an MDP by $\straas$, and the set of all memoryless deterministic strategies by $\straas^{\mathsf{MD}}$.
Fixing any stationary strategy $\straa$ induces a Markov chain.


\subsection{Example of Uniformization}
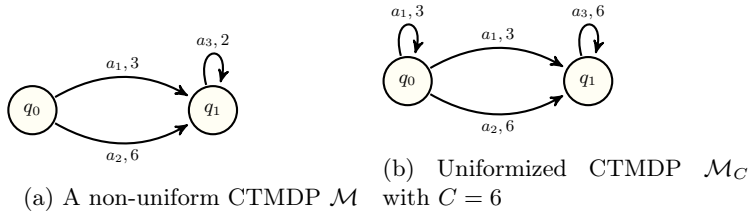
\begin{figure}[t]
    \centering
    \begin{subfigure}[b]{0.4\linewidth}
\begin{tikzpicture}[shorten >=1pt, node distance=3 cm, on grid, auto,thick,initial text=,scale=0.2,]
\begin{scope}[every node/.style={scale=.8}]
\node (l0) [state,fill=safecellcolor]  {$q_0$};
\node (l1) [state, fill=safecellcolor, right = of l0,xshift = -0.75cm]   {$q_1$};
\end{scope}
 \begin{scope}[every node/.style={scale=.7}]
\path [->]
    (l0) edge [bend left]  node [above] {$a_1,3$}   (l1)
    (l0) edge [bend right]  node [below] {$a_2,6$}   (l1)
    (l1) edge [loop above] node [above] {$a_3,2$}   ()
    ;
\end{scope}
\end{tikzpicture}
\caption{A non-uniform CTMDP $\Mdp$} \label{fig:PM1}
\end{subfigure}
\begin{subfigure}[b]{0.4\linewidth}
\begin{tikzpicture}[shorten >=1pt, node distance=3 cm, on grid, auto,thick,initial text=,scale=0.2,]
\begin{scope}[every node/.style={scale=.8}]
\node (l1) [state, fill=safecellcolor]  {$q_1$};
\node (l0) [state, fill=safecellcolor, left = of l1,xshift = 0.75cm]  {$q_0$};

\end{scope}
 \begin{scope}[every node/.style={scale=.7}]
\path [->]
     (l0) edge [loop above] node [above] {$a_1,3$}
    (l0) edge [bend left]  node [above] {$a_1,3$}   (l1)
    (l0) edge [bend right]  node [below] {$a_2,6$}   (l1)
    (l1) edge [loop above] node [above] {$a_3,6$}   ()
    ;
\end{scope}
\end{tikzpicture}
\caption{Uniformized CTMDP $\Mdp_{C}$ with $C = 6$} \label{fig:PM2}
\end{subfigure}
\caption{Uniformization of a CTMDP}
\label{fig:P1}
\end{figure}

We illustrate uniformization of CTMDPs here with an example.
In Figure~\ref{fig:P1}{\sf a} on the left we show a CTMDP, while on the right, in Figure~\ref{fig:P1}{\sf b}, we show its uniformized version.

\subsection{Hoeffding's inequality}
Hoeffding's inequality provides an upper bound $\delta$ on the probability that the sum of bounded independent random variables deviates from its expected value by more than a certain threshold $\epsilon$.
\begin{definition}[Hoeffding's inequality]
Let $X_1, \dots, X_n$ be independent random variables with domain bounded by the interval $[0,1]$, and let $X = \frac{1}{n}\sum_{1}^{n}X_i$.
Then for all $\varepsilon \geqslant 0$,
\[
\mathbb{P}(X - \mathbb{E}(X) \geqslant \epsilon) \leqslant e^{-2\varepsilon^2n}.
\]
\end{definition}
Note that each $X_i$ above can be independent Bernoulli random variables.

\section{Detailed Algorithms and pseudocode from Section \ref{sec:algo}} \label{app:algo}
In this section, we give the details of the procedures used in our blackbox algorithm~(Algorithm~\ref{alg:general}) that could not be provided in the main body of the paper due to lack of space.

The proceudre $\SIMULATE$ (Algorithm~\ref{alg:simulateNewMDP}) is called form Algorithm~\ref{alg:general}. Here, we simulate the MDP $\Mdp$ from the initial state, repeatedly till we encounter any of the sink states. During the simulation process, we update the number of times each transition is visited. We also, keep track of $r_{max}$, the maximum reward from a state (Assume $r_{max}$ is set to 0 before Algorithm~\ref{alg:general} starts). From a state, an action with the best upper bound is chosen to explore the MDP. During this process, we also check whether we are visiting a state too often, by calling $\mathsf{Appear(X,s)} \geqslant k$ and if so, we check for MEC in our partial model, by calling the procedure $\mathsf{LOOPING}$ (Algorithm~\ref{alg:stuck}). The procedure $\mathsf{LOOPING}$ checks for MECs in the partial model, and for all the MECs it identifies, it performs value iteration in them and adds a stay action to all the states in the MEC. If $\mathsf{LOOPING}$ returns true, then we call $\mathsf{BEST\_LEAVING\_ACTION}$, which gives the action with best upper bound, that has some transitions which goes out of the MEC.
\begin{algorithm}[H]
\caption{Simulation algorithm counting occurrences}\label{alg:simulateNewMDP}
		\hspace*{\algorithmicindent}\algorithmicrequire	MDP $\Mdp$, revisit threshold $k$.\\
		\hspace*{\algorithmicindent}\algorithmicensure Simulated path $X$
\begin{algorithmic}[1]
\Procedure{$\SIMULATE$}{}
    \State $X \gets \emptyset$ \Comment{Stack of states visited during this simulation}
    \State $s \gets \initstate$
    \Repeat
    	\State $X \gets X \cup \{s\}$ \Comment{Push $s$ to $X$}
        \State $r_{max} \gets \max(r_{max}, r(s))$
    	\State ${a} \gets $ sampled according to best$_{\ub, \lb}(s)$
    	\If {$a \:!\!\!= {\sf stay}$}
    	\State $t \gets$ sampled according to $\trans(s,{a})$
    	\Else \Comment{The {\sf stay} action simulates reaching the terminal states in $\Mdp'$}
    	\State $t \gets$ sampled according to $\trans'(s,{\sf stay})$
    	\EndIf
    	\State $\text{Increment }\#(s,{a},t)$
    	\State $s \gets t$
        \If {$\mathsf{Appear}(X, s) \geqslant k$}
            \If{$\STUCK(X, s)$}
                \State ${a} = $ $\mathsf{BEST\_LEAVING\_ACTION}(M)$ \Comment{$M$ is the MEC to which $s$ belongs}
                \State $s$ = origin of ${a}$
    	        \If {$a \:!\!\!= {\sf stay}$}
    	        \State $t \gets$ sampled according to $\trans(s,{a})$
    	        \Else \Comment{The {\sf stay} action simulates reaching the terminal states in $\Mdp'$}
    	        \State $t \gets$ sampled according to $\trans'(s,{\sf stay})$
    	    \EndIf
    	    \State $\text{Increment }\#(s,{a},t)$
            \State $s \gets t$
            \EndIf
        \EndIf
    \Until{$s \in \{s_+, s_-, s_?\}$}
    \State \textbf{return} $X \cup \set{s}$
\EndProcedure
\end{algorithmic}
\end{algorithm}

Algorithm~\ref{alg:stuck} finds all the $\deltasure EC$ in the partial model. This procedure returns true, if the given state $s$, is part of an $\deltasure$ EC. As we find ECs in the partial model, we also find its mean-payoff value by calling $\mathsf{UPDATE\_MEC\_VALUE}$ procedure. 

\begin{algorithm}[htbp]
	\caption{Check if we are probably looping and should stop the~simulation}\label{alg:stuck}
			\hspace*{\algorithmicindent}\algorithmicrequire	State set $X$, state $s$.\\
			\hspace*{\algorithmicindent}\algorithmicensure Boolean value (Yes/No).	
	\begin{algorithmic}[1]
		\Procedure{$\STUCK$}{}
		\State \textbf{return } 
		$\exists T \subseteq X \text{ s.t. } T \text{ is EC in partial model} \wedge s \in T \wedge \SUREEC(T)$
		\EndProcedure
	\end{algorithmic}
\end{algorithm}

Given a set of states $T$, Algorithm~\ref{alg:sureEC} returns true if all the state-action pairs formed by $T$, has been visited at least $requiredSamples$ number of times 

\begin{algorithm}[htbp]
	\caption{Check whether we are $\deltasure$ that $T$ is an EC}\label{alg:sureEC}
			\hspace*{\algorithmicindent}\algorithmicrequire	State set $T$, $\delta_{TP}$.\\
			\hspace*{\algorithmicindent}\algorithmicensure Boolean value (Yes/No).	
	\begin{algorithmic}[1]
		\Procedure{$\SUREEC$}{}
		\State $\stepsUntilSure = \frac{\ln(\delta_{TP})}{\ln(1-\pmin)}$
		\State $B \gets \set{(s,{a}) \mid s \in T \wedge \neg (s,{a}) \leaves T}$ \Comment{Set of staying state-action pairs}
		\If{$\bigwedge_{(s,{a}) \in B} \#(s,{a}) > \stepsUntilSure$}
		    \State \textbf{return} True
		\Else 
		    \State \textbf{return} False
		\EndIf
		\EndProcedure
	\end{algorithmic}
\end{algorithm}

In Algorithm \ref{alg:updateMEC}, given an $\deltasure$ EC, it starts by computing the number of samples needed to simulate the MEC, followed by simulating the MEC. This is done so as to better learn the transition probabilities inside the MEC, which helps to reduce \tpImprecision, which in turn will help in better convergence during the value iteration. Then we find the mean payoff of the MEC, by calling $\mathsf{VALUE\_ITERATION}$ procedure. 



\begin{algorithm}[H]
	\caption{Algorithm to update MEC reward value and stay action}\label{alg:updateMEC}
		\hspace*{\algorithmicindent}\algorithmicrequire	$\deltasure$ EC $M=(T, A)$.\\
		\hspace*{\algorithmicindent}\algorithmicensure Bounds on the mean-payoff value of $M$ are updated
	\begin{algorithmic}[1]
	\Procedure{$\mathsf{UPDATE\_MEC\_VALUE}$}{}
	    \State $\beta = (\widehat{v}^u_{M}- \widehat{v}^l_{M})/2$ \Comment{$\beta$ is the precision we aim to achieve after this algorithm.}
	    \State $nSamples = \mathsf{COMPUTE\_N\_SAMPLES}(M)$
	    
	    \State $\simulateMecHeuristic (M, nSamples)$
		\State $\widehat{v}^l_{M}, \widehat{v}^u_{M} = \mathsf{VALUE\_ITERATION} (M, \beta)$ \Comment{$\widehat{v}^l_{M}, \widehat{v}^u_{M}$ are updated}
    \EndProcedure
	\end{algorithmic}
\end{algorithm}

In Algorithm \ref{alg:computeNSamples} we return the number of times the given $\deltasure$ EC, needs to be sampled. Bounds on the mean payoff can be found for the same $\deltasure EC$ mutliple times, because of the iterative nature of Algorithm~\ref{alg:general}, each time making the bounds more and more precise. If the same $\deltasure$ EC, arrives later during the exploration, we increase its number of samples by a multiplicative factor of $5$. As we increase the number of samples for an EC, its \tpImprecision on the transitions gets reduced thereby providing more precise bounds on mean payoff on the EC.
\begin{algorithm}[H]
	\caption{Algorithm that gives nSamples based on MEC}\label{alg:computeNSamples}
		\hspace*{\algorithmicindent}\algorithmicrequire	$\delta_{TP}$-sure EC $M_i=(T_i, A_i)$.
	\begin{algorithmic}[1]
	\Procedure{$\mathsf{COMPUTE\_N\_SAMPLES}$}{}
	    \State $leastVisitCount \gets \min\{\#(s, a) \mid s \in T_i, a \in A_i\} $
	    
	    \State $initialSamples \gets 10^4$
	    \State $mFactor \gets 5$
	    \State $nSamples \gets initialSamples$
	    
	    \While{$nSamples < leastVisitCount$}
	        \State $nSamples = nSamples \cdot mFactor$
	        
	    \EndWhile
    \EndProcedure
	\end{algorithmic}
\end{algorithm}

Given an MEC $M$ and $nSamples$, Algorithm~\ref{alg:simulate_MEC_MDP_HEURISTIC} simulates the MEC \\ $nSamples * nTransitions$ number of times, where $nTransitions$ is the number of transitions inside $M$. Ideally we would like to simulate the MEC until all state-action pairs inside the MEC is visited at least $nSamples$ number of times. For a MEC, which has some states that are reachable only by a low probability from all the other states, then this algorithm will take too long to terminate. So we employ a heuristic, where we simulate only $nSample * nTransitions$ number of times, which will result in visiting the important states at least $nSamples$ number of times. Even though some states, which has very low reachability from other states, might not have been visited $nSamples$ number of times, their effect on the mean payoff of the MEC $M$, will be less. 

\begin{algorithm}[H]
\caption{Simulation algorithm for MDP MEC}\label{alg:simulate_MEC_MDP_HEURISTIC}
		\hspace*{\algorithmicindent}\algorithmicrequire	MDP MEC $M=(T, A)$, $nSamples$, $nTransitions$.\\
		\hspace*{\algorithmicindent}\algorithmicensure estimated transition probabilities for state-action pairs.
\begin{algorithmic}[1]
\Procedure{$\simulateMecHeuristic$}{}
\State $s \gets \initstate$
    \For{$nSamples \cdot nTransitions$ times}
    	\State ${a} \gets $ sampled uniformly from $s$
    	\State $t \gets$ sampled according to $\trans(s,{a})$
    	\State $\text{Increment }\#(s,{a},t)$
    	\State $s \gets t$ \Comment{Move from $s$ to $t$.}
    \EndFor
\EndProcedure
\end{algorithmic}
\end{algorithm}

Given a set of state-action pairs $T_{i}, A_{i}$, Algorithm~\ref{alg:vi} will find the mean payoff bounds associated with the state-action pairs, using value iteration. This algorithm assumes that the given set of state-action pairs is indeed an MEC and there are no other outgoing transitions. We initially set the lower bound $l_0$ and upper bound $u_0$ of all the states to be $0$. During iteration $i$, we set the value of $l_i(s)$ to be $r(s) + \widehat{L}(s, a)$, for an action $a$ which maximizes this value. Similarly we udpate the value of $u_i(s)$. We compute the value of $\widehat{L}(s, a)$ using the second set of bellman equations (otherwise known as greybox update equations), as described in Section~\ref{sec:algo}. We stop the value iteration when either the difference in the lower bound on two successive iterations or the difference in the upper bound values for two successive iterations, converges within the precision $\beta$. We say that, continue value iteration until sp$(\Delta_{n}^{L}) \leqslant \beta$ $or$ sp$(\Delta_{n}^{U}) \leqslant \beta$, where $\Delta_{n}$ is a function which is defined as $\Delta_{n}^{L} = l_n - l_{n-1}$ and similarly $\Delta_{n}^{U} = u_n - u_{n-1}$. At last we compute the lower bound on the mean payoff to be the maximum difference in the lower bound values, for the last two iterations. Similarly we compute the upper bound on the mean payoff.


\begin{algorithm}[H]
    \caption{Value Iteration algorithm for blackbox MECs}\label{alg:vi}
		\hspace*{\algorithmicindent}\algorithmicrequire	MEC $M_i = (T_i, A_i)$, precision $\beta > 0$. \\
		\hspace*{\algorithmicindent}\algorithmicensure mean-payoff lower bound $w_l$, mean-payoff upper bound $w_u$
	\begin{algorithmic}[1]
	    \Procedure{$\mathsf{VALUE\_ITERATION}$}{}
	    \State $l_0(\cdot) \gets 0, u_0(\cdot) \gets 0, n \gets 0$
		\Repeat
		    \State $n \gets n+1$
		    \For{$s \in T_i$}
		        \State $l_n(s) = \max\limits_{{a}\in\av(s)}(r(s) + \widehat{\lb}(s, {a}))$
		        \State $u_n(s) = \max\limits_{{a}\in\av(s)}(r(s) + \widehat{\ub}(s, {a}))$
		    \EndFor
        \Until{sp$(\Delta^{\lb}_n) \leqslant \beta$ or sp$(\Delta^{\ub}_n) \leqslant \beta$} \\
		\Return $\max\limits_{s\in T_i}(l_n(s)-l_{n-1}(s)), \max\limits_{s\in T_i}(u_n(s)-u_{n-1}(s))$
		\EndProcedure
	\end{algorithmic}
\end{algorithm}

In Algorithm~\ref{alg:updateNew}, for a given set of states, we compute the values of all states using the previous values of state-action pairs, which is given by $\widehat{f}(s,{a})$. For a state $s$ we set the value to be the maximum value among all of its actions.

\begin{algorithm}[htbp]
	\caption{New update procedure using the probability estimates}\label{alg:updateNew}
			\hspace*{\algorithmicindent}\algorithmicrequire	Set of States $S$.
	\begin{algorithmic}[1]
		\Procedure{$\UPDATE$}{}
		\For {$s\in S$}
    		\For {$f \in \set{\lb,\ub}$}  ~~~~~~\Comment{For both functions}
        		\State 
    			$ f(s) = \max_{{a} \in \Av(s)} \widehat{f}(s,{a})$
        	\EndFor
        \EndFor
		\EndProcedure
	\end{algorithmic}
\end{algorithm}

Algorithm~\ref{alg:deflate} performs the \emph{deflate} operation~\cite{KKKW18}. It deflates the upper bound of all the states inside the MEC. If there exists a state $s$ such that its upper bound, is greater than the upper bound of the best leaving action, we change the upper bound of $s$ to be equal to the upper bound of the best leaving action.



\begin{algorithm}[H]
    \caption{Deflate algorithm}\label{alg:deflate}
        \hspace*{\algorithmicindent}\algorithmicrequire MEC $M_i = (T_i, A_i)$
    \begin{algorithmic}[1]
        \Procedure{$\mathsf{DEFLATE}$}{}
            \For{$s \in T_i$}
                \State $\ub(s) = \min(\ub(s), \widehat{\ub}(\mathsf{BEST\_LEAVING\_ACTION}(M_i)))$
            \EndFor
        \EndProcedure
    \end{algorithmic}
\end{algorithm}

Algorithm~\ref{alg:findMecs} finds all $\deltasure$ MECS inside the current partial model $\Mdp'$. It works by removing actions that have not been sampled $\stepsUntilSure$ number of times to get the new partial model $\Mdp''$. Note that this in the accordance with the $\deltasure$ ECs definition in Section~\ref{sec:algo}. Finally, we return all the MECs that remain in $\Mdp''$.

\begin{algorithm}[H]
    \caption{Procedure to find $\deltasure$ MECs restricted to $\states'$}\label{alg:findMecs}
        \hspace*{\algorithmicindent}\algorithmicrequire State Set $\states'$
    \begin{algorithmic}[1]
        \Procedure{$\mathsf{FIND\_MECS}$}{}
            \State $\stepsUntilSure = \frac{\ln(\delta_{TP})}{\ln(1-\pmin)}$
            \State $deltaUnsureActions \leftarrow \{(s, \{a\in \Av(s) | \#(s,a)<\stepsUntilSure \} | s\in \states\cap \states' \}$
            \State $\Mdp'' \leftarrow \Mdp'$ with $\Av'$ replaced by $\Av'-deltaUnsureActions$
            \State \Return $\mec(\Mdp'')$
        \EndProcedure
    \end{algorithmic}
\end{algorithm}

\section{Proof of Theorem \ref{thm:algo}} \label{app:mdp_proof}
\noindent\textbf{Theorem~\ref{thm:algo}.}
\emph{Algorithm~\ref{alg:general} has the property that when it stops, it returns an interval for the mean-payoff value of the MDP that is PAC for the given \mpInconfidence
$\symbolMpInconfidence$ and the \mpImprecision $\symbolMpImprecision$.
}
\begin{proof}
After every iteration of the repeat-until loop, we say that the interval for mean-payoff is given by $[r_{max}\cdot\lb(\initstate), r_{max}\cdot\ub(\initstate)]$, where $r_{max}$ is the maximum observable reward over all states in the MDP. Assume that $v_\Mdp$ is the true mean-payoff value of the MDP. We now prove that $r_{max}\cdot\lb(\initstate)\leqslant v_\Mdp \leqslant r_{max}\cdot\ub(\initstate)$ with probability at least $1-\symbolMpInconfidence$.

We first reduce the mean-payoff problem to a reachability problem while assuming we know all transition probabilities. We then establish a notion of `correctness' with regard to our estimate of the given MDP, and show that if our estimate is correct, $r_{max}\cdot\lb(\initstate)\leqslant v_\Mdp \leqslant r_{max}\cdot\ub(\initstate)$ is true. Finally, we prove that our model is correct with probability at least $1-\symbolMpInconfidence$.

Section~\ref{sec:prelims} provides the following definition for the mean-payoff maximization problem.
\begin{equation*}
	\gain(s) = \max_{\straa} \sum_{M \in \mec(\Mdp)} \pr^\straa_s[\Diamond\Box M] \cdot \gain_M
\end{equation*}

Theorem 2 in \cite{cav17} proves that the mean-payoff maximization problem for an MDP $\Mdp$ can be reduced to a problem where we maximize the reachability probability $\pr^\straa_{\initstate}[\Diamond s_+]$ to $s_+$, in $\Mdp$'s weighted MEC Quotient $\Mdp^f$. From now on, for a given state $s$, we abbreviate $\pr^\straa_{s}[\Diamond s_+]$ as $v(s)$. See that $v(\initstate)$ in $\Mdp^f$ is equivalent to $v(\initstate)$ in a modified MDP $\Mdp'$, where instead of having the $\mathsf{stay}$ action from every collapsed state of an MEC $M = (T,A)$, we have the $\mathsf{stay}$ action from every state $s\in T$.
Further, by definition of the $\mathsf{stay}$ action, $s$ now has a transition to $s_+$ with probability $\frac{v_M}{r_{max}}$ and we have that $v(s)=\frac{v_M}{r_{max}}$.
Now, considering $s_i$ to be a state in $M_i$, we get the equation above,
\begin{align*}
v(\initstate) &= \max\limits_{\pi \in \Pi^{MD}} \sum\limits_{M_i\in\mec(\Mdp)}\pr^\straa_{\initstate}[\Diamond s_i]\cdot \pr^\straa_{s_i}[\Diamond s_+]\\
&=  \max\limits_{\pi \in \Pi^{MD}} \sum\limits_{M_i\in\mec(\Mdp)}\pr^\straa_{\initstate}[\Diamond s_i]\cdot \frac{v_M}{r_{max}}\\
&= \frac{v_\Mdp}{r_{max}}\\
\end{align*}

Thus, the mean-payoff maximization problem for an MDP $\Mdp$ is equivalent to the problem where we maximize $v(\initstate)$ in $\Mdp'$.

Note that, the above is valid when we know the actual transition probabilities on $\Mdp'$ correctly. Since, we are doing blackbox learning, we always have some inconfidence with respect to transition probabilities or the mean-payoff value.
Assume that we work with upper and lower bounds on $v_M$, such that $v^l_M\leqslant v_M\leqslant v^u_M$. Note that, if for all $M\in \mec(\Mdp)$, we use a lower bound $v^l_M$ on the mean-payoff value of $M$, we will get a lower bound $v^l_\Mdp$ on the overall mean-payoff value of the MDP $\Mdp$. 
We thus, slightly modify the $\mathsf{stay}$ action, such that for some state $s\in M$, we have $\lb(s) = \frac{v^l_M}{r_{max}}$ and $\ub(s) = \frac{v^u_M}{r_{max}}$. 
Here, $\lb(s)\leqslant v(s) \leqslant \ub(s)$, where we define $\lb(s)$ as $\pr^\straa_{s}[\Diamond s_+]$ and $\ub(s)$ as $\pr^\straa_{s}[\Diamond s_+\cup s_?]$. This modified {\sf stay} action now gives us the $\mathsf{stay}$-augmented MDP of Definition~\ref{def:stay_mec}. Further, we have that 
\begin{align*}
\lb(\initstate) &= \max\limits_{\pi \in \Pi^{MD}} \sum\limits_{M_i\in\mec(\Mdp)}\pr^\straa_{\initstate}[\Diamond s_i]\cdot \pr^\straa_{s_i}[\Diamond s_+]\\
&=  \max\limits_{\pi \in \Pi^{MD}} \sum\limits_{M_i\in\mec(\Mdp)}\pr^\straa_{\initstate}[\Diamond s_i]\cdot \frac{v^l_M}{r_{max}}\\
&= \frac{v^l_\Mdp}{r_{max}}\\
\end{align*}

Similarly, we get $\ub(\initstate) = \frac{v^u_\Mdp}{r_{max}}$. Thus, finally we have that, $r_{max}\cdot\lb(\initstate)\leqslant v_\Mdp \leqslant r_{max}\cdot\ub(\initstate)$ is true.

From now on, we call our estimated model `correct' if for all transitions $(s,a,t)$, our estimates $\widehat{\trans}(s,a,t)\le \trans(s,a,t)$.
We now prove that if our estimate of $\Mdp$ is correct, $r_{max}\cdot\lb(\initstate)\leqslant v_\Mdp \leqslant r_{max}\cdot\ub(\initstate)$ is true.

See that by the correctness of $\UPDATE$ (Lemma~1 in \cite{AKW19}; we also discuss it below.), for all states $s_i$ in some MEC $M_i$, the bounds on reaching $s_i$ from $\initstate$ are correct. We now show that, $\lb(s_i)\leqslant \frac{v_{M_i}}{r_{max}} \leqslant \ub(s_i)$.

Since, we have assumed that $\Mdp$ is correct, any MEC $M\in \mec(\Mdp)$ is also correct and we assume that we have seen all transitions. Hence, we treat all MECs as greybox models. We estimate $v^l_M$ and $v^u_M$ using the greybox Bellman update equations in value iteration. First, see that due to the correctness of $\UPDATE$ (Lemma 1 in \cite{AKW19}. Although, the given proof is for Blackbox update equations, we can follow the same sketch to prove that greybox $\UPDATE$ is correct for greybox models with the same $\tpInconfidence$.), we have that $v^l_M \leqslant v_M \leqslant v^u_M$. Now, value iteration further gives us bounds on $v^l_M$ and $v^u_M$. From Theorem 8.5.5 in \cite{Puterman}, we have that these bounds are correct and we set $\widehat{v}^l_M$ equal to the lower bound on $v^l_M$ and $\widehat{v}^u_M$ equal to upper bound on $v^u_M$. Thus, we have that if $M$ is correct, $\widehat{v}^l_M \leqslant v_M \leqslant \widehat{v}^u_M$.

Now, for some $s_i\in M_i$, $\lb(s_i)=\frac{\widehat{v}^l_M}{r_{max}}$ and $\ub(s_i) = \frac{\widehat{v}^u_M}{r_{max}}$. Clearly, we have that $\lb(s_i)\leqslant \frac{v_{M_i}}{r_{max}} \leqslant \ub(s_i)$.

Finally, we show that our estimate of $\Mdp$ is correct with probability at least $1-\symbolMpInconfidence$. This completes the proof.

During simulation, we try to learn the Bernoulli distribution $\trans(s,a)$ for many state-action pairs. We first bound the error on the learnt distribution $\widehat{\trans}(s,a)$ to a $\tpInconfidence$ $\symbolTpInconfidence := \dfrac{\symbolMpInconfidence\cdot \pmin}{\lvert\{{a} \vert\, \state \in \widehat{S} \wedge {a} \in \av(s) \}\rvert}$.
Here, we have essentially, distributed the $\mpInconfidence$ $\symbolMpInconfidence$ uniformly over all transitions.
Assuming that we have sampled the distribution $\trans(s,a)$ at least $\#(s,a)$ times, using the Hoeffding's inequality, we have that,

\[
\mathbb{P}\big[\widehat{\mathbb{T}}(s, {a}, t) \geqslant \mathbb{T}(s, {a}, t)\big] 
= \mathbb{P}\big [\dfrac{\#(s, {a}, t)}{\#(s, {a})} - \symbolTpImprecision \geqslant \mathbb{T}(s, {a}, t) \big ]
\leqslant e^{-2c^2\#(s,a)}
\leqslant \symbolTpInconfidence.
\]
where $\symbolTpImprecision = \sqrt{\frac{\ln (\delta_{\mathbb{TP}})}{-2 \#(s, {a})}}$.
Thus, at any point in the algorithm, we say that, $\widehat{\mathbb{T}}(s, {a}, t) \leqslant \mathbb{T}(s, {a}, t)$ is true over all state-action pairs in our model with probability at least $1-\symbolMpInconfidence$. Further, our procedure ensures that all MECs are $\deltasure$, or, all MECs are correct with probability at least $1-\symbolMpInconfidence$. Thus, we have that our estimate of $\Mdp'$ is correct with probability at least $1-\symbolMpInconfidence$.

Finally, given that $M$ is correct with a probability at least $1-\symbolMpInconfidence$, we have that $\widehat{v}^l_M \leqslant v_M \leqslant \widehat{v}^u_M$ with probability at least $1-\symbolMpInconfidence$.

\qed
\end{proof}

\subsection{Intuition about $\UPDATE$}
Below, we provide an intuition of the proof in Lemma 1 in \cite{AKW19}.
Recall that the values of the states outside the MECs are updated using Bellman update equations as follows:
\begin{equation*}
\widehat{\lb}(s, {a}) := \sum\limits_{\tsucc:\#(s, {a}, \tsucc)>0} \widehat{\mathbb{T}}(s, {a}, \tsucc) \cdot \lb(\tsucc)
\end{equation*}
\begin{equation*}
\widehat{\ub}(s, {a}) := \sum\limits_{\tsucc:\#(s, {a}, \tsucc)>0} \widehat{\mathbb{T}}(s, {a}, \tsucc) \cdot \ub(\tsucc) + (1-\sum\limits_{\tsucc:\#(s, {a}, \tsucc)>0} \widehat{\mathbb{T}}(s, {a}, \tsucc))
\end{equation*}
where
\begin{equation*}
\widehat{\mathbb{T}}(s, {a}, \tsucc) := \max (0, \dfrac{\#(s, a, t)}{\#(s, a)} - \symbolTpImprecision).
\end{equation*}
Since we only identify those end-components that are $\deltasure$, conditioned on this, it implies that we did not miss any of the successors of any state-action pair in the $\deltasure$ EC.
Thus inside a $\deltasure$ EC, we perform value iteration which uses the greybox update equations:
\begin{equation*}
\widehat{\lb}(s, {a}) := \sum\limits_{\tsucc:\#(s, {a}, \tsucc)>0} \widehat{\mathbb{T}}(s, {a}, \tsucc) \cdot \lb(\tsucc) + \min\limits_{\tsucc:\#(s, {a}, \tsucc)>0}\lb(\tsucc) \cdot (1-\sum\limits_{\tsucc:\#(s, {a}, \tsucc)>0} \widehat{\mathbb{T}}(s, {a}, \tsucc))
\end{equation*}
\begin{equation*}
\widehat{\ub}(s, {a}) := \sum\limits_{\tsucc:\#(s, {a}, \tsucc)>0} \widehat{\mathbb{T}}(s, {a}, \tsucc) \cdot \ub(\tsucc) + \max\limits_{\tsucc:\#(s, {a}, \tsucc)>0}\ub(\tsucc) \cdot (1-\sum\limits_{\tsucc:\#(s, {a}, \tsucc)>0} \widehat{\mathbb{T}}(s, {a}, \tsucc))
\end{equation*}
When using Bellman update, we use the \tpImprecision as computed based on the number of times action $a$ is chosen from state $s$.
Since the $\tpInconfidence$ $\symbolTpInconfidence$ is ensured for every individual transition, this further implies that an overall $\mpInconfidence$ of $\symbolMpInconfidence$ is ensured.

Thus whenever our algorithm is stopped, we ensure with probability at least $1-\symbolMpInconfidence$ that the true value is within the upper and lower bounds reported at the moment.

\section{The relation between \tpImprecision and \tpInconfidence for using greybox update equations} \label{app:error_tolerance}
Here we discuss the relation between \tpImprecision and \tpInconfidence when using greybox update equations.

Section~\ref{sec:algo} provided an explanation about how if we keep the TP-inconfidence fixed, the \tpImprecision increases. On the other hand, if we fix the overall TP-inconfidence $\symbolTpInconfidence$ \emph{a priori}, we have, $\symbolTpInconfidence = \symbolTpInconfidence' + (1- \pmin)^{\#(s, a)}$, where $\symbolTpInconfidence'$ is the \tpInconfidence on the transition probabilities. Then the \tpImprecision increases to a new value $\symbolTpImprecision' = \sqrt{\frac{\ln (\delta'_{TP})}{-2 \#(s, a)}}$. Note that the number of samples here remains the same.

We illustrate the above with an example.
Assume that $s$ has two successors for action $a$ that are ${t}_1$ and ${t}_2$ with actual transition probabilities $0.5$ and $0.5$ respectively, and let $\delta_{TP}=0.1$, and ${\pmin}=0.1$.
Let $\#(s,a)=30$, out of which the transition $(s,a,{t}_1)$ is visited $10$ times while the transition $(s,a,{t}_2)$ is visited $20$ times.
Thus assuming that the overall \tpInconfidence remains the same, that is $0.1$, by subtracting the error probability due to the use of greybox update equations from the overall \tpInconfidence, we get, $\delta'_{TP} = \delta_{TP} - (1-0.1)^{30} = 0.1-0.042 = 0.058$.
Hence $\symbolTpImprecision' = \sqrt{\frac{\ln 0.058}{-60}}=0.22$ which is the increased \tpImprecision obtained as a result of using the greybox update equations.
We note that the original \tpImprecision $\varepsilon_{TP} = \sqrt{\frac{\ln 0.1}{-60}}=0.196$.
Thus the $\widehat{\mathbb{T}}(s, a, {t}_1) = \frac{1}{3} - c' = 0.113$, while $\widehat{\mathbb{T}}(s, a, {t}_2) = \frac{2}{3} - c' = 0.447$.
On the other hand, if we keep the \tpImprecision $\varepsilon_{TP}$ unchanged, then the \tpInconfidence increases to $\delta_{TP} + (1- \pmin)^{\#(s, a)} = 0.1 + 0.042 = 0.142$.

\section{Proofs of Section \ref{sec:ctmdp2}}
\label{app:ctmdp_proofs2}
\noindent\textbf{Lemma~\ref{lem:ctmdp-pac}.}
Let $X_1, \dots, X_n$ be exponentially distributed i.i.d. random variables with mean $\frac{1}{\lambda}$. Then we have the following.
\[
\pr \Big [\abs{\frac{1}{n}\sum_{i=1}^n X_i - \frac{1}{\lambda} \geqslant \frac{1}{\lambda} \cdot \alpha_R)} \Big ] \leqslant \displaystyle{\inf_{-\lambda < t < 0}}
\Big (\frac{\lambda}{\lambda +t}\Big)^n \cdot e^{\frac{tn}{\lambda}(1+\alpha_R)} + \displaystyle{\inf_{ t > 0}} 
\Big (\frac{\lambda}{\lambda + t}\Big)^n \cdot e^{\frac{tn}{\lambda}(1-\alpha_R)} \leqslant \deltaR.
\]
\begin{proof}
If $X$ is a random variable, then for any $a \in \Reals$, using Chernoff bounds, we get that 
\begin{equation*}
\pr(X \geqslant a) \leqslant e^{-sa} \expected[e^{sX}], \text{ for all } s > 0
\end{equation*}
\begin{equation*}
\pr(X \leqslant a) \leqslant e^{-sa} \expected[e^{sX}], \text{ for all } s < 0.
\end{equation*}
Cramer's theorem states that if $X_1, X_2, ...$ are i.i.d. random variables with finite logarithmic moment generating function $\Lambda(s) = \log\expected[e^{sX_{i}}]$ then for every $x$ greater than mean, the sequence  $a_n = \frac{1}{n}\pr[S_n > xn]$ converges to infimum of $\Lambda(s) - sx$, where $S_n$ denotes partial sum $X_1+...+X_n$. Since exponential is a continuous and  monotone function, we get that probability actually converges to the Chernoff's bound and hence this bound is tight.

Since $M_X(s) = \expected[e^{sX}]$, where $M_X(s)$ is the \emph{moment generating function} of $X$,  and we want to obtain the best upper bound, this gives us:
\begin{equation*}
\pr(X \geqslant a) \leqslant e^{-sa} M_X(s), \text{ for all } s > 0
\end{equation*}
\begin{equation*}
\pr(X \leqslant a) \leqslant e^{-sa} M_X(s), \text{ for all } s < 0.
\end{equation*}

Now assuming we choose $n$ samples from an exponential distribution with a mean $\mu$, and we have $n$ exponentially distributed i.i.d random variables denoted by $X_1, \dots, X_n$, we want to find an upper bound for
\begin{equation*}
\pr \Big [\abs{\frac{1}{n}\sum_{i=1}^n X_i - \mu \geqslant \mu \cdot \alpha_R)} \Big ]
\end{equation*}
Let $X = \sum_{i=1}^n X_i$.
Now $\mu = \frac{1}{\lambda}$ since each $X_i$ is an exponentially distributed i.i.d random variable, we have that $\expected[e^{sX}] = \expected[e^{s(X_1+ \dots +X_n)}]$, and $M_{X_1}(s) = \frac{\lambda}{\lambda-s}$ for $s < \lambda$.

Thus $M_X(s) = (M_{X_1}(s))^n = (\frac{\lambda}{\lambda-s})^n$ for $s < \lambda$.

Now 
\begin{align*}
\pr \Big [\abs{\frac{1}{n}\sum_{i=1}^n X_i - \mu \geqslant \mu \cdot \alpha_R)} \Big ] = 
\pr \Big [ (X - n\mu) \geqslant n \mu \alpha_R) \Big ] + \pr \Big [ (n\mu - X) \geqslant n \mu \alpha_R) \Big ] \\
=\pr \Big [ X \geqslant n\mu (1 + \alpha_R)) \Big ] + \pr \Big [ X \leqslant n \mu (1 - \alpha_R) \Big ]
\end{align*}

We start with $\pr \Big [ X \geqslant  n \mu(1  + \alpha_R) \Big ]$.
Using Chernoff's bound,
\begin{align*}
\pr \Big [X \geqslant  n\mu(1 + \alpha_R) \Big ] \leqslant e^{-sn\mu (1+\alpha_R)}M_X(s) \text{ for all } s > 0 \\
=e^{\frac{-sn}{\lambda} (1+\alpha_R)} \Big (\frac{\lambda}{\lambda -s}\Big)^n \text{ for all } 0 < s < \lambda
\end{align*}
Replacing $s$ with $-t$ and $\mu$ with $\frac{1}{\lambda}$, we get
\begin{align*}
\pr \Big [X \geqslant  \frac{n}{\lambda}(1 + \alpha_R) \Big ] \leqslant 
\Big (\frac{\lambda}{\lambda +t}\Big)^n \cdot e^{\frac{tn}{\lambda}(1+\alpha_R)} \text{ for all } -\lambda < t < 0
\end{align*}
Thus
\begin{align*}
\pr \Big [X \geqslant  \frac{n}{\lambda}(1 + \alpha_R) \Big ] \leqslant 
\Big (\frac{\lambda}{\lambda +t}\Big)^n \cdot e^{\frac{tn}{\lambda}(1+\alpha_R)} \text{ for all } -\lambda < t < 0
\end{align*}

Thus
\begin{align*}
\pr \Big [X \geqslant  \frac{n}{\lambda}(1 + \alpha_R) \Big ] \leqslant 
\displaystyle{\inf_{-\lambda < t < 0}}
\Big (\frac{\lambda}{\lambda +t}\Big)^n \cdot e^{\frac{tn}{\lambda}(1+\alpha_R)}
\end{align*}

For $\pr \Big [ X \leqslant  n \mu(1  - \alpha_R) \Big ]$ case, we again use Chernoff's bound,
\begin{align*}
\pr \Big [X \leqslant  n\mu(1 - \alpha_R) \Big ] \leqslant e^{-sn\mu (1-\alpha_R)}M_X(s) \text{ for all } s < 0 \\
\text{Since } \lambda > 0,
\pr \Big [X \geqslant  \frac{n}{\lambda}(1 - \alpha_R) \Big ] \leqslant e^{\frac{-sn}{\lambda} (1-\alpha_R)} \Big (\frac{\lambda}{\lambda -s}\Big)^n \text{ for all } s < 0\\
\end{align*}
Replacing $s$ with $-t$, we get
\begin{align*}
\pr \Big [X \geqslant  \frac{n}{\lambda}(1 - \alpha_R) \Big ] \leqslant e^{\frac{tn}{\lambda} (1-\alpha_R)} \Big (\frac{\lambda}{\lambda +t}\Big)^n \text{ for all }  t > 0\\ 
\pr \Big [X \geqslant  \frac{n}{\lambda}(1 - \alpha_R) \Big ] \leqslant 
\displaystyle{\inf_{ t > 0}} 
\Big (\frac{\lambda}{\lambda + t}\Big)^n \cdot e^{\frac{tn}{\lambda}(1-\alpha_R)}
\end{align*}
\end{proof} \qed

Below we also provide an enlarged version of Table~\ref{tab:nSamplesTable} including larger values of $\alpha_R$.
\begin{table}
\begin{center}
\begin{tabular}{ | m{1cm} | m{1cm}| m{1cm} | m{1cm} | m{1.6cm} | } 
\hline
$\alpha_R$\textbackslash $\deltaR$ & 10\% & 5\% & 0.01\% & 0.00001\%\\ 
\hline
3\% & 7000 & 9000 & 23000 & 60000\\
\hline
5\% & 2500 & 3100 & 8000 & 13400 \\ 
\hline
10\% & 650 & 800 & 2100 & 3500\\ 
\hline
20\% & 160 & 200 & 530 & 920\\ 
\hline
\end{tabular}
\caption{Lookup table for number of samples based on $\alpha_R$ and $\deltaR$}
\label{tab:nSamplesTable2}
\end{center}
\end{table}

\begin{example} \label{example:CTMDP_samples}
We show the computation for the case of $\deltaR=0.1$ and $\alpha_R=0.05$.
We want $\pr \Big [| X \geqslant  \frac{n}{\lambda}(1 + \alpha_R) \Big |] \leqslant \deltaR$.

This gives $\inf_{-\lambda < t < 0} \Big (\frac{\lambda}{\lambda +t}\Big)^n \cdot e^{\frac{tn}{\lambda}(1+\alpha_R)} + \displaystyle{\inf_{t > 0}} \Big (\frac{\lambda}{\lambda +t}\Big)^n \cdot e^{\frac{tn}{\lambda}(1-\alpha_R)}$

We split $\deltaR=0.1$ such that $\displaystyle{\inf_{-\lambda < t < 0}} \Big (\frac{\lambda}{\lambda +t}\Big)^n \cdot e^{\frac{tn}{\lambda}(1+\alpha_R)} \leqslant 0.05$ and
$\displaystyle{\inf_{t > 0}} \Big (\frac{\lambda}{\lambda +t}\Big)^n \cdot e^{\frac{tn}{\lambda}(1-\alpha_R)} \leqslant 0.05$.

Let $\lambda=10$. Note that since $\alpha_R$ denotes a fraction by which the estimated mean is away from the true mean, the particular value of $\lambda$ does not have an effect on the value of $n$.
Now $\displaystyle{\inf_{-\lambda < t < 0}} \Big (\frac{\lambda}{\lambda +t}\Big)^n \cdot e^{\frac{tn}{\lambda}(1+\alpha_R)} \leqslant 0.05$.
This is satisfied for $n=2500$ and the minimum occurs for $t=-0.477$.

Then $\displaystyle{\inf_{t > 0}} \Big (\frac{\lambda}{\lambda +t}\Big)^n \cdot e^{\frac{tn}{\lambda}(1-\alpha_R)} \leqslant 0.05$ is minimized at $t=0.526$, for $n=2320$.
The maximum of the two values of $n$ suffice and thus we need ~2500 samples.
\end{example}

\noindent\textbf{Lemma~\ref{lem:CTMDP_bounds}}.
\emph{Given a CTMDP $\Mdp$ with rates known $\alpha_R$-precisely, with transition probabilities known precisely, and with maximum reward per unit time over all states $r_{max}$, we have $v_{\Mdp}(\frac{1-\alpha_R}{1+\alpha_R}) \leq \widehat{v}_{\Mdp} \leq v_{\Mdp}(\frac{1+\alpha_R}{1-\alpha_R})$ and $|\widehat{v}_{\Mdp} - v_{\Mdp}| \leq r_{max} \frac{2 \alpha_R}{1 - \alpha_R}$.}
\begin{proof}
Since the mean payoff is a prefix-independent objective, it depends only on the value inside the MECs. We thus first estimate the value in a particular MEC $M$.\todo{Some changes in this paragraph; please check.}
Note that optimal positional strategies exist in MDPs for expected mean-payoff objective~\cite{Puterman}.
Assume a positional strategy $\sigma$ in the CTMDP such that under this strategy, the states of the MEC that are visited are $s_1, \dots, s_n$. 
Considering $\pi_{1}, ... , \pi_{n}$ to be the fraction of times (stationary distribution of the underlying discrete time Markov chain) these states are entered under strategy $\sigma$, the expected time spent in a state $s_{i} \in M$ is given by $\frac{\pi_{i}}{\lambda_{i}}$, where $\lambda_{i} = \lambda (s_{i}, a_{i})$ and $a_{i}$ is the action chosen from $s_{i}$ under strategy $\sigma$.

The expected mean-payoff value in $M$ is given by
\begin{equation*}
   v_M = \frac{\sum\limits_{s_i \in M} \frac{r(s_i) \pi_i}{\lambda_i}}{\sum\limits_{s_i \in M} \frac{\pi_i}{\lambda_i}}
\end{equation*}

Now we do not have the exact rates $\lambda_i$ for state $s_i$, but assuming an ${\alpha_R}_i$ multiplicative error on the mean of the exponential distributions, we have the estimate of the mean $\frac{1}{\widehat{\lambda}_i} \in \lbrack \frac{1}{\lambda_i}(1 - {\alpha_R}_i), \frac{1}{\lambda_i}(1 + {\alpha_R}_i) \rbrack$,
or $\widehat{\lambda}_i \in \lbrack \frac{\lambda_i}{(1 + {\alpha_R}_i)}, \frac{\lambda_i}{(1 - {\alpha_R}_i)} \rbrack$.
Let $\alpha_R = \max\limits_{s_i \in S} \{ {\alpha_R}_i \}$.

Thus the estimated mean cost of the MEC is given by

\begin{align*}
    \widehat{v}_M = \frac{\sum\limits_{s_i \in M} \frac{r(s_i) \pi_i}{\widehat{\lambda}_i}}{\sum\limits_{s_i \in M} \frac{\pi_i}{\widehat{\lambda}_i}} = \frac{\sum\limits_{s_i \in M} \frac{r(s_i) \pi_i}{\lambda_i} \frac{\lambda_i}{\widehat{\lambda}_i}}{\sum\limits_{s_i \in M} \frac{\pi_i}{\lambda_i} \frac{\lambda_i}{\widehat{\lambda}_i}}
\end{align*}

Now $\widehat{v}_M$ is maximized when the numerator is maximized and the denominator is minimized. Since $1-\alpha_R \leq \frac{\lambda_i}{\widehat{\lambda}_i} \leq 1+\alpha_R$

\begin{align*}
    \widehat{v}_M \leq \frac{\sum\limits_{s_i \in M} \frac{r(s_i) \pi_i}{\lambda_i} \left( 1+\alpha_R \right)}{\sum\limits_{s_i \in M} \frac{\pi_i}{\lambda_i} \left( 1-\alpha_R \right)}
    = v_M \left( \frac{1+\alpha_R}{1-\alpha_R} \right)
\end{align*}

Similarly, $\widehat{v}_M$ is minimized when the numerator is minimized and the denominator is maximized. Then

\begin{equation*}
    \widehat{v}_M \geq v_M \left( \frac{1-\alpha_R}{1+\alpha_R} \right).
\end{equation*}

Thus we have that,

\begin{align*}
    v_M \left( \frac{1-\alpha_R}{1+\alpha_R} \right) \leq \widehat{v}_M \leq v_M \left( \frac{1+\alpha_R}{1-\alpha_R} \right).
\end{align*}

Using the lower bound on $\widehat{v}_M$,

\begin{align*}
    v_M - \widehat{v}_M \leq v_M - v_M \left( \frac{1-\alpha_R}{1+\alpha_R} \right) = v_M \left ( \frac{2 \alpha_R}{1+\alpha_R} \right ) \leq r_{max_{M}} \left ( \frac{2 \alpha_R}{1+\alpha_R} \right ).
\end{align*}

where $r_{max_{M}} = \max\limits_{s_i \in M} \{ r(s_i) \}$

Similarly using the upper bound on $\widehat{v}_M$, we have

\begin{align*}
    \widehat{v}_M - v_M \leq v_M \left( \frac{1+\alpha_R}{1-\alpha_R} \right) - v_M = v_M \left ( \frac{2 \alpha_R}{1-\alpha_R} \right ) \leq r_{max_{M}} \left ( \frac{2 \alpha_R}{1-\alpha_R} \right ).
\end{align*}

Now let $v_{\Mdp}$ be the mean-payoff value for the whole CTMDP $\Mdp$. Let the MECs of $\Mdp$ are $M_1, M_2, ..., M_k$ and the probability to reach them under strategy $\sigma$ be $p_1, p_2, ..., p_k$. Then $v_{\Mdp} = \sum\limits_{i=1}^{k} p_i \cdot v_{M_i}$.

Since the probability of reaching each MEC depends solely on the transition probabilities of the embedded MDP, and not on the rates of the states of the CTMDP $\Mdp$ in the transient part, we have that $\widehat{v}_{\Mdp} = \sum\limits_{i=1}^{k} p_i \cdot \widehat{v}_{M_i}$.

Now,

\begin{align*}
    \sum\limits_{i=1}^{k} p_i v_{M_i} \left( \frac{1-\alpha_R}{1+\alpha_R} \right) &\leq \widehat{v}_{\Mdp} \leq \sum\limits_{i=1}^{k} p_i v_{M_i} \left( \frac{1+\alpha_R}{1-\alpha_R} \right)\\[3pt]
    \text{ or } v_{\Mdp} \left( \frac{1-\alpha_R}{1+\alpha_R} \right) &\leq \widehat{v}_{\Mdp} \leq v_{\Mdp} \left( \frac{1+\alpha_R}{1-\alpha_R} \right).
\end{align*}

Also,

\begin{align*}
    v_{\Mdp} - \widehat{v}_{\Mdp} = \sum\limits_{i=1}^{k} p_i v_{M_i} - \sum\limits_{i=1}^{k} p_i \widehat{v}_{M_i}
    = \sum\limits_{i=1}^{k} p_i \left ( v_{M_i} - \widehat{v}_{M_i} \right )
    \leq \sum\limits_{i=1}^{k} p_i \cdot r_{max_{m_i}} \left ( \frac{2 \alpha_R}{1+\alpha_R} \right ).
\end{align*}

Let $r_{max} = \max\limits_{M_i \in MECs(\Mdp)} \{r_{max_{M_i}}\}$. Then we have that,

\begin{align*}
    v_{\Mdp} - \widehat{v}_{\Mdp} \leq r_{max} \left ( \frac{2 \alpha_R}{1+\alpha_R} \right ).
\end{align*}

since $\sum\limits_{i=1}^{k} p_i = 1$.

Similarly we have that
\begin{align*}
    \widehat{v}_{\Mdp} - v_{\Mdp} \leq r_{max} \left ( \frac{2 \alpha_R}{1-\alpha_R} \right ).
\end{align*}

Thus 
    $| \widehat{v}_{\Mdp} - v_{\Mdp}| \leq r_{max} \left ( \frac{2 \alpha_R}{1-\alpha_R} \right )$.
\end{proof} \qed
\section{More Algorithms from Section \ref{sec:ctmdp2}} \label{app:ctmdp_algo2}

In this section we show the pseudocode of the procedures used in our CTMDP learning algorithm and their detailed description. Most of the CTMDP algorithms are similar to that of their MDP counterparts, except for a few. During simulation in addition to learning the transitions probabilities, in the case of CTMDP, we also learn about the rates associated with every state-action pair. Instead of listing all the algorithms of CTMDP, we only list the algorithms that have differences with their MDP counterpart. In some of the CTMDP algorithms, we refer to functions that are already defined for MDPs.

Algorithm~\ref{alg:ctmdp2} is similar Algorithm~\ref{alg:general}, except for two changes. One is, instead of calling $\mathsf{UPDATE\_MEC\_VALUE}$ function, here we call $\mathsf{FIND\_MEC\_MEANPAYOFF\_BOUNDS}$, which computes the bounds of a CTMDP MEC. Another is, instead of splitting $\symbolMpInconfidence$ among all transitions as in the case of MDP, here we split $\symbolMpInconfidence$ as $\deltaMPOne$ and $\deltaMPTwo$, such that $\symbolMpInconfidence=\deltaMPOne + \deltaMPTwo$. $\deltaMPOne$ denotes the inconfidence in the estimation of transition probabilties and $\deltaMPTwo$ denotes the inconfidence in the estimation of rates. We say that $\deltaMPOne=\left( \frac{1}{\pmin + 1} \right) \cdot \symbolMpInconfidence$ and $\deltaMPTwo=\left( \frac{\pmin}{\pmin + 1} \right) \cdot \symbolMpInconfidence$. The reason for this particular split of $\deltaMPOne$ and $\deltaMPTwo$ is that, the value of $\symbolTpInconfidence$ equals the value of $\deltaR$ for this particular split. Say if $\symbolTpInconfidence$ is less than $\deltaR$, then we may need to perform more samples on the transitions. By balancing the values of $\symbolTpInconfidence$ and $\deltaR$, we can perform less samples and still maintain the overall inconfidence of $\symbolMpInconfidence$. As in the case of MDP, for every $n$ times we simulate the CTMDP, we backpropagate the learned bounds through $\mathsf{UPDATE}$ and $\mathsf{DEFLATE}$ methods, and we check for convergence. If the bounds converge then we stop our algorithm, else we continue to simulate the CTMDP another $n$ times. 

\begin{algorithm}[tbp]
    \caption{Blackbox mean-payoff learning for CTMDPs}\label{alg:ctmdp2}
    \algorithmicrequire CTMDP $\Mdp$, imprecision $\varepsilon_{MP} > 0$, $\mpInconfidence$ $\symbolMpInconfidence > 0$, 
    lower bound $\pmin$ on transition probabilities in $\Mdp$  \\
    \textbf{Parameters:} revisit threshold $k \geq 2$, episode length $n\geq 1$\\
    \algorithmicensure at every moment an upper and lower bound on the maximum mean payoff for $\Mdp$ (with a computable confidence); upon termination $\symbolMpImprecision$-precise estimate of the maximum mean payoff for $\Mdp$ with confidence $1-\symbolMpInconfidence$, i.e.\ $(\symbolMpImprecision,1-\symbolMpInconfidence)$-PAC estimate
    \begin{algorithmic}[1]
        \Procedure{$\mathsf{ON\_DEMAND\_BVI}$}{}
        
            \emph{//Initialization} 
            \State Set $\lb(s_+)=\ub(s_+)=\ub(s_?)=1$, $\lb(s_-)=\ub(s_-)=\lb(s_?)=0$ \Comment{Augmentation}
            \State ${\states}^\prime = \emptyset$ \Comment{States of learnt model}
            \Repeat
            \For{$n$ times}
            
                \emph{//Get $n$ simulation runs and update MP of MECs where they end up}
                \State $\fpath \gets \SIMULATE(k)$ \Comment{Simulation run}
                \State ${\states}^\prime \leftarrow {\states}^\prime \cup w$ \Comment{Add states to the model}
                \State $\deltaMPOne \gets \left( \frac{1}{\pmin + 1} \right) \cdot \symbolMpInconfidence$
                \State $\deltaMPTwo \gets \left( \frac{\pmin}{\pmin + 1} \right) \cdot \symbolMpInconfidence$
                \State $\symbolTpInconfidence \gets \frac{\deltaMPOne\cdot \pmin}{\lvert\{{a} \vert\, s \in {\states}^\prime \wedge {a} \in {\av}^\prime(s) \}\rvert}$ \Comment{Split inconfidence among all transitions}
                \State $\deltaR \gets \frac{\deltaMPTwo}{\lvert\{{a} \vert\, s \in {\states}^\prime \wedge {a} \in {\av}^\prime(s) \}\rvert}$
                \If{last state of $\fpath$ is $s_+$ or $s_?$} \Comment{Probably entered a good MEC $M$}
                    \State $M \gets$ MEC from which we entered the last state of $\fpath$ 
                    \State $(l(M),u(M))\gets\mathsf{FIND\_MEC\_MEANPAYOFF\_BOUNDS}$($M, (\widehat{\gain}^{u}_M-\widehat{\gain}^{l}_M$))
                    \State Update ${\trans}^\prime(s, \mathsf{stay})$ according to Definition \ref{def:stay_mec} for all $s\in M$
                \EndIf
            \EndFor
            
            \emph{//Identify $\delta_{TP}$-sure MECs and propagate their MP by VI for reachability}
            \State $\mathit{ProbableMECs}\gets\mathsf{FIND\_MECS}$  \Comment{$\delta_{TP}$-sure MECs}
            \State $\INITIALIZE$
            \Repeat     
                \State $\UPDATE({\states}^\prime)$  
                \Comment{Single iteration of VI (one Bellman update per state)}
                \For{ $T \in \mathit{ProbableMECs}$}
                    \State $\mathsf{DEFLATE}(T)$
                \EndFor
            \Until{No change in $\lb$, $\ub$ for all states, respectively}
            \State Increment $i$
            \Until{$\ub(\initstate)$ - $\lb(\initstate) < \frac{2\symbolMpImprecision}{\rmax}
            $}\Comment{$\symbolMpImprecision$ is the absolute error; we use ``$<\frac{2\symbolMpImprecision}{\rmax}$'' for relative difference between upper and lower values, where $\rmax = \max\limits_{s\in{\states}^\prime} r(s)$.}
            \State\Return{$(\lb(\initstate)$ + $\ub(\initstate))/2 $}\Comment{ }
        \EndProcedure
    \end{algorithmic}
\end{algorithm}

In Algorithm~\ref{alg:simulateNewCtmdp}, we simulate the CTMDP $\Mdp$ from the initial state, repeatedly till we encounter any of the sink states. As in the case of MDP, an action from a state is chosen according to the best upper bound. During this process, we check whether we are visiting a state too often, and if so, we check for existence of MECs in our partial model, in Lines~\ref{fcall:appear_ctmdp}-\ref{fcall:looping_ctmdp}. In addition to learning the transition probabilities during the simulation, we also learn the mean of the exponential distribution associated with each state in the CTMDP, as shown in Line~\ref{algline:learn_rate_ctmdp}.

\begin{algorithm}[t]
\caption{New simulation algorithm counting occurrences for CTMDP}\label{alg:simulateNewCtmdp}
		\hspace*{\algorithmicindent}\algorithmicrequire	CTMDP $\Mdp$, revisit threshold $k$.\\
		\hspace*{\algorithmicindent}\algorithmicensure Simulated path $X$
\begin{algorithmic}[1]
\Procedure{$\SIMULATE{\sf \_CTMDP}$}{}
    \State $X \gets \emptyset$ \Comment{Stack of states visited during this simulation}
    \State $\state \gets \initstate$
    \Repeat
    	\State $X \gets X \cup \state$ \Comment{Push $\state$ to $X$}
    	\State ${\action} \gets $ sampled according to best$_{\ub, \lb}(\state)$
    	\State $\tau \gets $ sampled from the exponential distribution with rate $\lambda(\state, \action)$
    	\State $\tau_{(\state, \action)} \gets \tau_{(\state, \action)} \cup \tau$
    	\label{algline:learn_rate_ctmdp}
    	\If {$a \:!\!\!= {\sf stay}$}
    	\State $t \gets$ sampled according to $\trans(s,{a})$ \Comment{sampled after staying in $\state$ for time $\tau$}
    	\Else \Comment{The {\sf stay} action simulates reaching the terminal states in $\Mdp'$}
    	\State $t \gets$ sampled according to $\trans'(s,{\sf stay})$
    	\EndIf
    	\State $\text{Increment }\#(\state,{\action},\tsucc)$
    	\State $\state \gets \tsucc$ \Comment{Move from $s$ to $t$ after time $\tau$.}
        \If {$\mathsf{Appear}(X, \state) \ge k$} \label{fcall:appear_ctmdp}
            \If{$\STUCK(X, \state)$}
            \label{fcall:looping_ctmdp}
                \State ${\action} = $ $\mathsf{BEST\_LEAVING\_ACTION}(T)$ \Comment{$T$ is the MEC to which $\state$ belongs; actions are ranked according to best$_{\ub, \lb}(\state)$}
                \State $\state$ = origin of ${\action}$
                \State $\tau \gets $ sampled from the exponential distribution with rate $\lambda(\state, \action)$
            	\If {$a \:!\!\!= {\sf stay}$}
    	            \State $t \gets$ sampled according to $\trans(s,{a})$ \Comment{sampled after staying in $\state$ for time $\tau$}
    	        \Else \Comment{The {\sf stay} action simulates reaching the terminal states in $\Mdp'$}
    	            \State $t \gets$ sampled according to $\trans'(s,{\sf stay})$
    	        \EndIf
            	\State $\text{Increment }\#(\state,{\action},\tsucc)$
            	\State $\state \gets \tsucc$ \Comment{Move from $s$ to $t$ after time $\tau$.}
            \EndIf
        \EndIf
    \Until{$\state \in \{s_+, s_-, s_?\}$}
    \State \textbf{return} $X \cup \set{\state}$
\EndProcedure
\end{algorithmic}
\end{algorithm}

Algorithm~\ref{alg:updateMECCtmdp}, finds the mean-payoff bounds for a CTMDP MEC $M_{i}$. To find the bounds of the mean payoff, we first convert the CTMDP MEC $M_{i}$ to an MDP MEC $M_{i}^\prime$, using the uniformization process (Line~\ref{algline:uniformization_ctmdp}). Then we find the mean payoff of $M_{i}^\prime$ using the value iteration algorithm for MDP MEC (Line~\ref{algline:vi_ctmdp}), as described in Algorithm~\ref{alg:vi}.

\begin{algorithm}[t]
	\caption{Algorithm to update MEC reward value and stay action}\label{alg:updateMECCtmdp}
		\hspace*{\algorithmicindent}\algorithmicrequire	$\deltasure$ EC $M=(T, A)$, required precision $\beta$.
	\begin{algorithmic}[1]
	\Procedure{$\mathsf{UPDATE\_MEC\_VALUE\_CTMDP}$}{}
	    
	   
	    \State $nSamples = \mathsf{COMPUTE\_N\_SAMPLES}(M_i)$
	    \State $\mathsf{SIMULATE\_MEC}(M, nSamples)$
		\State $M' = \mathsf{UNIFORMIZE}(M)$
		\label{algline:uniformization_ctmdp}
		\State $\widehat{v}^l_{M'}, \widehat{v}^u_{M'} = \mathsf{VALUE\_ITERATION} (M', \beta)$
		\label{algline:vi_ctmdp}
	\State \Return $\widehat{v}^l_{M'}, \widehat{v}^u_{M'}$
    \EndProcedure
	\end{algorithmic}
\end{algorithm}

Algorithm~\ref{alg:simulate_MEC_Ctmdp} is similar to Algorithm~\ref{alg:simulate_MEC_MDP_HEURISTIC}, except that, here in addition to learning transition probabilities, we also learn the mean of the exponential distribution of the states inside the MEC (Line~\ref{algline:sample_exp_dist}-\ref{algline:append_exp_dist}).

\begin{algorithm}[h]
\caption{Simulation algorithm for CTMDP MEC}\label{alg:simulate_MEC_Ctmdp}
		\hspace*{\algorithmicindent}\algorithmicrequire	CTMDP MEC $M=(T, A)$, $nSamples$.\\
		\hspace*{\algorithmicindent}\algorithmicensure estimated transition probabilities and rates for state-action pairs.
\begin{algorithmic}[1]
\Procedure{$\SIMULATE{\sf \_MEC}$}{}
\State $\state \gets \initstate$
     \For{$nSamples * nTransitions$ times}
    	\State ${\action} \gets $ sampled uniformly from $\state$
    	\State $\tsucc \gets$ sampled according to $\trans(\state,{\action})$
    	\State $\text{Increment }\#(\state,{\action},\tsucc)$
    	\State $\tau \gets $ sampled from the exponential distribution with rate
    	$\lambda(\state, \action)$ \Comment{sampled after staying in $\state$ for time $\tau$}
    	\label{algline:sample_exp_dist}
    	\State $\tau_{(\state, \action)} \gets \tau_{(\state, \action)} \cup \tau$
    	\label{algline:append_exp_dist}
    	\State $\state \gets \tsucc$ \Comment{Move from $s$ to $t$.}
    \EndFor
    \State $\tau_{avg(\state, \action)} \gets \dfrac{\sum_{\tau \in \tau_{(\state, \action)}} \tau}{\lvert \tau_{\state, \action} \rvert}$
    \State $\widehat{\lambda}_{(\state, \action)} \gets \dfrac{1}{\tau_{avg(\state, \action)}}$
    \State \textbf{return} $\widehat{\lambda}_{(\state, \action)}$ for all $(\state, \action)$.
\EndProcedure
\end{algorithmic}
\end{algorithm}


In Algorithm~\ref{alg:mecMeanPayoffBoundsTrueAlgorithm}, we compute the bounds on $v_{M}$.

\begin{algorithm}[t]
    \caption{Algorithm to find the bounds on the value of $M$}\label{alg:mecMeanPayoffBoundsTrueAlgorithm}
		\hspace*{\algorithmicindent}\algorithmicrequire	CTMDP MEC $M$, Target precision $\beta$. \\
		\hspace*{\algorithmicindent}\algorithmicensure Bounds on $v_{M}$
	\begin{algorithmic}[1]
	    \Procedure{$\mathsf{FIND\_MEC\_MEANPAYOFF\_BOUNDS}$}{}
	    
	    \State $\widehat{v}^u_M \gets \mathsf{FIND\_MAXIMAL\_MEANPAYOFF}(M, \beta)$
	    \State $\widehat{v}^l_M \gets \mathsf{FIND\_MINIMAL\_MEANPAYOFF}(M, \beta)$\\
		\Return $\widehat{v}^l_M, \widehat{v}^u_M$
		\EndProcedure
	\end{algorithmic}
\end{algorithm}

Let $M_j$ be an MEC same as that of $M$, but with the rates chosen according to the function, 

\begin{equation*}
    \lambda_{i} = 
    \begin{cases}
    \widehat{\lambda_i} \left(1-\alphaR\right), & \text{if}\ i \leq j \\
      \widehat{\lambda_i} \left(1+\alphaR\right), & \text{otherwise}
    \end{cases}
\end{equation*}

From $\deltaR$ and the number of samples, we can get the value of $\alphaR$ from Table~\ref{tab:nSamplesTable2}. We also assume that the states in $M_j$ are sorted by the descending order of their rewards. That is, $r(s_1) \geq r(s_2) \geq r(s_3) \geq ... \geq r(s_{m})$.

That is for the first $j$ states, we choose the lower value of the rate, and for the remaining states we choose the upper value of the rate. Given a CTMDP MEC $M$, the following algorithm finds the upper bound on $v_{M}$. 

\begin{algorithm}[t]
    \caption{Algorithm to find the upper bound of $\widehat{v}_{M}$}\label{alg:maximumMecMeanPayoffTrueAlgorithm}
		\hspace*{\algorithmicindent}\algorithmicrequire	CTMDP MEC $M$, Target precision $\alpha$. \\
		\hspace*{\algorithmicindent}\algorithmicensure Upper bound of $\widehat{v}_{M}$
	\begin{algorithmic}[1]
	    \Procedure{$\mathsf{FIND\_MAXIMAL\_MEANPAYOFF}$}{$M$, $\beta$}
	    \State $v \gets \mathsf{UPDATE\_MEC\_VALUE\_CTMDP}(M_1, \beta)$
		  \For{$i=2$ to $m$}
		      \State $v^\prime \gets \mathsf{UPDATE\_MEC\_VALUE\_CTMDP}(M_i, \beta)$
		      \If {$v^\prime < v$}
		        \State \Return $v$
		      \EndIf
		      \State $v=v^\prime$
		  \EndFor

		\Return $v$
		\EndProcedure
	\end{algorithmic}
\end{algorithm}

Similarly we can find a lower bound on $v_M$, by making $M_j$ to have the rates, 

\begin{equation*}
    \lambda_{i} = 
    \begin{cases}
    \widehat{\lambda_i} \left(1+\alphaR\right), & \text{if}\ i \leq j \\
      \widehat{\lambda_i} \left(1-\alphaR\right), & \text{otherwise}
    \end{cases}
\end{equation*}

and modifying Algorithm \ref{alg:maximumMecMeanPayoffTrueAlgorithm} to obtain $\mathsf{FIND\_MINIMAL\_MEANPAYOFF}$. We can use these values as our new bounds on the mean payoff of $M$, as shown in Algorithm~\ref{alg:mecMeanPayoffBoundsTrueAlgorithm}.

Since from Algorithm~\ref{alg:maximumMecMeanPayoffTrueAlgorithm}, the number of calls made to Algorithm~\ref{alg:updateMECCtmdp} is proportional to the number of states of the MEC $M$, we provide a more efficient heuristic in Algorithm~\ref{alg:maximumMecMeanPayoffPracticalAlgorithm} to the bounds on mean payoff of a MEC. Here we approximately compute the upper and lower bounds on $v_M$. Let $f_{L}$ and $f_{U}$ be two functions that are defined as follows,

\begin{equation*}
    f_{L}\left(i\right) = 
    \begin{cases}
    \widehat{\lambda}_i \left(1+\alphaR\right), & \text{if}\ r(s_{i}) \geqslant \widehat{v}_{M} \\
      \widehat{\lambda}_i \left(1-\alphaR\right), & \text{otherwise}
    \end{cases}
\end{equation*}

\begin{equation*}
    f_{U}\left(i\right) = 
    \begin{cases}
    \widehat{\lambda}_i \left(1-\alphaR\right), & \text{if}\ r(s_{i}) \geq \widehat{v}_{M} \\
      \widehat{\lambda}_i \left(1+\alphaR\right), & \text{otherwise}
    \end{cases}
\end{equation*}
where $\widehat{v}_M$ is the mean-payoff value of $M$, obtained by using the estimated rate $\widehat{\lambda}$, ignoring the error $\alphaR$. In Algorithm~\ref{alg:maximumMecMeanPayoffPracticalAlgorithm}, we call the Algorithm~\ref{alg:updateMECCtmdp} only 3 times. For the first time, to find the mean payoff of the MEC $M$ using the estimated values,for the second time to find the lower bound on the mean payoff and for the last time, to find the upper bound on the mean payoff of $M$. 

\begin{algorithm}[t]
    \caption{Algorithm to find the upper bound of $\widehat{v}_{M}$}\label{alg:maximumMecMeanPayoffPracticalAlgorithm}
		\hspace*{\algorithmicindent}\algorithmicrequire	CTMDP MEC $M$, Target precision $\beta$.
	\begin{algorithmic}[1]
	    \Procedure{$\mathsf{FIND\_MEC\_MEANPAYOFF\_BOUNDS}$}{}
	    \State $\widehat{v}_M \gets \mathsf{UPDATE\_MEC\_VALUE\_CTMDP}(M, \beta)$
	    \State $\widehat{v}^l_M \gets \mathsf{UPDATE\_MEC\_VALUE\_CTMDP}(M_{L}, \beta)$
	    \Comment{$M_{L}$ is same as $M$, except it has its rates chosen according to function $f_{L}$}
	    \State $\widehat{v}^u_M \gets \mathsf{UPDATE\_MEC\_VALUE\_CTMDP}(M_{U}, \beta)$
	    \Comment{$M_{U}$ is same as $M$, except it has its rates chosen according to function $f_{U}$}
		\State \Return $\widehat{v}^l_M, \widehat{v}^u_M$
		\EndProcedure
	\end{algorithmic}
\end{algorithm}
\section{More on Experimental Results and benchmarks} \label{app:moreExp}
In this section we provide our findings on several additional experiments apart from the ones shown in Section~\ref{sec:results}. 
We provide descriptions of the benchmarks, and also add more plots for our results of benchmarks.

\subsection{More about MDP benchmarks} \label{app:benchmarks}

\paragraph*{virus} \cite{kwiatkowska2009probabilistic} models a virus spreading through a network. Each attack carried out by a machine is rewarded. This model is comprised of $809$ states where the entire set of states form a single EC. There are no other ECs in the model.

\paragraph*{cs\_nfail} \cite{komuravelli2012assume} models a client-server mutual exclusion protocol with probabilistic failures of the clients. A reward is given for each successfully handled connection. In our experiments, we use the model with 3 clients.

\paragraph*{investor} \cite{MM02,MM07} models an investor operating in a stock market. The investor can decide to sell his stocks and keep their value as a reward or hold them and wait to see how the market evolves. The rewards correspond to the value of the stocks when the investor decides to sell them, so maximizing the average reward corresponds to maximizing the expected selling value of the stocks.

\paragraph*{zeroconf} \cite{KNPS06} models a network protocol designed for assigning IP addresses to clients without the need for a dedicated server. The scenario involves a set number of hosts ($N$) that are already using a set of IP addresses and a new host trying to pick an IP address. The host sends out a fixed number of probes ($K$) to verify if the IP address is not being used by some other host and then, finally proceeds to use the address. This model is comprised of several hundred thousand transient states and a few `final' states that have self-loops (and are thus single state MECs). This is essentially a reachability problem where we find a path to one of the `final' state. We assign a reward of 1 for the desired `final' state to convert this to a mean-payoff problem. In our experiments, we use the model with $N=40$ and $K=10$.

\paragraph*{sensors} \cite{komuravelli2012assume} models a network of sensors sending packets to a central processor over a lossy connection. Every successfully processed packet counts for a reward. In our experiments, we use the model that assumes that 3 packets have to be sent.

\paragraph*{consensus} \cite{AH90} models a randomised consensus protocol between N asynchronous processes that communicate with each other. The processes proceed through possibly an unbounded number of rounds until an agreement is reached between the 2. We convert this reachability problem to a mean-payoff problem by adding a reward of 1 on the self-loop of the goal state. In our experiments, we use the model that assumes 2 asynchronous processes.

\paragraph*{ij} \cite{IJ90} models a self-stabilizing protocol for a network of processes which, when started from some possibly illegal start configuration, returns to a legal/stable configuration without any outside intervention within some finite number of steps. We convert this reachability problem to a mean-payoff problem by adding a reward of 1 on the self-loop of the goal state. In our experiments, we use the model that assumes 2 asynchronous processes. In our experiments, we use a model that assumes 3 processes and another model that assumes 10 processes.

\paragraph*{pacman} \cite{jansen2019safe} encodes a small variant of the arcade game Pac-Man in which the ghosts behave randomly based on observed behavior. In some sense, it is a grid-world with many obstacles: still largely connected, but far fewer actions than states. The process ends when pacman eats all the fruits. We convert this reachability problem to a mean-payoff problem by adding a reward of 1 on the self-loop of the goal state.


\paragraph*{wlan}
\cite{KNS02} models two stations colliding - trying to send messages at the same time over one channel - and then entering the randomised exponential backoff procedure. We add a reward $1$ on the goal state, to convert this into a mean-payoff problem from reachability problem. 

\paragraph*{blackjack}~\cite{SB98} is a popular card game played in casinos. In {\sf blackjack} model, the objective is to win money by obtaining a point total higher than the dealer's without exceeding 21.

\paragraph*{counter}~\cite{HPSSTW21} This model has a counter whose value lies between $0$ and $7$. 
The middle value is $\lfloor \frac{0+7}{2} \rfloor = 3$.
If the counter value is less than the maximum value, then one can increment the counter, and if its value is more than the minimum value, then one can decrement the counter.
If we try to increment the counter when it has already the maximum value of $7$, then the value remains the same with probability $\frac{1}{2}$, while the value is reset to $0$ with probability $\frac{1}{2}$.
Similarly, if the value of the counter is $0$, and we try to decrement its value, it remains $0$ with probability $\frac{2}{3}$, and changes to $7$ with probability $\frac{1}{3}$.
A positive reward of $10$ is obtained when the counter is incremented and its value is more than $3$, otherwise, while incrementing the counter a negative reward of $10$ is obtained.

\paragraph*{recycling}~\cite{SB98} is a \emph{recycling robot} example where there is a robot that runs on a battery, and its task is to collect empty cans. When it collects one, it gets a positive reward.
When performing its task, with certain probability, it can run out of its battery or the battery can change state from {\sf high} to {\sf low}.
If the battery drains out, it gets a negative reward.
In each state, the robot can decide whether to actively {\sf search} for an empty can, or it can {\sf wait} till someone gets it a can, or it can go back to its `home' and {\sf recharge} its battery. 

\paragraph*{busyRing, busyRingMc}~\cite{HTWW86,Kinniment2007} examples describe classical busy-ring asynchronous arbiters that decide which request to grant and they are based on the use of a MUTEX circuit.

\subsection{More about CTMDP benchmarks} \label{app:CTMDP_benchmarks}
 \paragraph{Dynamic Power Management (DPM)} models encode dynamic power management problem based on~\cite{QiuWP99}.
\paragraph{Queuing System (QS)} \cite{HatefiH12} models are based on a CTMDP modelling of queuing systems with arrival rate, service rate, and jump rate as the key parameters.
\paragraph{Polling System} \cite{GuckHHKT13} examples consist of $j$ stations and $1$ server. Here, the incoming requests of $j$ types are buffered in queues of size $k$ each, until they are processed by the server and delivered to their station. The system starts in a state with all the queues being nearly full. We consider 2 goal conditions: (i) all the queues are empty and (ii) one of the queues is empty.
\paragraph{Erlang Stages} \cite{ZhangN10} models have two different paths to reach the goal state: a fast but risky path or a slow but sure path. The slow path is an Erlang chain of length $k$ and rate $r$.
\paragraph{Stochastic Job Scheduling (SJS)} \cite{QuatmannJK17} models multiprocessors with a sequence of independent jobs with the goal being the completion of the jobs. 

\subsection{Parameters used in benchmarks}
\label{app:benchmarks_parameters}

Table~\ref{tab:ctmdp_benchmarks_parameters} and Table~\ref{tab:mdp_benchmarks_parameters} shows the value of parameters that we set for CTMDP and MDP benchmarks in our experiments. The benchmarks and the description of the parameters shown in this table are from the Quantitative Verification Benchmark Set~\cite{HKPQR19}.

\begin{table}[]
\centering
\scalebox{0.95}{
\begin{tabular}{|l|l|l|}
\hline
\multicolumn{1}{|c|}{Benchmarks} & \multicolumn{1}{c|}{Values used} & \multicolumn{1}{c|}{Description}              \\ \hline
zeroconf &
  \begin{tabular}[c]{@{}l@{}}N=40\\ K=10\\ reset=false\end{tabular} &
  \begin{tabular}[c]{@{}l@{}}N is the number of existing nodes\\ K is the number of probes sent\\ reset denotes whether or not to \\ clear messages on restart\end{tabular} \\ \hline
sensors                          & K=3                              & K is the number of chunks of data             \\ \hline
consensus &
  \begin{tabular}[c]{@{}l@{}}N=2\\ K=2\end{tabular} &
  \begin{tabular}[c]{@{}l@{}}N is the number of processes\\ K denotes the bound on random walk\end{tabular} \\ \hline
ij10 &
  num\_tokens\_var=10 &
  \multirow{2}{*}{\begin{tabular}[c]{@{}l@{}}num\_tokens\_var is the number of tokens\\ and processes\end{tabular}} \\ \cline{1-2}
ij3                              & num\_tokens\_var=3               &                                               \\ \hline
pacman                           & MAXSTEPS=3                       & MAXSTEPS is the number of steps we plan ahead \\ \hline
wlan                             & COL=0                            & COL denotes the maximum number of collisions  \\ \hline
\end{tabular}
}
\caption{Parameters for MDP benchmarks}
\label{tab:mdp_benchmarks_parameters}
\end{table}

\begin{table}[!ht]
\scalebox{0.95}{
\centering
\begin{tabular}{|l|l|l|}
\hline
\multicolumn{1}{|c|}{Benchmarks} &
  \multicolumn{1}{c|}{Values used} &
  \multicolumn{1}{c|}{Description} \\ \hline
DynamicPM &
  \begin{tabular}[c]{@{}l@{}}N=3\\C=2\end{tabular} &
  \begin{tabular}[c]{@{}l@{}}N is the number of different types of tasks\\C is maximum queue size for each task\end{tabular} \\ \hline
ErlangStages &
  \begin{tabular}[c]{@{}l@{}}K=500\\R=10\end{tabular} &
  \begin{tabular}[c]{@{}l@{}}K is the shape parameter of the Erlang\\ distribution\\R is the rate parameter of the Erlang \\ distribution\end{tabular} \\ \hline
PollingSystem1 &
  \begin{tabular}[c]{@{}l@{}}JOB\_TYPES=1\\C=1\end{tabular} &
  \multirow{3}{*}{\begin{tabular}[c]{@{}l@{}}\\JOB\_TYPES is the number of job types\\C is the capacity of each queue\end{tabular}} \\ \cline{1-2}
PollingSystem2 &
  \begin{tabular}[c]{@{}l@{}}JOB\_TYPES=1\\C=4\end{tabular} &
   \\ \cline{1-2}
PollingSystem3 &
  \begin{tabular}[c]{@{}l@{}}JOB\_TYPES=1\\C=7\end{tabular} &
   \\ \hline
QueueingSystem &
  \begin{tabular}[c]{@{}l@{}}JOB\_TYPES=2\\C\_LEFT=1\\C\_RIGHT=1\end{tabular} &
  \begin{tabular}[c]{@{}l@{}}JOB\_TYPES is the number of job types\\C\_LEFT denotes the capacity of the left queue\\C\_RIGHT denotes the capacity of the right queue\end{tabular} \\ \hline
SJS1 &
  \begin{tabular}[c]{@{}l@{}}N=2\\K=2\end{tabular} &
  \multirow{3}{*}{\begin{tabular}[c]{@{}l@{}}\\N is the number of jobs to schedule\\K is the number of processes\end{tabular}} \\ \cline{1-2}
SJS2 &
  \begin{tabular}[c]{@{}l@{}}N=2\\K=6\end{tabular} &
   \\ \cline{1-2}
SJS3 &
  \begin{tabular}[c]{@{}l@{}}N=6\\K=2\end{tabular} &
   \\ \hline
\end{tabular}
}
\caption{Parameters for CTMDP benchmarks}
\label{tab:ctmdp_benchmarks_parameters}
\end{table}

\subsection{Comparison of $\pmin$ and Max successors in computing $\delta_{TP}$}
\label{app:pminandmaxsuccessorcomparison}

The $\mpInconfidence$ $\symbolMpInconfidence$ distributed over all transitions in a model, is defined as $\symbolTpInconfidence := \dfrac{\symbolMpInconfidence\cdot \pmin}{\lvert\{{a} \vert\, s \in \widehat{S} \wedge {a} \in \av(s) \}\rvert}$. The quantity $\frac{1}{\pmin}$ gives an upper bound on the maximum number of possible successors for an available action from a state. 
Instead of using $\frac{1}{\pmin}$, if we additionally assume the knowledge of $\max\limits_{s \in ss,a \in \Av(s)} |\post(s,a)|$, that is, the maximum number of successors over all states $s$ and all actions available from $s$, this can be used to obtain $\symbolTpInconfidence'$, a better estimate of $\symbolTpInconfidence$.
Here $\symbolTpInconfidence' := \dfrac{\symbolMpInconfidence}{\lvert\{{a} \vert\, s \in \widehat{S} \wedge {a} \in \av(s) \}\rvert \cdot m}$. The quantity $\frac{1}{\pmin}$. 
Clearly, $\delta_{TP}'\le\delta_{TP}$ and we observe a little improvement in the convergence as shown in Table~\ref{tab:pminAndMaxSuccessorComparisonTable}, when we use $\symbolTpInconfidence'$ instead of $\symbolTpInconfidence$.

\begin{table}[!htb]
\scalebox{0.9}{
\centering
\begin{tabular}{|l|llllll|llllll|}
\hline
\multicolumn{1}{|c|}{\multirow{3}{*}{Benchmarks}} &
  \multicolumn{6}{c|}{Blackbox} &
  \multicolumn{6}{c|}{\begin{tabular}[c]{@{}c@{}}Blackbox with \\ greybox update equations\end{tabular}} \\ \cline{2-13} 
\multicolumn{1}{|c|}{} &
  \multicolumn{3}{c|}{pMin} &
  \multicolumn{3}{c|}{\begin{tabular}[c]{@{}c@{}}Max \\ successors\end{tabular}} &
  \multicolumn{3}{c|}{pMin} &
  \multicolumn{3}{c|}{\begin{tabular}[c]{@{}c@{}}Max \\ successors\end{tabular}} \\ \cline{2-13} 
\multicolumn{1}{|c|}{} &
  \multicolumn{1}{c|}{\begin{tabular}[c]{@{}c@{}}Lower  \\ bound\end{tabular}} &
  \multicolumn{1}{c|}{\begin{tabular}[c]{@{}c@{}}Upper \\ bound\end{tabular}} &
  \multicolumn{1}{c|}{\begin{tabular}[c]{@{}c@{}}Time \\ (s)\end{tabular}} &
  \multicolumn{1}{c|}{\begin{tabular}[c]{@{}c@{}}Lower  \\ bound\end{tabular}} &
  \multicolumn{1}{c|}{\begin{tabular}[c]{@{}c@{}}Upper \\ bound\end{tabular}} &
  \multicolumn{1}{c|}{\begin{tabular}[c]{@{}c@{}}Time \\ (s)\end{tabular}} &
  \multicolumn{1}{c|}{\begin{tabular}[c]{@{}c@{}}Lower \\ bound\end{tabular}} &
  \multicolumn{1}{c|}{\begin{tabular}[c]{@{}c@{}}Upper \\ bound\end{tabular}} &
  \multicolumn{1}{c|}{\begin{tabular}[c]{@{}c@{}}Time \\ (s)\end{tabular}} &
  \multicolumn{1}{c|}{\begin{tabular}[c]{@{}c@{}}Lower \\ bound\end{tabular}} &
  \multicolumn{1}{c|}{\begin{tabular}[c]{@{}c@{}}Upper \\ bound\end{tabular}} &
  \multicolumn{1}{c|}{\begin{tabular}[c]{@{}c@{}}Time \\ (s)\end{tabular}} \\ \hline
virus &
  \multicolumn{1}{l|}{0.0} &
  \multicolumn{1}{l|}{0.5319} &
  \multicolumn{1}{l|}{TO} &
  \multicolumn{1}{l|}{0.0} &
  \multicolumn{1}{l|}{0.519} &
  TO &
  \multicolumn{1}{l|}{0.0} &
  \multicolumn{1}{l|}{0.008} &
  \multicolumn{1}{l|}{273.01} &
  \multicolumn{1}{l|}{0.0} &
  \multicolumn{1}{l|}{0.007} &
  250.09 \\ \hline
cs\_nfail &
  \multicolumn{1}{l|}{0.3275} &
  \multicolumn{1}{l|}{0.3618} &
  \multicolumn{1}{l|}{TO} &
  \multicolumn{1}{l|}{0.3275} &
  \multicolumn{1}{l|}{0.3508} &
  TO &
  \multicolumn{1}{l|}{0.332} &
  \multicolumn{1}{l|}{0.337} &
  \multicolumn{1}{l|}{126.77} &
  \multicolumn{1}{l|}{0.332} &
  \multicolumn{1}{l|}{0.337} &
  149.04 \\ \hline
investor &
  \multicolumn{1}{l|}{0.8458} &
  \multicolumn{1}{l|}{0.9559} &
  \multicolumn{1}{l|}{TO} &
  \multicolumn{1}{l|}{0.8552} &
  \multicolumn{1}{l|}{0.9554} &
  TO &
  \multicolumn{1}{l|}{0.945} &
  \multicolumn{1}{l|}{0.954} &
  \multicolumn{1}{l|}{620.23} &
  \multicolumn{1}{l|}{0.945} &
  \multicolumn{1}{l|}{0.955} &
  539.27 \\ \hline
zeroconf &
  \multicolumn{1}{l|}{0.923} &
  \multicolumn{1}{l|}{1.0} &
  \multicolumn{1}{l|}{TO} &
  \multicolumn{1}{l|}{0.921} &
  \multicolumn{1}{l|}{1.0} &
  TO &
  \multicolumn{1}{l|}{0.990} &
  \multicolumn{1}{l|}{1.0} &
  \multicolumn{1}{l|}{116.04} &
  \multicolumn{1}{l|}{0.990} &
  \multicolumn{1}{l|}{1.0} &
  95.73 \\ \hline
sensors &
  \multicolumn{1}{l|}{0.3299} &
  \multicolumn{1}{l|}{0.3513} &
  \multicolumn{1}{l|}{TO} &
  \multicolumn{1}{l|}{0.3298} &
  \multicolumn{1}{l|}{0.3463} &
  TO &
  \multicolumn{1}{l|}{0.332} &
  \multicolumn{1}{l|}{0.336} &
  \multicolumn{1}{l|}{64.64} &
  \multicolumn{1}{l|}{0.332} &
  \multicolumn{1}{l|}{0.336} &
  84.005 \\ \hline
consensus &
  \multicolumn{1}{l|}{0.0936} &
  \multicolumn{1}{l|}{0.1605} &
  \multicolumn{1}{l|}{TO} &
  \multicolumn{1}{l|}{0.0935} &
  \multicolumn{1}{l|}{0.1602} &
  TO &
  \multicolumn{1}{l|}{0.103} &
  \multicolumn{1}{l|}{0.113} &
  \multicolumn{1}{l|}{190.32} &
  \multicolumn{1}{l|}{0.103} &
  \multicolumn{1}{l|}{0.113} &
  174.92 \\ \hline
ij10 &
  \multicolumn{1}{l|}{0.3626} &
  \multicolumn{1}{l|}{1.0} &
  \multicolumn{1}{l|}{TO} &
  \multicolumn{1}{l|}{0.3558} &
  \multicolumn{1}{l|}{1.0} &
  TO &
  \multicolumn{1}{l|}{0.999} &
  \multicolumn{1}{l|}{1.0} &
  \multicolumn{1}{l|}{26.822} &
  \multicolumn{1}{l|}{0.999} &
  \multicolumn{1}{l|}{1.0} &
  23.273 \\ \hline
ij3 &
  \multicolumn{1}{l|}{0.990} &
  \multicolumn{1}{l|}{1.0} &
  \multicolumn{1}{l|}{15.92} &
  \multicolumn{1}{l|}{0.990} &
  \multicolumn{1}{l|}{1.0} &
  14.23 &
  \multicolumn{1}{l|}{0.999} &
  \multicolumn{1}{l|}{1.0} &
  \multicolumn{1}{l|}{0.7127} &
  \multicolumn{1}{l|}{0.999} &
  \multicolumn{1}{l|}{1.0} &
  0.5243 \\ \hline
pacman &
  \multicolumn{1}{l|}{0.535} &
  \multicolumn{1}{l|}{0.575} &
  \multicolumn{1}{l|}{TO} &
  \multicolumn{1}{l|}{0.535} &
  \multicolumn{1}{l|}{0.574} &
  TO &
  \multicolumn{1}{l|}{0.5477} &
  \multicolumn{1}{l|}{0.5577} &
  \multicolumn{1}{l|}{215.36} &
  \multicolumn{1}{l|}{0.548} &
  \multicolumn{1}{l|}{0.558} &
  201.72 \\ \hline
wlan &
  \multicolumn{1}{l|}{0.6577} &
  \multicolumn{1}{l|}{1.0} &
  \multicolumn{1}{l|}{TO} &
  \multicolumn{1}{l|}{0.6693} &
  \multicolumn{1}{l|}{1.0} &
  TO &
  \multicolumn{1}{l|}{1.0} &
  \multicolumn{1}{l|}{1.0} &
  \multicolumn{1}{l|}{16.924} &
  \multicolumn{1}{l|}{1.0} &
  \multicolumn{1}{l|}{1.0} &
  13.75 \\ \hline
blackjack &
  \multicolumn{1}{l|}{0.0} &
  \multicolumn{1}{l|}{0.3014} &
  \multicolumn{1}{l|}{TO} &
  \multicolumn{1}{l|}{0.0} &
  \multicolumn{1}{l|}{0.2659} &
  TO &
  \multicolumn{1}{l|}{0.0} &
  \multicolumn{1}{l|}{0.006} &
  \multicolumn{1}{l|}{91.503} &
  \multicolumn{1}{l|}{0.0} &
  \multicolumn{1}{l|}{0.008} &
  90.767 \\ \hline
counter &
  \multicolumn{1}{l|}{0.4998} &
  \multicolumn{1}{l|}{0.5} &
  \multicolumn{1}{l|}{30.37} &
  \multicolumn{1}{l|}{0.4999} &
  \multicolumn{1}{l|}{0.5} &
  26.755 &
  \multicolumn{1}{l|}{0.4999} &
  \multicolumn{1}{l|}{0.5} &
  \multicolumn{1}{l|}{15.215} &
  \multicolumn{1}{l|}{0.4999} &
  \multicolumn{1}{l|}{0.5} &
  21.791 \\ \hline
recycling &
  \multicolumn{1}{l|}{0.726} &
  \multicolumn{1}{l|}{0.727} &
  \multicolumn{1}{l|}{1.309} &
  \multicolumn{1}{l|}{0.726} &
  \multicolumn{1}{l|}{0.727} &
  1.046 &
  \multicolumn{1}{l|}{0.726} &
  \multicolumn{1}{l|}{0.727} &
  \multicolumn{1}{l|}{0.927} &
  \multicolumn{1}{l|}{0.726} &
  \multicolumn{1}{l|}{0.727} &
  0.787 \\ \hline
busyRing &
  \multicolumn{1}{l|}{0.706} &
  \multicolumn{1}{l|}{1.0} &
  \multicolumn{1}{l|}{TO} &
  \multicolumn{1}{l|}{0.739} &
  \multicolumn{1}{l|}{1.0} &
  TO &
  \multicolumn{1}{l|}{0.999} &
  \multicolumn{1}{l|}{1.0} &
  \multicolumn{1}{l|}{34.86} &
  \multicolumn{1}{l|}{0.999} &
  \multicolumn{1}{l|}{1.0} &
  39.95 \\ \hline
busyRingMC &
  \multicolumn{1}{l|}{0.969} &
  \multicolumn{1}{l|}{1.0} &
  \multicolumn{1}{l|}{1773.25} &
  \multicolumn{1}{l|}{0.9775} &
  \multicolumn{1}{l|}{1.0} &
  1729.34 &
  \multicolumn{1}{l|}{0.999} &
  \multicolumn{1}{l|}{1.0} &
  \multicolumn{1}{l|}{114.50} &
  \multicolumn{1}{l|}{0.999} &
  \multicolumn{1}{l|}{1.0} &
  102.48 \\ \hline
\end{tabular}
}
\captionsetup{justification=centering}
\caption{\label{tab:pminAndMaxSuccessorComparisonTable} Results on MDP benchmarks comparing use of $\pmin$ and maximum number of successors over all state-action pairs in the computation of $\symbolTpInconfidence$.}
\end{table}

\subsection{Experiments on increase in error tolerance for using greybox equations in blackbox models}
\label{app:greyboxerrorprobability}

The greybox update equations, assumes that we know all the successors of a state $s$ for an action $a$, while updating the values of $\widehat{\lb}(s, {a})$ and $\widehat{\ub}(s, {a})$. However, in blackbox this is not the case, that is, we do not know the actual number of successors for a state-action pair. So there is a probability that we might not have visited all the successors for a state-action pair, but still we update the values using these equations. Here we find the probability of using the greybox udpate equations, without visiting all the successors. 


Suppose for state $s$ and action $a$, there are $l$ successors ${{s}_1}, {{s}_2} ... {{s}_l}$. Then the probability that ${s}_1$ might not be visited after visiting the action $n$ times, is given by $(1 - \trans{(s, a, {s}_1)})^n$. We compute the probability of not visiting each such successor and add them up, to get the probability of not visiting a successor after $n$ visits. So for a state ${s}$ and action $a$, the probability that one of its successors has not been visited is given by $t(s, a)=(1 - \trans{(s, a, {s}_1)})^n + (1 - \trans{(s, a, {s}_2)})^n + ... + (1 - \trans{(s, a, {s}_l)})^n$.

We multiply this probability with $p\_reach\_s$, the probability the state $s$ is reachable from the initial state. That is $p(s, a)=p\_reach\_s \cdot t(s, a)$.
We multiply with $p\_reach\_s$, because the states which are closer to the initial state, will have more impact on the mean-payoff value than the states which are farther. We ignore the probabilities of missing more than one successor since those probabilities are much smaller.
There can be multiple paths that lead to state $s$ from the initial state $\initstate$. Here we take the reachability probability from the path, that we take to reach $s$ the first time, while computing the error probability.

Here the error probabilities of all the state-action pairs, that belongs to the partial model constructed during value iteration, are computed, except for the state-action pairs that belong to an MEC since the MECs are already identified as $\symbolTpInconfidence$-sure. At last, we compute the total error probability $p^\prime$ by the following equation.

\begin{equation*}
    p^\prime = 1 - \prod_{\set{(s, a) \mid s \in S \wedge a \in A}} (1 - p(s, a))
\end{equation*}

where $p(s, a)$ denotes the error probability on the state-action pair ($s$, $a$). $p(s,a)$ will be $0$ when $(s, a)$ pair belongs to an MEC. The quantity $p^\prime$ denotes the probability that for at least one of the state-action pair, all the successors of that pairs, has not been visited. The error probabilities for the benchmarks can be found in Table~\ref{tab:errorProb2}.

\begin{table}
\scalebox{0.9}{
\centering
\begin{minipage}[b]{0.4\hsize}
    \begin{tabular}{|l|l|}
\hline
Benchmarks & Increase in inconfidence \\ \hline
virus               & 0.000003                   \\ \hline
cs\_nfail          & 0.7459                     \\ \hline
investor            & 0.0                        \\ \hline
zeroconf            & 0.3861                     \\ \hline
sensors             & 0.0007                     \\ \hline
consensus          & 0.0                        \\ \hline
ij10                & 0.009                      \\ \hline
ij3                 & 0.0                        \\ \hline
pacman              & 0.007                      \\ \hline
wlan               & 0.2909                     \\ \hline
blackjack           & 0.0001                     \\
\hline
counter             & 0                          \\
\hline
recycling           & 0                          \\
\hline
busyRing           & 0.1399                     \\
\hline
busyRingMC4           & -\footnotemark           \\
\hline
\end{tabular}
\captionsetup{justification=centering}
\caption{Increase in inconfidence due to the use of greybox update equations in blackbox learning}
\label{tab:errorProb2}
\end{minipage}
\quad \quad \quad
\begin{minipage}[b]{0.4\hsize}
    \begin{tabular}{|l|l|l|l|l|}
\hline
\multicolumn{1}{|c|}{Benchmarks} &
  \multicolumn{1}{c|}{\begin{tabular}[c]{@{}c@{}}States \\ explored\end{tabular}} &
  \multicolumn{1}{c|}{\begin{tabular}[c]{@{}c@{}}Lower \\ bound\end{tabular}} &
  \multicolumn{1}{c|}{\begin{tabular}[c]{@{}c@{}}Upper \\ bound\end{tabular}} &
  \multicolumn{1}{c|}{\begin{tabular}[c]{@{}c@{}}Time \\ (s)\end{tabular}} \\ \hline
virus             & 809  & 0.0    & 0.008 & 443.76 \\ \hline
cs\_nfail         & 184  & 0.330  & 0.34  & 91.44  \\ \hline
investor          & 5837 & 0.945  & 0.954 & 460.43 \\ \hline
zeroconf          & 370  & 0.990  & 1.0   & 98.205 \\ \hline
sensors           & 188  & 0.332  & 0.337 & 18.259 \\ \hline
consensus         & 272  & 0.103  & 0.113 & 211.15 \\ \hline
ij10              & 1023 & 0.999  & 1.0   & 16.891 \\ \hline
ij3               & 7    & 0.999  & 1.0   & 0.820  \\ \hline
pacman            & 496  & 0.547  & 0.557 & 160.05 \\ \hline
wlan              & 2935 & 1.0    & 1.0   & 14.658 \\ \hline
blackjack         & 3829 & 0.0    & 0.004 & 94.776 \\ \hline
counter           & 8    & 0.4999 & 0.5   & 17.898 \\ \hline
recycling         & 5    & 0.726  & 0.727 & 1.069  \\ \hline
busyRing          & 1059 & 0.999  & 1.0   & 110.64 \\ \hline
busyRingMC        & 1667 & 0.999  & 1.0   & 529.27 \\ \hline
\end{tabular}
\captionsetup{justification=centering}
\caption{Results for greybox learning on MDP benchmarks}
\label{tab:mdpGreyboxResults}
\end{minipage}
}
\end{table}
\footnotetext{Our script failed to compute the increase in inconfidence for this benchmark since it gave a stack overflow error resulting from long paths in this benchmark.}

The model {\sf ij3} has error probability of $0$. This is because, the least transition probability in the {\sf ij3} model, we have ${\pmin}=0.5$. Also for {\sf consensus}, and {\sf ij10} models, the ${\pmin}$ equals $0.5$. So the error probability of not visiting a state is bounded by, $(1-0.5)^n$ which is $(0.5)^n$, and for $n = 50$, this value is $\approx 8E^{-16}$. 

This is also the case with {\sf consensus}. Each transition is visited more than $300$ times by our algorithm as observed by running the algorithm multiple times. So it also has error probability close to $0$. 

For {\sf ij10}, some of the states has low visit counts such as $1$ or $2$. This causes some error probability. The effect of the error is reduced since we are multiplying it with reachability probability $p\_reach\_s$. 

\subsection{Greybox learning} \label{app:greybox}
In the greybox setting, for every state-action pair $(s,a)$, we additionally know the exact number of successors $\abs{\post(s,a)}$.
This gives an improved EC-detection mechanism.
In blackbox EC-detection, while detecting $\deltasure$-EC, we try to have many samples of the same action so that the probability that a transition might have been missed would be reduced. In greybox, on the other hand, we know the exact number of transitions for every state-action pair, and thus usually we don't need many samples to make sure that there aren't any unseen transitions.

Further, we use the greybox Bellman update equations for estimating the mean-payoff values corresponding to state-action pairs where all the successors are visited, while if all the successors are not visited, we use blackbox update equations. These improvements have considerable impact on performance as demonstrated In Table~\ref{tab:mdpGreyboxResults}. The parameters used to run experiments for greybox learning are exactly the same as that of the parameters used for blackbox learning, as detailed in Section~\ref{sec:results}.
For most benchmarks, other than {\sf virus, consensus, busyring}, and {\sf busyringMC}, the results for greybox learning are marginally better than the results of blackbox learning with greybox update equations.

Blackbox learning with greybox update equations always uses the greybox update equations while computing the upper and the lower bounds for a state-action pair. However, in greybox learning, greybox update equations are only used for a state-action pair, when we are sure that all the successors of that state-action pair have been visited at least once. This may result in blackbox learning with greybox update equations to converge slightly faster than greybox learning.
On the other hand, once an MEC is detected in the partial model, it is surely known to be an MEC in the case of greybox learning, but in the case of blackbox learning it is only known to be a $\deltasure$ MEC.
Thus in the case of blackbox learning, an MEC may get modified over the time in the partial model and its value gets recomputed. Thus stabilization of the MECs and computing their values may take more time for blackbox learning even when greybox update equations are used.

\subsection{Number of samples required for CTMDP learning}
\label{app:numberSamplesCTMDP2}
Let $\symbolMpInconfidence=0.1$, $\pmin=0.05$ and number of state-action pairs is $1000$. Let $\delta_{MP1} = \left(\frac{1}{\pmin + 1} \cdot \symbolMpInconfidence \right)$ and $\delta_{MP2} =\left(\frac{\pmin}{\pmin + 1} \cdot \symbolMpInconfidence \right)$. From this we get $\delta_{MP1} = 0.09523$ and $\delta_{MP2} = 0.00476$. We already know that $\symbolTpInconfidence := \dfrac{{\symbolMpInconfidence}_{1}\cdot \pmin}{\lvert\{{a} \vert\, \state \in \widehat{S} \wedge {a} \in \av(s) \}\rvert}$, and $\delta_R := \dfrac{{\symbolMpInconfidence}_{2}}{\lvert\{{a} \vert\, \state \in \widehat{S} \wedge {a} \in \av(s) \}\rvert}$. 
Substituting the known values we get $\symbolTpInconfidence = \frac{0.09523 \cdot 0.05}{1000} = 4 \cdot 10^{-6}$ and $\delta_R = \frac{0.00476}{1000} = 4 \cdot 10^{-6}$. From Table~\ref{tab:nSamplesTable}, the number of samples required to ensure $\alpha_{R}=3\%$, on the mean of the exponential distribution on a state-action pair is only around $60,000$. Using Hoeffding's bound we have that $\symbolTpImprecision \geqslant \sqrt{\dfrac{\ln \symbolTpInconfidence}{-2\#(s, a)}}$ on $\trans(s,a,t)$. From this we get the number of samples required to ensure $\symbolTpImprecision=0.01$ is equal to $124292$.\todo{For one good benchmark, add that $\epsilon_{MP}$ that we obtain, and the number of samples collected after 30 minutes.}

\subsection{More plots of MDP Benchmarks}
\label{app:more_plots_mdp_benchmarks}
Here we give the plots for the remaining MDP benchmarks showing the convergence of the upper and lower values with time.

\begin{figure}[!htb]
    \centering
    \begin{minipage}{.5\textwidth}
        \centering
        \includegraphics[width=1\linewidth]{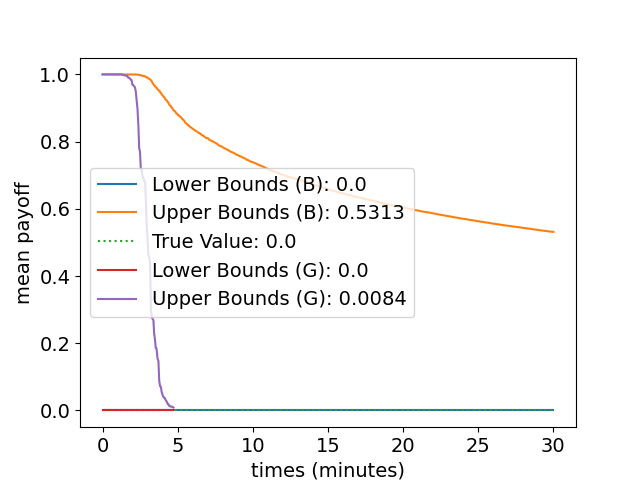}
        \caption{virus}
        \label{fig:virus}
    \end{minipage}%
    \begin{minipage}{0.5\textwidth}
        \centering
        \includegraphics[width=1\linewidth]{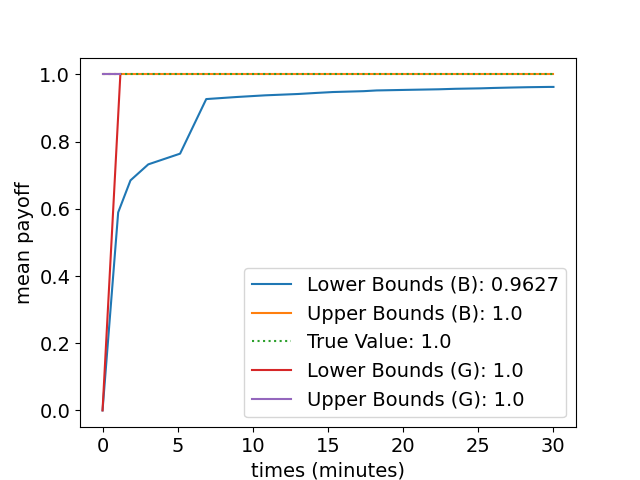}
        \caption{busyRingMC}
        \label{fig:busyRingMC4}
    \end{minipage}
\end{figure}

\begin{figure}[!htb]
    \centering
    \begin{minipage}{.5\textwidth}
        \centering
        \includegraphics[width=1\linewidth]{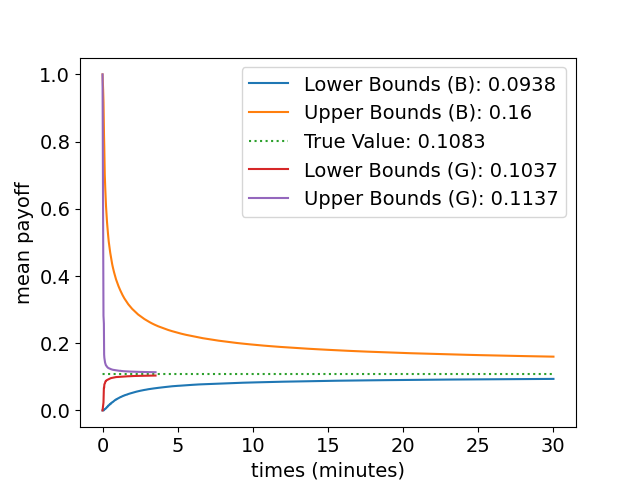}
        \caption{consensus}
        \label{fig:consensus2}
    \end{minipage}%
    \begin{minipage}{0.5\textwidth}
        \centering
        \includegraphics[width=1\linewidth]{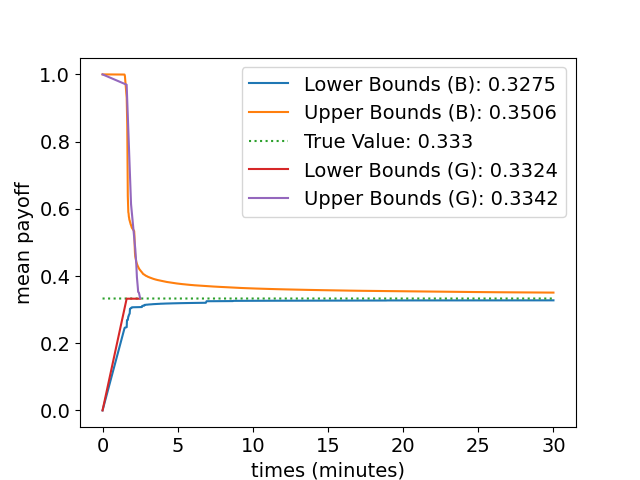}
        \caption{cs\_nfail}
        \label{fig:cs_nfail3}
    \end{minipage}
\end{figure}

\begin{figure}[!htb]
    \centering
    \begin{minipage}{.5\textwidth}
        \centering
        \includegraphics[width=1\linewidth]{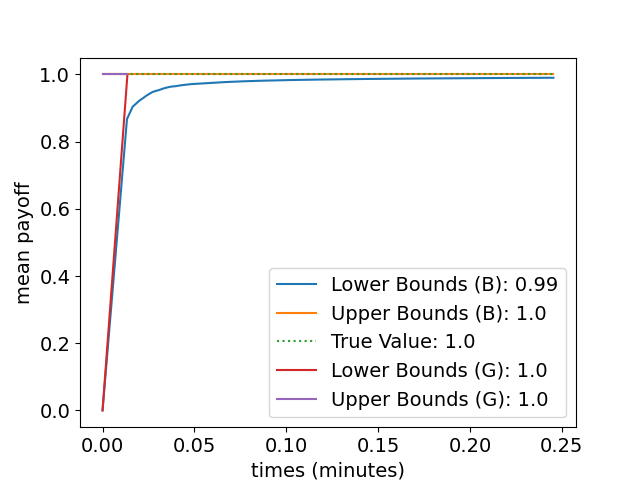}
        \caption{ij3}
        \label{fig:ij3}
    \end{minipage}%
    \begin{minipage}{0.5\textwidth}
        \centering
        \includegraphics[width=1\linewidth]{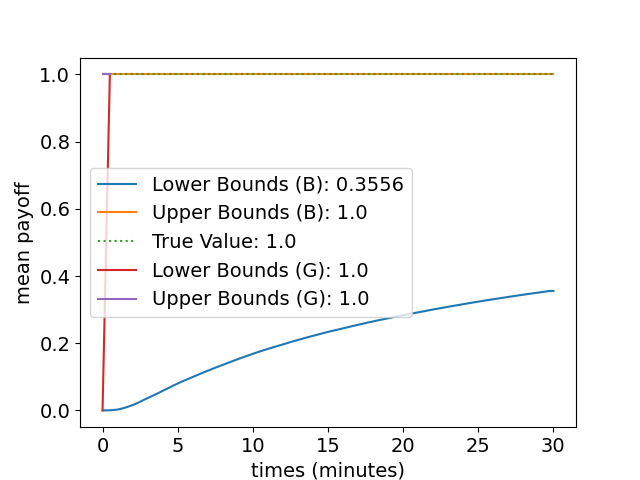}
        \caption{ij10}
        \label{fig:ij10}
    \end{minipage}
\end{figure}

\begin{figure}[!htb]
    \centering
    \begin{minipage}{.5\textwidth}
        \centering
        \includegraphics[width=1\linewidth]{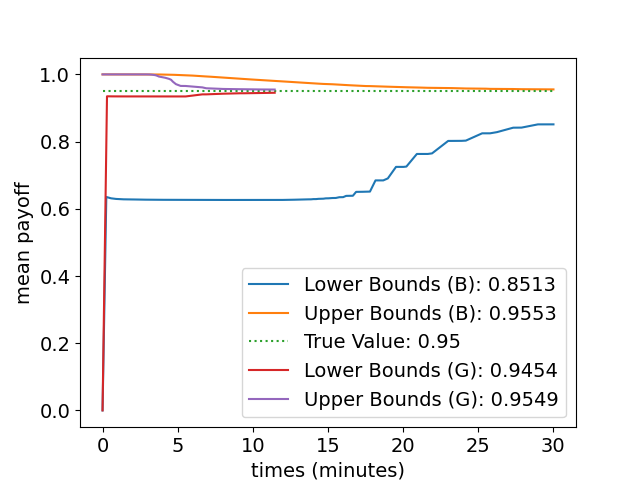}
        \caption{investor}
        \label{fig:investor}
    \end{minipage}%
    \begin{minipage}{.5\textwidth}
        \centering
        \includegraphics[width=1\linewidth]{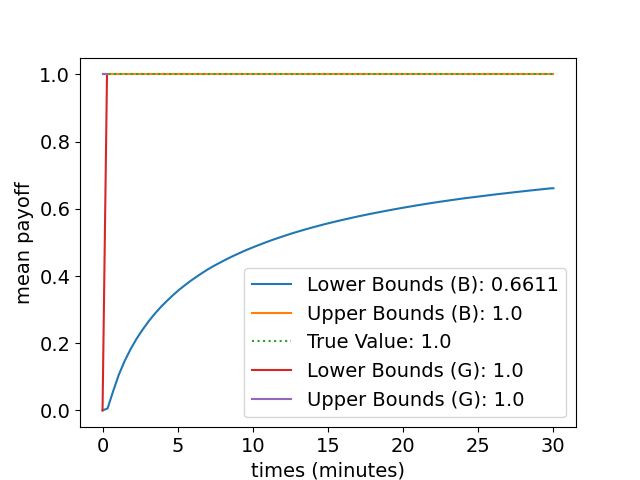}
        \caption{wlan0}
        \label{fig:wlan0}
    \end{minipage}%
\end{figure}

\begin{figure}[!htb]
    \begin{minipage}{0.5\textwidth}
        \centering
        \includegraphics[width=1\linewidth]{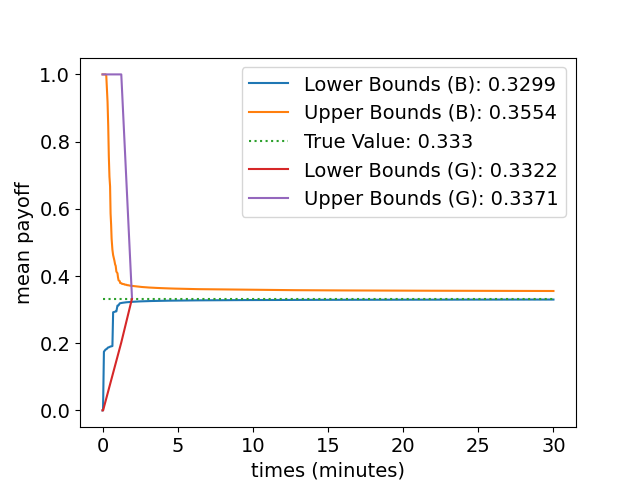}
        \caption{sensors}
        \label{fig:sensors}
    \end{minipage}
    \begin{minipage}{0.5\textwidth}
        \centering
        \includegraphics[width=1\linewidth]{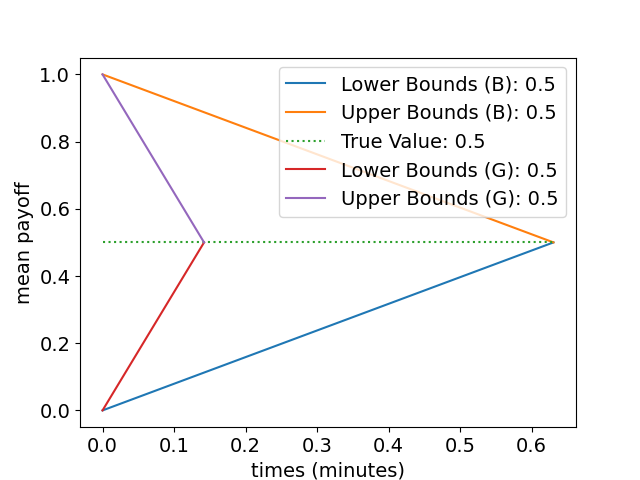}
        \caption{counter}
        \label{fig:counter}
    \end{minipage}
\end{figure}

\begin{figure}[!htb]
    \centering
    \begin{minipage}{.5\textwidth}
        \centering
        \includegraphics[width=1\linewidth]{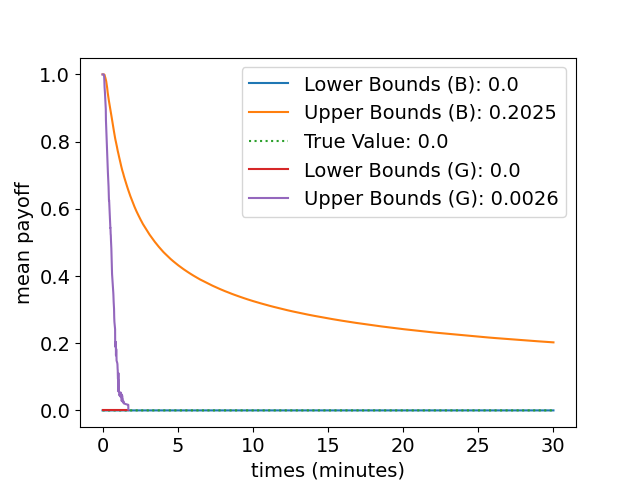}
        \caption{blackjack}
        \label{fig:blackjack}
    \end{minipage}%
\end{figure}

\begin{figure}[!htb]
    \centering
    \begin{minipage}{.5\textwidth}
        \centering
        \includegraphics[width=1\linewidth]{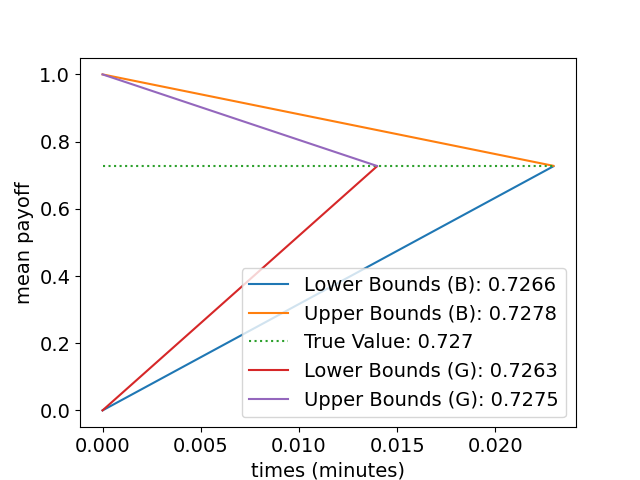}
        \caption{recycling}
        \label{fig:recycling}
    \end{minipage}%
    \begin{minipage}{0.5\textwidth}
        \centering
        \includegraphics[width=1\linewidth]{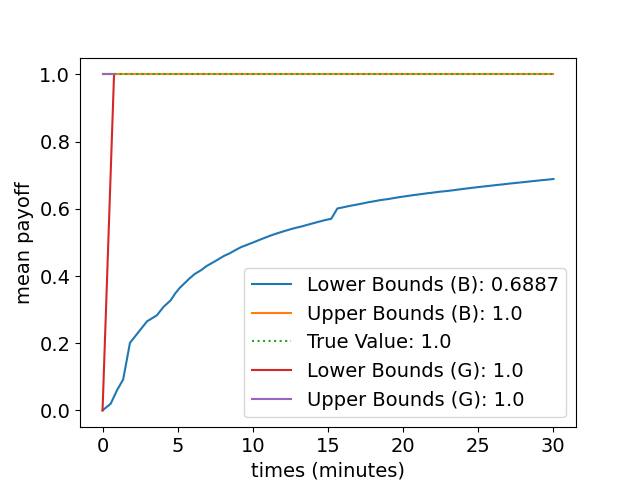}
        \caption{busyRing}
        \label{fig:busyRing4}
    \end{minipage}
\end{figure}

\FloatBarrier
\subsection{More plots of CTMDP benchmarks}
Here we give the plots for the remaining CTMDP benchmarks showing the convergence of the upper and lower values with time.

\label{app:more_plots_ctmdp_benchmarks}

\begin{figure}[!htb]
    \centering
    \begin{minipage}{.5\textwidth}
        \centering
        \includegraphics[width=1\linewidth]{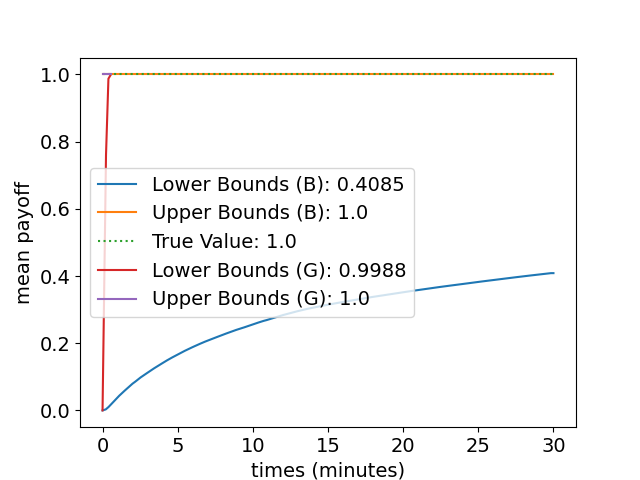}
        \caption{DynamicPM}
        \label{fig:DynamicPM-tt_3_qs_2}
    \end{minipage}%
    \begin{minipage}{0.5\textwidth}
        \centering
        \includegraphics[width=1\linewidth]{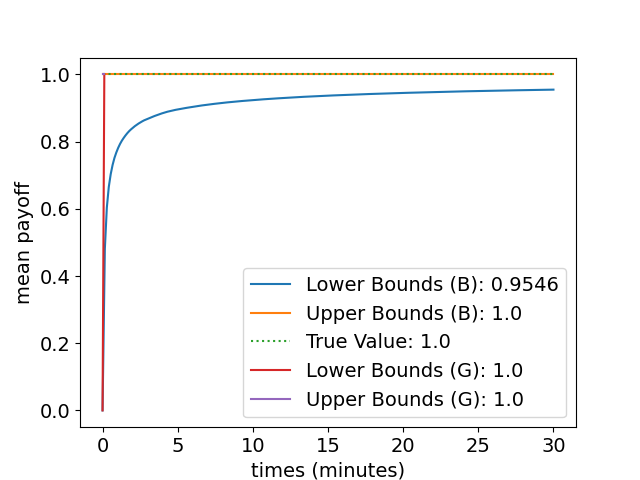}
        \caption{ErlangStages}
        \label{fig:ErlangStages-k500_r10}
    \end{minipage}
\end{figure}

\begin{figure}[!htb]
    \centering
    \begin{minipage}{.5\textwidth}
        \centering
        \includegraphics[width=1\linewidth]{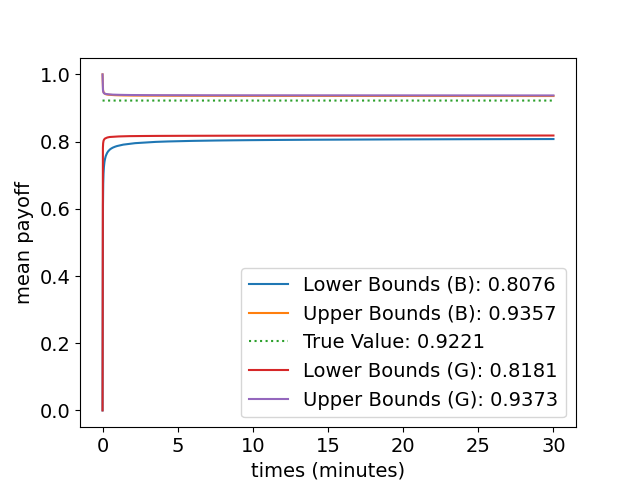}
        \caption{PollingSystem1}
        \label{fig:PollingSystem-jt1_qs1_sctmdp}
    \end{minipage}%
    \begin{minipage}{0.5\textwidth}
        \centering
        \includegraphics[width=1\linewidth]{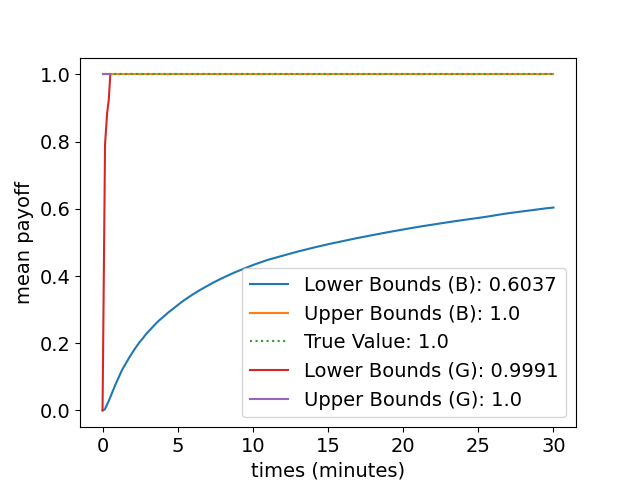}
        \caption{PollingSystem2}
        \label{fig:PollingSystem-jt1_qs4_sctmdp}
    \end{minipage}
\end{figure}

\begin{figure}[!htb]
    \centering
    \begin{minipage}{.5\textwidth}
        \centering
        \includegraphics[width=1\linewidth]{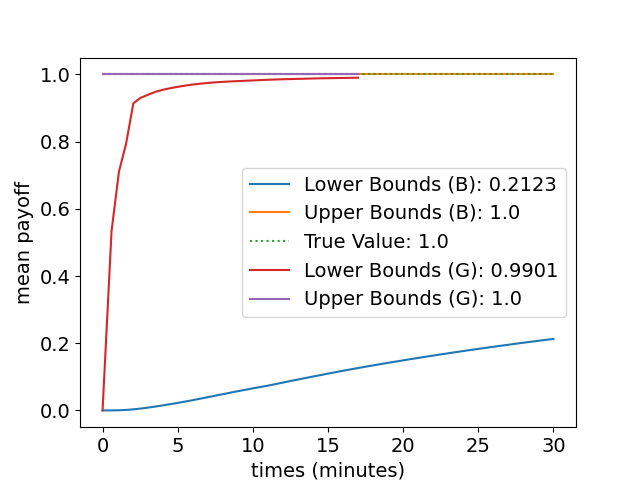}
        \caption{PollingSystem3}
        \label{fig:PollingSystem-jt1_qs7_sctmdp}
    \end{minipage}%
    \begin{minipage}{0.5\textwidth}
        \centering
        \includegraphics[width=1\linewidth]{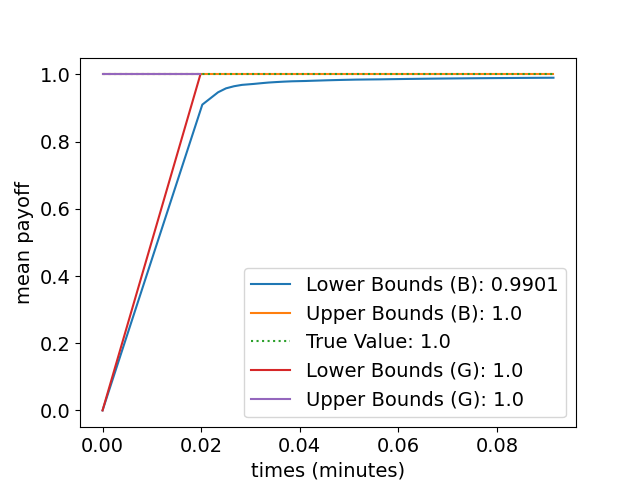}
        \caption{toy}
        \label{fig:toy}
    \end{minipage}
\end{figure}

\begin{figure}[!htb]
    \centering
    \begin{minipage}{.5\textwidth}
        \centering
        \includegraphics[width=1\linewidth]{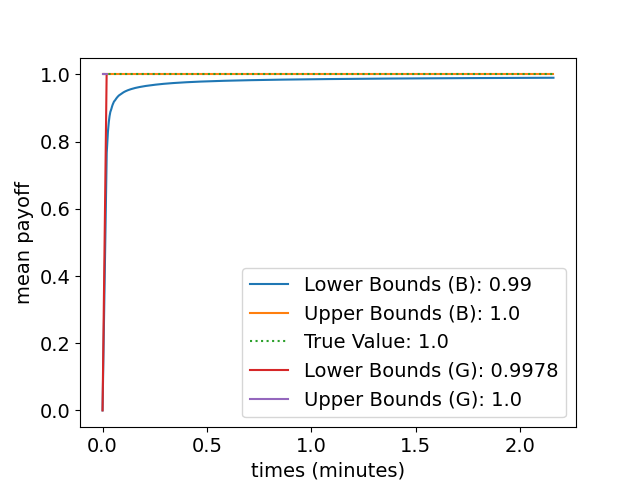}
        \caption{SJS1}
        \label{fig:SJS-procn_2_jobn_2_sctmdp}
    \end{minipage}%
    \begin{minipage}{0.5\textwidth}
        \centering
        \includegraphics[width=1\linewidth]{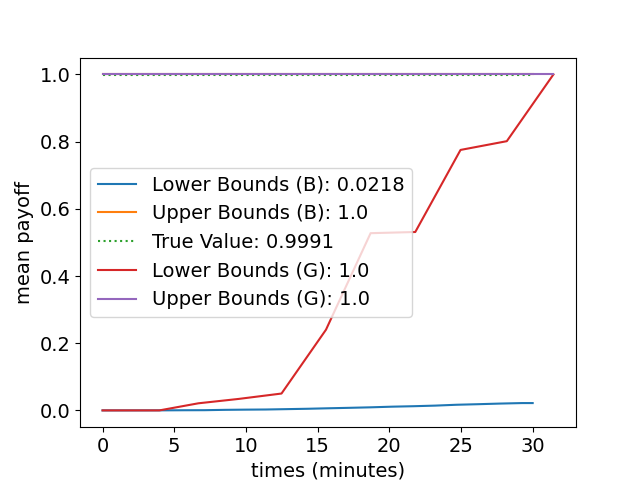}
        \caption{SJS2}
        \label{fig:SJS-procn_2_jobn_6_sctmdp}
    \end{minipage}
\end{figure}

\end{document}